\documentclass[acmsmall,screen,nonacm]{acmart}

\usepackage{enumitem}
\usepackage{listings}
\usepackage{amsmath}
\usepackage{subcaption}
\usepackage{wrapfig}
\usepackage{stmaryrd}
\usepackage{tikz}
\usetikzlibrary{positioning,fit,arrows.meta}
\usepackage{quantikz}

\usepackage{pifont}
\newcommand{\cmark}{\ding{51}}%
\newcommand{\xmark}{\ding{55}}%

\newcommand{\polylog}{\mathrm{polylog}}
\newcommand{\comp}[1]{{[|}\,#1\,{|]}}
\newcommand{\mmul}{\mathbin{\times}}
\usepackage{mathpartir}
\usepackage{placeins}
\newcommand{\kw}[1]{\ensuremath{\mathtt{#1}}}

\newcommand{\den}[1]{\mathcal{D}_{\rho}(#1)}

\usepackage[most]{tcolorbox}
\usepackage{xcolor}
\usepackage{longtable}
\usepackage{array}

\usepackage{forest}
\colorlet{tyindex}{blue!65!black}
\colorlet{tyvalue}{orange!75!black}
\colorlet{tybool}{green!45!black}
\newcommand{\tI}{\ensuremath{\textcolor{tyindex}{\texttt{Index}}}}
\newcommand{\tV}{\ensuremath{\textcolor{tyvalue}{\texttt{Value}}}}

\colorlet{mbtype}{teal!60!black}
\lstdefinelanguage{marble}{
  morekeywords=[1]{param,const,type,table,def,use,library,as},
  morekeywords=[2]{sum,prod,dsum,kron,if,then,else,mod,div,xor,and,or,not},
  morekeywords=[3]{Range,Qubits,Unit},
  morekeywords=[4]{Index},
  morekeywords=[5]{Value},
  sensitive=true,
  morecomment=[l]{\#},
}
\lstdefinestyle{marblecode}{
  language=marble,
  basicstyle=\ttfamily\footnotesize,
  columns=fixed, basewidth=0.5em, keepspaces=true, showstringspaces=false, upquote=true,
  keywordstyle=[1]\bfseries,
  keywordstyle=[2]\bfseries,
  keywordstyle=[3]\color{mbtype},
  keywordstyle=[4]\color{tyindex},
  keywordstyle=[5]\color{tyvalue},
  emph={Digit,Trit,Velocity}, emphstyle=\color{mbtype},
  commentstyle=\itshape\color{gray!60!black},
  escapeinside={(*}{*)},
  aboveskip=0pt, belowskip=0pt,
}
\newtcblisting[auto counter]{marble}[2][]{
  listing only, enhanced, listing options={style=marblecode},
  colback=black!2, colframe=black!22, boxrule=0.4pt, arc=2pt,
  left=2pt, right=2pt, top=0pt, bottom=0pt,
  before skip=8pt, after skip=8pt, after app={\noindent},
  #1
}

\tcbset{marble rejected/.style={
  colframe=red!35, coltitle=red!60!black, boxed title style={colframe=red!35},
  listing and comment, segmentation style={red!35, dashed},
  comment={\ttfamily\scriptsize\color{red!70!black}#1}}}

\usepackage{pgfplots}
\usepgfplotslibrary{groupplots}
\usetikzlibrary{calc,arrows.meta}
\pgfplotsset{compat=1.18}
\definecolor{oiblue}{HTML}{0072B2}\definecolor{oiverm}{HTML}{D55E00}\definecolor{oigreen}{HTML}{009E73}
\definecolor{oipurple}{HTML}{CC79A7}\definecolor{oisky}{HTML}{56B4E9}
\pgfplotsset{every axis/.append style={font=\footnotesize, tick label style={font=\scriptsize},
  label style={font=\footnotesize}, legend style={font=\scriptsize, draw=none, fill=none}, axis line style={gray!70},
  every tick/.style={gray!70}, major grid style={gray!20}}}

\begin{document}

\title{Compiling Matrix Formulas to Quantum Block Encodings}

\author{Daksh Shami}
\orcid{0009-0008-5938-7621}
\email{dshami@wisc.edu}

\author{Charles Yuan}
\orcid{0000-0002-4918-4467}
\email{charlesyuan@cs.wisc.edu}
\affiliation{
  \institution{University of Wisconsin-Madison}
  \country{USA}
}

\settopmatter{}
\authorsaddresses{}

\begin{abstract}
Realizing the promise of quantum applications such as chemistry or fluid dynamics requires building quantum circuits that efficiently encode and manipulate exponentially large matrices.
This process forces researchers to grapple with the reality that a matrix expressible as one equation on paper can map to thousands of bit-level quantum logic gates, historically written by hand.
Even worse, it often ends with a discovery that the needed resources grow so quickly with matrix size that the quantum algorithm loses its asymptotic speedup.

In this work, we present MARBLE, a language that automatically synthesizes an efficient circuit from the publication-style formula for the entries of a matrix.
MARBLE algebraically rewrites programs into symbolic combinations of simpler matrices that have known circuit realizations.
It analyzes costs of possible representations, notably the subnormalization term that dominates time complexity.
It then emits an OpenQASM circuit and certifies whether its cost is poly-logarithmic in problem size, as required by quantum algorithms.

When evaluated on quantum applications ranging from fluid dynamics to physical Hamiltonians, MARBLE matches or surpasses 37 of 39 published baselines.
On the low end, it automatically matches many theoretically optimal circuits, and on the high end, it improves on a published construction in fluid dynamics by 10,487$\times$ in subnormalization.
MARBLE produces exponentially smaller circuits than the leading tool FABLE, and it successfully compiles matrices whose dimensions exceed $2^{100}$ within five seconds on a laptop.
\end{abstract}

\maketitle

\section{Introduction}
\label{sec:intro}

Quantum computers promise exponential speedups for problems in linear algebra.
Since $n$ qubits hold a vector of $2^n$ entries, quantum algorithms can invert a matrix~\cite{harrowQuantumAlgorithmLinear2009} or simulate a Hamiltonian~\cite{childsHamiltonianSimulation2012} in time poly-logarithmic in the size of the system.
This concept forms the basis for applications across chemistry~\cite{babbushEncodingElectronicSpectra2018}, biology~\cite{goingsReliablyAssessingElectronic2022}, fusion energy~\cite{vaszarySolvingNonlinearVlasov2026}, and fluid dynamics~\cite{liuEfficientQuantumAlgorithm2021}.

As quantum hardware moves closer to fault tolerance~\cite{bluvsteinFaulttolerantNeutralatomArchitecture2026,google2025}, realizing the promised speedups in practice becomes a problem of programming.
These algorithms operate over matrices that are exponentially large: with dimension >$10^{25}$ in an algorithm for fluid dynamics~\citep{penuelDetailedAssessmentCalculating2025}.
The algorithm cannot simply read such a large matrix entry by entry, because data loading would take as long as solving the problem classically.
Instead, to avoid this input bottleneck, algorithms access the matrix via the concept of \emph{block encoding}.

\paragraph{Block Encodings}
Every quantum circuit has a unitary matrix as its semantics, which enables the matrix to be encoded implicitly by that circuit.
Since most matrices $A$ of interest are not unitary, a \emph{block encoding} is defined as a circuit whose matrix contains $A$ in its top-left block.
This construction lets an algorithm apply $A$ to a quantum state, and circuits to perform arithmetic on block encodings form the basis of now-celebrated quantum algorithms~\cite{gilyenQuantumSingularValue2019,martynGrandUnificationQuantum2021}.

Many algorithms treat a block encoding of their input matrix as given.
For example, \citet{kalantzisEfficientSolutionsGeneral2025} present quantum algorithms for structured Markov processes whose speedup assumes an \emph{efficient} block encoding of the input matrix --- one whose circuit grows only poly-logarithmically with the size of the matrix.
The paper provides no explicit quantum circuit for this matrix.

\paragraph{Structure}
To encode a matrix using asymptotically fewer gates than its dimension, a circuit must exploit its structure.
The field has long measured the structure of a matrix by its sparsity~\cite{nguyenBlockencodingDenseFullrank2022,campsExplicitQuantumCircuits2024}.
Indeed, approaches such as FABLE~\citep{campsFABLEFastApproximate2022} that enumerate the entries of dense matrices yield asymptotically inefficient circuits.
Sparsity is a default assumption made by programming frameworks such as Qualtran~\cite{harriganExpressingAnalyzingQuantum2024} and Qrisp~\cite{petricBlockencodingsProgrammingAbstractions2026}, which provide pre-built circuits for specific sparse matrices, and Cobble~\citep{yuanCobbleCompilingBlock2026}, which can compose block encodings but not encode arbitrary matrices.

Sparsity, however, is ultimately an imprecise proxy for existence of an efficient block encoding.
For instance, a diagonal matrix containing all random entries has only one nonzero entry per row, yet requires one gate to load each entry --- as many as the matrix's size.
On the other hand, there exist dense matrices, for example the quantum Fourier transform, that have efficient circuits.

We propose a different metric for structure in a matrix via its \emph{multiplicity}: the number of distinct formulas to compute the entries, not the number of nonzero entries.
Indeed, the quantum Fourier transform has constant multiplicity --- every entry is computed by a uniform function of the row and column index, whereas a random diagonal matrix loads a different constant every row and thus has linear multiplicity.
Notably, multiplicity is a syntactic property, which a compiler can check from the definition of a matrix, and a language can constrain for programs by construction.

\paragraph{Matrix Compilation}
Leveraging multiplicity as the source of structure, we introduce MARBLE (\underline{M}atrix \underline{AR}ithmetic \underline{BL}ock \underline{E}ncodings), a language and compiler to turn the research paper-style formula for the entries of a matrix into an efficient quantum circuit for its block encoding.

MARBLE achieves for block encodings what NumPy does for classical arrays: the programmer writes the mathematics, and the system supplies the implementation.
A program in MARBLE is an \emph{entry comprehension} that defines each entry of a matrix as a formula of its row and column indices.
The type checker guarantees that every entry is computed by a fixed number of distinct operations over indices, however large the matrix --- a property we call \emph{static multiplicity}.
The compiler then lowers these computations to circuits without ever enumerating the entries of the matrix.

\paragraph{Subnormalization}
To find the cost of the resulting circuit, one can count its gates and qubits.
But equally important is its \emph{subnormalization}, the number of times that the quantum algorithm must repeat the circuit.
Many matrices in quantum applications~\citep{rubin2024} incur relatively few gates per iteration but also a large subnormalization that dominates the running time.
Computing subnormalization requires detailed accounting for each matrix.
Even worse, a matrix may possess multiple constructions whose subnormalizations differ by orders of magnitude~\citep{yuanCobbleCompilingBlock2026}.
The prior tools above cannot determine whether a given block encoding is close to optimal for a matrix, or whether any block encoding exists to realize the speedup of the quantum algorithm at all.

\paragraph{Optimization and Cost Certification}
MARBLE automatically computes and reduces the subnormalization of the program.
It uses algebraic identities to rewrite the program into \emph{symbolic matrix forms}, expressions that combine simpler matrices with known circuits.
Under the static multiplicity property, these rewrites act only on compact expressions whose size is independent of the matrix dimension rather than huge circuits.
Upon emitting the circuit, MARBLE determines whether its per-iteration gate and qubit counts are efficient, how close the achieved subnormalization is to the theoretical optimum, and whether that optimal subnormalization scales poly-logarithmically.

\paragraph{Evaluation}
We evaluate MARBLE on 74 matrices drawn from applications such as fluid dynamics, partial differential equations, and Hamiltonians in physics.
MARBLE matches or surpasses 37 of 39 published circuit baselines in subnormalization --- ranging from automatically rediscovering known theoretically optimal constructions to a $10{,}487\times$ improvement for one fluid dynamics application.
MARBLE also constructs explicit block encodings for structured Markov processes, such as those assumed by \citet{kalantzisEfficientSolutionsGeneral2025} --- the first explicit circuits to our knowledge.
Compared to the general-purpose tool FABLE, MARBLE produces exponentially smaller circuits in significantly less time; it compiles matrices whose dimensions exceed $2^{100}$ in under five seconds on a laptop.

\paragraph{Contributions}
In this work, we present the following contributions:
\begin{enumerate}[labelwidth=!, labelsep=0.5em, leftmargin=3em, itemindent=0pt, listparindent=\parindent, align=parleft]
  \item[Sec.~\ref{sec:language}:] MARBLE, a language for expressing matrices as entry comprehensions, whose type system ensures the static multiplicity property that gives rise to efficient block encodings;
  \item[Sec.~\ref{sec:compilation}:] A compiler that rewrites a program into symbolic matrix forms via sound algebraic rewrite rules, computes costs such as subnormalization, and emits an efficient quantum circuit;
  \item[Sec.~\ref{sec:certification}:] An analysis that symbolically certifies whether the block-encoding circuit generated for a program is asymptotically efficient, both in gate count and in subnormalization; and
  \item[Sec.~\ref{sec:evaluation}:] An evaluation on 74 matrices from the quantum algorithms literature, in which MARBLE matches or surpasses 37 of 39 published baseline circuits in subnormalization.
\end{enumerate}

Our technique for compiling matrices via their structure as programs sheds light on the long-standing problem of loading data into a quantum computer.
Using MARBLE, algorithm designers can estimate hardware requirements of proposed algorithms, quickly find whether an application may achieve a practical speedup, and prototype matrix representations for new applications.

\section{Mathematical Background}
\label{sec:background}

In this section, we review the necessary concepts from quantum algorithms for linear algebra.
We omit definitions from introductory quantum information, including qubits, quantum logic gates, and measurement; \citet{nielsenQuantumComputationQuantum2010} explain these fundamental concepts in detail.

\paragraph{Encodings}
Quantum algorithms for linear algebra rely on the central idea that a quantum state encodes an exponentially large vector, and similarly a quantum circuit encodes an exponentially large matrix.
Based on these encodings, quantum algorithms construct arithmetic operations over matrices that can compute polynomials~\citep{gilyenQuantumSingularValue2019} and hence approximations of arbitrary functions on encoded matrices, including matrix inversion and Hamiltonian simulation.

\begin{definition}
\label{def:amplitude-encoding}
Given a vector $x \in \mathbb{C}^N$ with $N = 2^n$, its \emph{amplitude encoding} is the $n$-qubit state
{\setlength{\belowdisplayskip}{0pt} \setlength{\belowdisplayshortskip}{0pt}
\setlength{\abovedisplayskip}{4pt} \setlength{\abovedisplayshortskip}{0pt}
\[
  |x\rangle = \frac{1}{\|x\|} \sum_{j=0}^{N-1} x_j\, |j\rangle .
\]}%
\end{definition}

In general, encoding a vector with arbitrary entries takes time linear in $N$~\citep{babbushEncodingElectronicSpectra2018}.
However, a vector with structure can be far cheaper to encode than an arbitrary one.
For example, applying $H$ to each of $n$ qubits prepares the normalized all-ones vector $x_j = 1$, which corresponds to a uniform superposition $\smash{H^{\otimes n} |0\rangle^{\otimes n} = \frac{1}{\sqrt{N}} \sum_{j=0}^{N-1} |j\rangle}$ and takes only $n$ gates.
Thus, it is often the structure of the vector, rather than its dimension, that gives the cost of amplitude encoding.

Encoding a matrix is less straightforward, because a quantum circuit can implement only unitary matrices, whereas most matrices relevant to applications are not unitary.
To solve this problem, such a matrix must instead be embedded as a component of a larger unitary matrix:
\begin{definition}
\label{def:block-encoding}
Given a matrix $A \in \mathbb{C}^{N \times N}$, a \emph{block encoding} of $A$ is an $(m+n)$-qubit unitary
\[
  U = \begin{bmatrix} A/\alpha & \cdot \\ \cdot & \cdot \end{bmatrix},
\]
whose top-left $N \times N$ block is $A$ divided by a scalar $\alpha > 0$ that is called the \emph{subnormalization}.
\end{definition}

In the definition of $U$, the values other than the top-left corner can be arbitrary as long as $U$ is unitary.
Equivalently, given any $n$-qubit state $|x\rangle$, $U$ acts on $\ket{x}$ plus a set of ancilla qubits by
\begin{equation}
\label{eq:block-encoding-action}
  U \bigl( |0\rangle^{\otimes m} \otimes |x\rangle \bigr) = |0\rangle^{\otimes m} \otimes \tfrac{1}{\alpha} A |x\rangle + |\bot\rangle ,
\end{equation}
where $|\bot\rangle$ has zero amplitude on every basis vector whose $m$ ancilla qubits are all $|0\rangle$.

\paragraph{Subnormalization}
A matrix $A$ can only fit inside a unitary $U$ if its spectral norm $\sigma_1 = \|A\|_2$ is no greater than 1.
Otherwise, the subnormalization $\alpha \ge \sigma_1$ shrinks $A$ to size.
For example, the all-ones matrix dilates the all-ones vector by a factor of $N$, so it fits only after dividing by $\alpha \ge N$.

But subnormalization incurs a cost --- it shrinks the probability of observing the desired $A|x\rangle/\alpha$ of Eq.~\eqref{eq:block-encoding-action} and not the unwanted $\ket{\bot}$.
To discard $\ket{\bot}$, one measures the $m$ ancillas and keeps the result
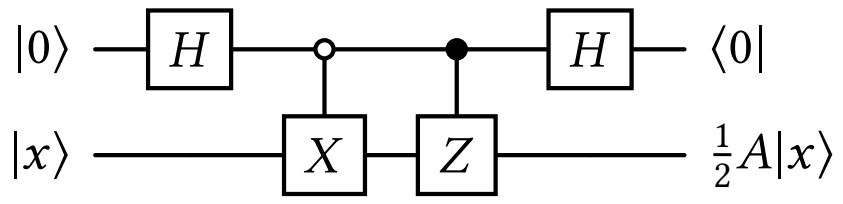
\begin{wrapfigure}{r}{0.41\textwidth}
\vspace{-3ex}%
\centering
\begin{quantikz}[row sep={0.7cm,between origins}, column sep=0.35cm]
\lstick{$|0\rangle$} & \gate{H} & \octrl{1} & \ctrl{1} & \gate{H} & \rstick{$\langle 0|$} \\
\lstick{$|x\rangle$} & & \gate{X} & \gate{Z} & & \rstick{$\tfrac{1}{2} A |x\rangle$}
\end{quantikz}
\setlength{\abovecaptionskip}{0.5ex}%
\caption{Block encoding of $A = X + Z$.}
\label{fig:lcu-circuit}%
\vspace{-2.5ex}%
\end{wrapfigure}%
only when all $m$ qubits are 0, a step called \emph{post-selection}.
Invoking oblivious amplitude amplification~\citep{berry2014}, the expected number of trials grows as $O(\alpha)$.
In simpler words, the quantum algorithm runs in time proportional to the subnormalization $\alpha$~\citep{yuanCobbleCompilingBlock2026}.

\begin{example}
\label{ex:lcu}
The matrix $A = X + Z$ is not unitary, but the circuit in Figure~\ref{fig:lcu-circuit} block-encodes it using one ancilla qubit.
Hollow $\circ$ denotes control on $|0\rangle$, filled $\bullet$ controls on $|1\rangle$, and $\langle 0|$ post-selects for 0.
This \emph{linear combination of unitaries}~\citep{childsHamiltonianSimulation2012} implements the matrix
\[
  U = \tfrac{1}{2} \begin{bmatrix} X + Z & X - Z \\ X - Z & X + Z \end{bmatrix},
\]
whose top-left block is $A / 2$.
Thus, $U$ is a block encoding of $A$ with subnormalization $\alpha = 2$.
\end{example}

\section{Overview and Examples}
\label{sec:overview}

In this section, we walk through the problem of encoding matrices into a quantum computer.
Using examples drawn from the quantum algorithms literature, we illustrate how the core concepts in the MARBLE language enable developers to express matrices using familiar mathematical notation, reduce the required hardware resources, and synthesize correct and efficient quantum circuits.

\subsection{Encoding Matrices into a Quantum Computer}

Quantum algorithms for scientific applications including chemistry~\cite{babbushEncodingElectronicSpectra2018}, biology~\cite{goingsReliablyAssessingElectronic2022}, fusion energy~\cite{vaszarySolvingNonlinearVlasov2026}, and fluid dynamics~\cite{liuEfficientQuantumAlgorithm2021} center on linear algebra: using a quantum computer to solve linear systems, differential equations, and eigenvalue problems.
An algorithm typically presents a matrix, represented as a mathematical formula, that must be encoded within the semantics of a quantum circuit of logic gates.

\paragraph{Running Example}
\citet{penuelDetailedAssessmentCalculating2025} propose a quantum algorithm to simulate fluid flow for ship hull design via the lattice Boltzmann method, which simulates a fluid as particles of a lattice grid as they move and collide over time.
In this process, a dense matrix called $F_1$ governs how the particles in a cell redistribute.
Its entries are given by \citet[Eqs. (10)--(11)]{kuklinskiEfficientBlockencodingsRequire2025} as
\begin{equation} \label{eq:f1}
  F_1[i, j] = w(i) \, (1 + 3\, c(i) \cdot c(j)) - \delta_{ij} \quad \text{for} \quad i, j \in \{0, 1, \dots, 26\},
\end{equation}
where in full detail: $\delta_{ij} = 1$ iff $i=j$, $c(i)$ is the $i$-th velocity vector of a $3 \times 3 \times 3$ cubic cell, and $w(i)$ is its weight.
Each index $i = (i_1, i_2, i_3) \in \{0, 1, 2\}{}^3$ decomposes into a trit for each dimension of velocity. Each trit picks a component from $\bar{c} = [1, -1, 0]$ and a weight factor from $\bar{w} = [\frac{1}{4}, \frac{1}{4}, 1]$:
\[
  c(i) = [\bar{c}_{i_1}, \bar{c}_{i_2}, \bar{c}_{i_3}], \qquad w(i) = \bar{w}_{i_1}\, \bar{w}_{i_2}\, \bar{w}_{i_3}.
\]

\paragraph{Representation}
The matrix above handles one lattice cell.
To simulate fluid flow in a grid around a full-scale ship hull, the needed $F_1$ matrix would have dimension on the order of $10^{25}$~\citep[Table 9]{penuelDetailedAssessmentCalculating2025}, which takes more than one million exabytes if stored on a classical computer.

A quantum computer can in principle represent this matrix in exponentially less space --- a matrix of dimension $10^{25}$ requires only a hundred data qubits.
The challenge is that the programmer must painstakingly construct a quantum circuit whose semantics implicitly represent, or \emph{block-encode}, the matrix.
Figure~\ref{fig:f1-kuklinski} depicts a simplified version of the circuit that \citet{kuklinskiEfficientBlockencodingsRequire2025} construct to block-encode the one-cell $F_1$ matrix from above.
Even for a $27 \times 27$ matrix, the full circuit contains hundreds of gates, and its expert construction is non-obvious when compared to Equation~\eqref{eq:f1}.

\colorlet{padfill}{gray!15}
\colorlet{prepfill}{blue!10}
\colorlet{termfill}{orange!14}
\colorlet{basisfill}{green!12}
\tikzset{
  roledraw/.style={draw=gray!55!black},
  padblock/.style={roledraw, fill=padfill, inner ysep=1pt},
  prepblock/.style={roledraw, fill=prepfill, minimum height=0.4cm},
  termblock/.style={roledraw, fill=termfill, inner ysep=1pt},
  basisblock/.style={roledraw, fill=basisfill},
  stage/.style={draw=none, inner xsep=0pt, inner ysep=2pt},
  stagebelow/.style={label position=below, anchor=north, yshift=2pt, font=\scriptsize, text=gray!45!black},
}
\providecommand{\blk}[2]{\text{\smash{\raisebox{-2.5pt}{\scriptsize\begin{tabular}{@{}c@{}}$#1$\tabularnewline[-1pt]{\tiny\sffamily\color{gray!35!black}#2}\end{tabular}}}}}
\newcommand{\stg}[1]{\text{\scriptsize\begin{tabular}{@{}c@{}}#1\end{tabular}}}
\providecommand{\prep}[1]{\text{\scriptsize\scshape #1}}
\newsavebox{\boxa}\newsavebox{\boxb}
\sbox{\boxa}{%
\begin{quantikz}[row sep={0.45cm,between origins}, column sep=0.18cm]
& \qwbundle{6} & & \gate[2, style=padblock]{\blk{P}{padding}}
  & & & & \gate[2, style=padblock]{\blk{P}{padding}} & \\
& \qwbundle{6} & &
  & \gate[2, style=termblock]{\blk{W}{weights}}
  & \gate[2, style=termblock]{\blk{J}{all-ones}} & \gate[2, style=termblock]{\blk{c(i) \cdot c(j)}{dot product}} & & \\
& \qwbundle{12} & & & & & & & \\[0cm]
& & & \gate[3, style=prepblock]{\prep{prep}}
  & \octrl{-1} & \octrl{-1} & \octrl{-1} & \gate[3, style=prepblock]{\prep{prep}^{\smash{\dagger}}} & \\[-0.2cm]
& & & & & \octrl{-1} & \ctrl{-1} & & \\[-0.2cm]
& & & & & \octrl{-1} & & &
\end{quantikz}}
\sbox{\boxb}{%
\begin{quantikz}[row sep={0.45cm,between origins}, column sep=0.18cm]
& \qwbundle{3} & & & & & & & \gate[2, style=padblock]{\blk{P}{padding}} & \\
& \qwbundle{6} & & \gate[style=termblock]{\stg{rotate}}
  & \gate[2, style=termblock]{\blk{\text{rotate}}{per trit}}
  & \gate[2, style=termblock]{\blk{\text{scale}}{4 states}} & \gate[2, style=termblock]{\blk{\text{rotate}^{-1}}{per trit}}
  & \gate[style=termblock]{\stg{rotate$^{-1}$}} & & \\
& \qwbundle{10} & & & & & & & & \\[0cm]
& & & \gate[style=prepblock]{\prep{prep}}
  & \octrl{-1} & \octrl{-1} & \octrl{-1} & \gate[style=prepblock]{\prep{prep}^{\mkern-2mu\smash{\raisebox{-0.2ex}{$\scriptscriptstyle\dagger$}}}} & &
\end{quantikz}}
\pgfmathsetmacro{\fscale}{min(1, 0.99*\the\linewidth/(\the\wd\boxa+\the\wd\boxb))}
\newlength{\wda}\setlength{\wda}{\fscale\wd\boxa}
\newlength{\wdb}\setlength{\wdb}{\fscale\wd\boxb}
\newlength{\hta}\setlength{\hta}{\fscale\dimexpr\ht\boxa+\dp\boxa\relax}
\newlength{\htb}\setlength{\htb}{\fscale\dimexpr\ht\boxb+\dp\boxb\relax}
\newlength{\hmax}\setlength{\hmax}{\hta}\ifdim\htb>\hmax\setlength{\hmax}{\htb}\fi
\begin{figure}[t]
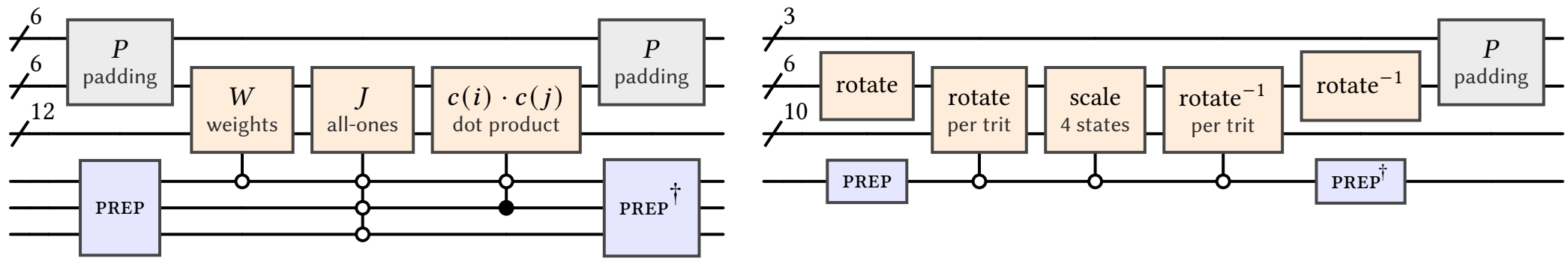

\centering
\footnotesize
\begin{subfigure}[t]{\wda}
\centering
\raisebox{-\height}[0pt][\hmax]{\scalebox{\fscale}{\usebox{\boxa}}}
\setlength{\abovecaptionskip}{0.5ex}
\subcaption{Built manually by \citet[Fig.~14]{kuklinskiEfficientBlockencodingsRequire2025}.}
\label{fig:f1-kuklinski}
\end{subfigure}\hfill
\begin{subfigure}[t]{\wdb}
\centering
\raisebox{-\height}[0pt][\hmax]{\scalebox{\fscale}{\usebox{\boxb}}}
\setlength{\abovecaptionskip}{0.5ex}
\subcaption{Synthesized automatically by MARBLE.}
\label{fig:f1-marble}
\end{subfigure}
\setlength{\abovecaptionskip}{1.5ex}%
\caption{Quantum circuits that block-encode the one-cell fluid flow $F_1$ matrix as defined by Equation~\eqref{eq:f1}.}
\label{fig:f1-circuits}
\vspace{-1pt}%
\end{figure}

To bridge the gap between mathematical notation and executable circuits, we introduce MARBLE, a programming language for matrices encoded into a quantum computer.
In the MARBLE language, a developer can realize the dense one-cell $F_1$ matrix using syntax that matches Equation~\eqref{eq:f1}:
\vspace{-.8ex}%
\begin{marble}[label={prog:f1}]{$F_1$}
type Trit = Range 3                      # a trit is 0, 1, or 2
type Velocity = Trit * Trit * Trit       # three trits, one each for x, y, and z axis
table C : Trit -> Index = (1, -1, 0)     # a trit's velocity component
table W : Trit -> Value = (1/4, 1/4, 1)  # its weight in the LBM lattice
def c(i) = (C[i.1], C[i.2], C[i.3])
def w(i) = W[i.1] * W[i.2] * W[i.3]
def dot(i, j) = i.1 * j.1 + i.2 * j.2 + i.3 * j.3
[| w(i) * (1 + 3 * dot(c(i), c(j))) - [i == j] | i, j : Velocity |]
\end{marble}

The syntax $\comp{\texttt{e} \mid \texttt{i}, \texttt{j} : \texttt{T}}$ is called an \textit{entry comprehension}.
Like a list comprehension in Python, it defines a matrix via an expression \verb+e+ for each entry at row \verb+i+ and column \verb+j+.
The syntax $\kw{Range}\ N$ means that an index takes the values $0, \ldots, N-1$.
The \verb|type| keyword defines a type alias, which can be a tuple type.
Tuples like \verb|Velocity| can be used lexicographically to index rows and columns.
The \verb|table| keyword defines an array of coefficients indexed by integers.
Built-in types $\tI$ and $\tV$ distinguish integer indices from other data values, which are complex numbers.

\paragraph{Compilation and Optimization}
Given the program above, the MARBLE compiler takes 0.13 s to synthesize the circuit in Figure~\ref{fig:f1-marble}, whose semantics is a block encoding of the $F_1$ matrix.

Moreover, the high-level abstractions of MARBLE enable the compiler to significantly reduce the hardware costs of a program.
The two runtime costs of a block encoding are its \emph{subnormalization}~$\alpha$, which governs how many iterations that the circuit must be run before it produces a correct output (see Section~\ref{sec:background}), and the numbers of logic gates and qubits per iteration of the circuit.

The handwritten baseline by \citet{kuklinskiEfficientBlockencodingsRequire2025} in Figure~\ref{fig:f1-kuklinski} has $\alpha = 146$ and uses \textasciitilde{}196 $T$ gates, 22 other non-Clifford rotation gates, and 27 qubits.
By contrast, MARBLE's optimized output in Fig.~\ref{fig:f1-marble} has $\alpha = 7.2$ and uses 60 $T$, 16 other non-Clifford rotations, and 20 qubits.
Taking total time cost to be subnormalization $\times$ non-Clifford gate count~\citep{sunderhaufBlockencodingStructuredMatrices2024}, MARBLE yields a $58.2\times$ speedup and a $25.9\%$ space reduction over the baseline.
In general, the circuits that minimize subnormalization, gate count, or their product can be distinct; MARBLE can produce them all.

\subsection{Compiling a Simple Matrix via Symbolic Form and Algebraic Rewrites}
To understand the core concepts of the language and compiler that build up $F_1$, we will first walk through the compilation of a toy example matrix, which we will call $M_\textrm{simple}$ and define as
\begin{equation} \label{eq:m_simple}
  M_\textrm{simple}[i,j] = i + j, \quad \text{where} \quad i,j \in \{0, 1, \ldots, 2^{100}-1\}.
\end{equation}

The corresponding program closely follows the notation of Equation~\eqref{eq:m_simple}:
\vspace{-.8ex}%
\begin{marble}{$M_\textrm{simple}$}
[| i + j | i, j : Range N |]
\end{marble}

\paragraph{Representation}
In principle, a programmer can build a circuit for this matrix via FABLE~\citep{campsFABLEFastApproximate2022}, which explicitly enumerates each entry of the matrix in time $O(N^2 \log N)$ and does not scale beyond \textasciitilde{}$2^{12} \times 2^{12}$ matrices.
Even if it could, the output circuit would have $O(N^2)$ gates, which is \textasciitilde{}$2^{201}$ gates for $M_\textrm{simple}$.
The underlying problem is that a quantum algorithm delivers an exponential advantage over classical methods only if its total time cost is poly-logarithmic and not polynomial in the dimension $N$ of the matrix.
If a matrix contains fully arbitrary entries, then a quantum circuit that loads each entry explicitly would nullify any quantum speedup.

For a matrix to have an $O(\polylog\ N)$-sized circuit, it should have low \emph{multiplicity} --- the number of distinct formulas to compute its entries.
For example, though $M_\textrm{simple}$ has $2^{200}$ entries, every entry is computed by $\texttt{i} + \texttt{j}$ and its multiplicity is 1.
We say that a matrix has \emph{static multiplicity} if its multiplicity is $O(1)$ independent of $N$.
Every matrix expressible in MARBLE has static multiplicity, because any entry comprehension contains a fixed number of expressions by construction.
MARBLE exploits this property to synthesize a circuit for $M_\textrm{simple}$ within 0.3 seconds.
This circuit, in Figure~\ref{fig:msimple-pieces}, contains 9,536 non-Clifford gates and 1,103 qubits, and both counts scale as $O(\polylog\ N)$.

Aside from the gate count per circuit iteration, total time cost also scales with the subnormalization $\alpha$, the number of iterations needed.
Subnormalization is lower-bounded by the spectral norm of the encoded matrix --- a property of the matrix itself rather than its implementation.
In fact, the spectral norm of $M_\textrm{simple}$ is not poly-logarithmic in $N$.
We thus use this toy example only to explain MARBLE before returning to the asymptotically efficient $F_1$ example in the next subsection.

\colorlet{prepfill}{blue!10}
\colorlet{termfill}{orange!14}
\tikzset{
  roledraw/.style={draw=gray!55!black},
  prepblock/.style={roledraw, fill=prepfill, minimum height=0.3cm, inner ysep=0pt},
  termblock/.style={roledraw, fill=termfill, inner ysep=-2pt},
  irbox/.style={draw, dotted, line width=0.8pt, rounded corners=3pt},
  irJ/.style={irbox, yshift=2pt, draw=black!50, fill=black!4, inner xsep=2.5pt, inner ysep=2pt},
  irDi/.style={irbox, yshift=2pt, draw=oiblue, fill=oiblue!6, inner xsep=2.5pt, inner ysep=2pt},
  irDj/.style={irbox, yshift=2pt, draw=oiverm, fill=oiverm!6, inner xsep=2.5pt, inner ysep=2pt},
  irprod/.style={irbox, draw=black!70, inner xsep=5.5pt, inner ysep=7.4pt, yshift=5.6pt},
  irtag/.style={label position=north west, anchor=west, xshift=5pt, yshift=-0.2cm, fill=white, inner sep=1.2pt,
    font=\scriptsize},
}
\providecommand{\prep}[1]{\text{\scriptsize\scshape #1}}
\providecommand{\blk}[2]{\text{\begin{tabular}{@{}c@{}}$#1$\tabularnewline[-1pt]{\tiny\sffamily\color{gray!35!black}#2}\end{tabular}}}
\newcommand{\irin}[2]{\makebox[1.5em][l]{\scriptsize\color{black!55}$#1$}\,$#2$\hspace{3pt}}
\newcommand{\irout}[1]{\hspace{3pt}$#1$}
\newcommand{\irprepdag}{\prep{prep}^{\mkern-2mu\smash{\raisebox{-0.2ex}{$\scriptscriptstyle\dagger$}}}}
\begin{figure}[t]
\centering
\footnotesize
\makebox[\linewidth][c]{%
\begin{quantikz}[row sep={0.5cm,between origins}, column sep=0.06cm]
\lstick{\irin{m_J}{\ket{0}}} &[0.03cm]
  \qwbundle{100}&[0.15cm]
  \gate[style=prepblock]{\prep{prep}}\gategroup[4, steps=2, style=irJ, background, label style={irtag, text=black!60}]{$J$}
  &[0.12cm] \swap{1} &[0.4cm] \gategroup[4, steps=3, style=irDi, background, label style={irtag, text=oiblue}]{$D_i$} &[0.08cm] &[0.08cm] &[0.42cm]
  \gategroup[4, steps=3, style=irDj, background, label style={irtag, text=oiverm}]{$D_j$} &[0.08cm] &[0.08cm] &[0.42cm]
  \gate[style=prepblock]{\prep{prep}}\gategroup[4, steps=2, style=irJ, background, label style={irtag, text=black!60}]{$J$} &[0.12cm] \swap{1} &[0.12cm] &[0.03cm] \rstick{\irout{\langle 0|}} \\
\lstick{\irin{x}{\ket{j}}} & \qwbundle{100} & \gate[style=prepblock]{\irprepdag} & \targX{} & &
  \gate[2, style=termblock]{\text{\scriptsize$m_D \ge i$}} & & & \gate[2, style=termblock]{\text{\scriptsize$m_D \ge j$}} & &
  \gate[style=prepblock]{\irprepdag} & \targX{} & & \rstick{\irout{\langle i|}} \\
\lstick{\irin{m_D}{\ket{0}}} & \qwbundle{100} & & & \gate[style=prepblock]{\prep{prep}} & &
  \gate[style=prepblock]{\irprepdag} & \gate[style=prepblock]{\prep{prep}} & &
  \gate[style=prepblock]{\irprepdag} & & & & \rstick{\irout{\langle 0|}} \\[-0.05cm]
\lstick{\irin{f}{\ket{0}}} & & & & & \targ{}\wire[u][2]{q} & & & \targ{}\wire[u][2]{q} & & & & & \rstick{\irout{\langle 0|}} \\[0.05cm]
\lstick{\irin{c}{\ket{0}}} & \gate[style=prepblock]{\prep{prep}}
  & \octrl{-4} & \octrl{-3} & \octrl{-2} & \octrl{-1} & \octrl{-2}
  & \ctrl{-2} & \ctrl{-1} & \ctrl{-2} & \ctrl{-4} & \ctrl{-3} & \gate[style=prepblock]{\irprepdag} & \rstick{\irout{\langle 0|}}
\end{quantikz}%
}
\setlength{\abovecaptionskip}{1ex}
\caption{The synthesized block encoding circuit for $M_\textrm{simple}$ at $N = 2^{100}$.
Its structure is recursively defined from the symbolic matrix representation of $M_\textrm{simple} = \textcolor{oiblue}{D_i} \mmul J + J \mmul \textcolor{oiverm}{D_j}$.
This drawing is a simplification: the full circuit has 1,103 qubits, because preparation and comparison blocks each use their own helper qubits.}
\label{fig:msimple-pieces}
\end{figure}
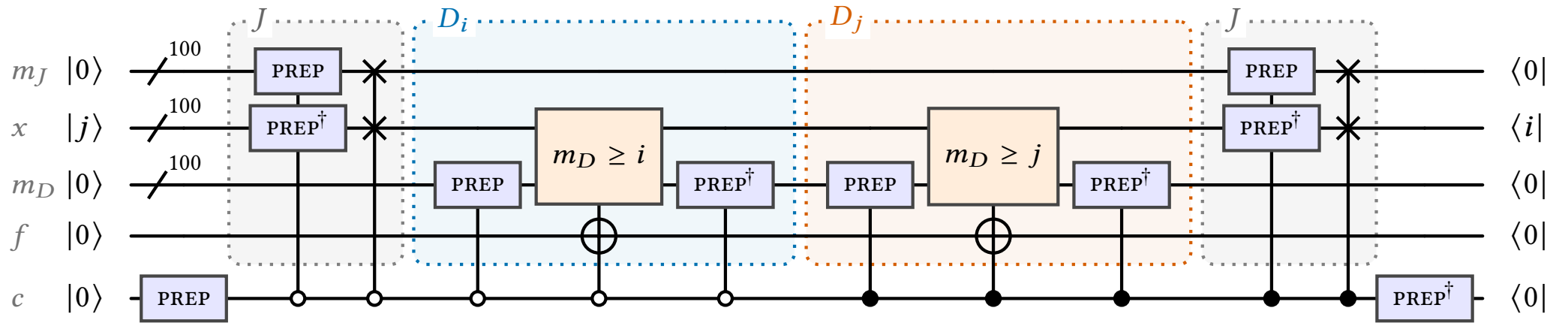

\paragraph{Symbolic Matrix Form}
To compile the $M_\textrm{simple}$ program to a circuit, MARBLE first translates it to a novel IR known as the \emph{symbolic matrix form}.
Each atomic expression in this IR is one of:

\begin{enumerate}[topsep=2pt]
  \item An entry comprehension $\comp{f(i,j)}$.
  For $M_\textrm{simple}$, this term is $\comp{f(i,j)} = \comp{i + j}$.
  \item A \emph{mask formula} $\comp{u(i) \cdot g(i, j) \cdot v(j)}$.
  Here, $u$ and $v$ can scale, flip a sign of, apply a phase on, or zero out a row or column by its index.
  The function $g$ is a \emph{mask} that maps a column index to the nonzero row indices in that column.
  For example, $g = 1$ maps every column to every row (the all-ones matrix), $g = [i \mathrel{\texttt{==}} j]$ maps each column to its own row (the identity matrix), $g = [i \mathrel{\texttt{==}} j + 1]$ is a shift matrix, and $g = \operatorname{phase}(i \cdot j, N)$ is a Fourier transform.
  \item A specific named matrix, for example the all-ones matrix $J$, the diagonal matrix $D$, and the identity matrix $I$, for which a succinct block-encoding circuit is known.
\end{enumerate}

\paragraph{Algebraic Rewriting}
Starting with the entry comprehension of the original program, MARBLE applies a set of rewrites to lower it to a mask formula.
For $M_\textrm{simple}$, it performs two rewrites:
\begin{equation}
  \comp{f(i,j)} = \comp{i+j} \rightarrow \comp{i} + \comp{j} \rightarrow \comp{i \cdot 1 \cdot 1} + \comp{1 \cdot 1 \cdot j},
  \label{eq:m-simple-rewrite}
\end{equation}
in which the first mask formula $\comp{i \cdot 1 \cdot 1}$ has three components: the \emph{row scale} $u_1(i) = i$, the \emph{mask} $g_1 = 1$, and the \emph{column scale} $v_1 = 1$. Similarly, in the second term, $u_2 = 1$, $g_2 = 1$, and $v_2(j) = j$.

\colorlet{rwrow}{oiblue}
\colorlet{rwcol}{oiverm}
\colorlet{rwnone}{black}
\tikzset{
  rwline/.style={draw=black!65, line width=0.3pt},
  rwfaint/.style={draw=black!25, line width=0.3pt},
  rwplus/.style={draw=black!80, line width=0.5pt},
  rwlbl/.style={anchor=south, inner sep=2pt, font=\scriptsize},
  rwname/.style={anchor=north, inner sep=2pt, font=\scriptsize},
  rwhead/.style={anchor=base, inner sep=1pt, font=\fontsize{7.5}{9}\selectfont\sffamily, text=black},
  rwrole/.style={anchor=south, inner sep=1pt, font=\fontsize{5.75}{6}\selectfont, text=black!55, align=center},
  rwop/.style={font=\small},
}
\providecommand{\rwGridStyled}[2]{%
  \foreach \k in {0,...,#1} {\draw[#2] (\k,0) -- (\k,#1); \draw[#2] (0,\k) -- (#1,\k);}}
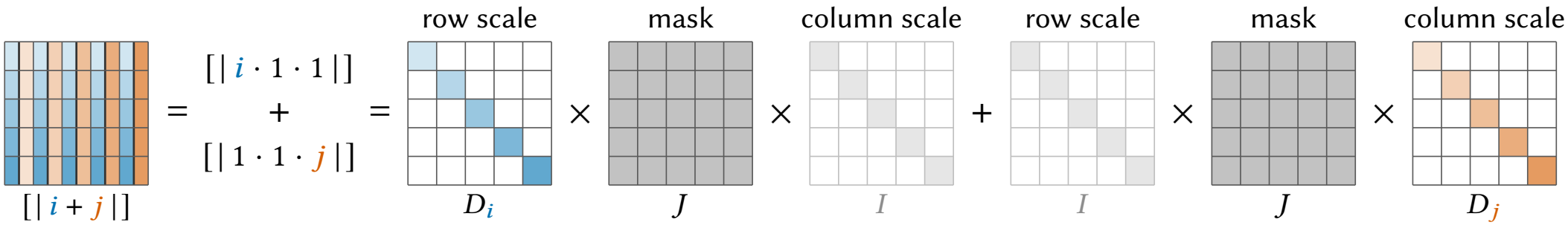
\begin{figure}[t]
\centering
\footnotesize
\resizebox{\textwidth}{!}{%
\begin{tikzpicture}
  \begin{scope}[x=2.5mm, y=-2.5mm]
    \foreach \r in {0,...,4} {\foreach \c in {0,...,4} {
      \pgfmathtruncatemacro\pr{18+11*\r}
      \pgfmathtruncatemacro\pc{18+11*\c}
      \fill[rwrow!\pr!white] (\c,\r) rectangle ++(0.5,1);
      \fill[rwcol!\pc!white] (\c+0.5,\r) rectangle ++(0.5,1);
      \draw[rwplus] (\c+0.5,\r) -- ++(0,1);
    }}
    \rwGridStyled{5}{rwline}
    \node[rwname] at (2.5,5) {$\comp{\textcolor{rwrow}{i} + \textcolor{rwcol}{j}}$};
  \end{scope}
  \node[rwop] at (1.5,-0.625) {$=$};
  \node[inner sep=0pt, anchor=west, align=center] (stk) at (1.7,-0.625)
    {$\comp{\textcolor{rwrow}{i} \cdot 1 \cdot 1}$\\[1pt]{\small $+$}\\[1pt]$\comp{1 \cdot 1 \cdot \textcolor{rwcol}{j}}$};
  \begin{scope}[shift={(stk.east)}, xshift=-0.7mm]
    \node[rwhead] at (1.125,0.75) {row scale};
    \node[rwhead] at (2.875,0.75) {mask};
    \node[rwhead] at (4.625,0.75) {column scale};
    \node[rwhead] at (6.375,0.75) {row scale};
    \node[rwhead] at (8.125,0.75) {mask};
    \node[rwhead] at (9.875,0.75) {column scale};
    \node[rwop] at (0.25,0) {$=$};
    \begin{scope}[shift={(0.5,0.625)}, x=2.5mm, y=-2.5mm]
      \foreach \r in {0,...,4} {
        \pgfmathtruncatemacro\p{18+11*\r}
        \fill[rwrow!\p!white] (\r,\r) rectangle ++(1,1);
      }
      \rwGridStyled{5}{rwline}
      \node[rwname] at (2.5,5) {$D_{\textcolor{rwrow}{i}}$};
    \end{scope}
    \node[rwop] at (2,0) {$\mmul$};
    \begin{scope}[shift={(2.25,0.625)}, x=2.5mm, y=-2.5mm]
      \fill[rwnone!24!white] (0,0) rectangle (5,5);
      \rwGridStyled{5}{rwline}
      \node[rwname] at (2.5,5) {$J$};
    \end{scope}
    \node[rwop] at (3.75,0) {$\mmul$};
    \begin{scope}[shift={(4,0.625)}, x=2.5mm, y=-2.5mm]
      \foreach \r in {0,...,4} {\fill[rwnone!10!white] (\r,\r) rectangle ++(1,1);}
      \rwGridStyled{5}{rwfaint}
      \node[rwname, text=black!45] at (2.5,5) {$I$};
    \end{scope}
    \node[rwop] at (5.5,0) {$+$};
    \begin{scope}[shift={(5.75,0.625)}, x=2.5mm, y=-2.5mm]
      \foreach \r in {0,...,4} {\fill[rwnone!10!white] (\r,\r) rectangle ++(1,1);}
      \rwGridStyled{5}{rwfaint}
      \node[rwname, text=black!45] at (2.5,5) {$I$};
    \end{scope}
    \node[rwop] at (7.25,0) {$\mmul$};
    \begin{scope}[shift={(7.5,0.625)}, x=2.5mm, y=-2.5mm]
      \fill[rwnone!24!white] (0,0) rectangle (5,5);
      \rwGridStyled{5}{rwline}
      \node[rwname] at (2.5,5) {$J$};
    \end{scope}
    \node[rwop] at (9,0) {$\mmul$};
    \begin{scope}[shift={(9.25,0.625)}, x=2.5mm, y=-2.5mm]
      \foreach \c in {0,...,4} {
        \pgfmathtruncatemacro\p{18+11*\c}
        \fill[rwcol!\p!white] (\c,\c) rectangle ++(1,1);
      }
      \rwGridStyled{5}{rwline}
      \node[rwname] at (2.5,5) {$D_{\textcolor{rwcol}{j}}$};
    \end{scope}
  \end{scope}
\end{tikzpicture}%
}
\setlength{\abovecaptionskip}{1ex}%
\caption{Algebraic rewriting of the symbolic matrix form of $M_\textrm{simple}$, turning an entry comprehension into a sum of mask formulas each containing a row scale, a mask, and a column scale, and then into explicit matrices.}
\label{fig:msimple-mask}
\vspace{-1ex}%
\end{figure}

As shown in Fig.~\ref{fig:msimple-mask}, scaling row $i$ of a matrix by $u(i)$ is the same as multiplying it on the left by the diagonal matrix $D_u$ defined as $D_u[i,i] = u(i)$. Similarly, scaling its columns by $v$ is the same as multiplying it on the right by $D_v$. Thus, the following rewrite rule preserves the matrix:
\begin{equation}
  \comp{u(i) \cdot g(i,j) \cdot v(j)} \rightarrow D_u \mmul \comp{g(i,j)} \mmul D_v,
\end{equation}
where $\mmul$ denotes the standard matrix product.
In $M_\textrm{simple}$, both masks $g$ are $1$ and yield the all-ones matrix $J$. A row or column scale that is always $1$ gives $D_1 = I$, and $u_1(i) = i$ and $v_2(j) = j$ give $D_i$ and $D_j$ respectively. Thus, MARBLE continues from Equation~\eqref{eq:m-simple-rewrite} to yield named matrices:
\begin{equation}
  \comp{i \cdot 1 \cdot 1} + \comp{1 \cdot 1 \cdot j} \rightarrow (D_i \mmul J \mmul I) + (I \mmul J \mmul D_j) = D_i \mmul J + J \mmul D_j.
  \label{eq:msimple-dj-jd}
\end{equation}

\paragraph{Rewrite Rules}
In Equation~\eqref{eq:m-simple-rewrite}, the rule to convert $\comp{i+j}$ to $\comp{i} + \comp{j}$ is known as \emph{entry sum}, and turns an entry comprehension containing sums into a sum of entry comprehensions.
Figure~\ref{fig:rewrite-rules} depicts a sample of other rewrite rules employed by MARBLE, which correspond to linear-algebraic identities such as the distributivity of scalar and tensor multiplication over matrix addition.
\colorlet{rwrow}{oiblue}
\colorlet{rwcol}{oiverm}
\colorlet{rwnone}{black}
\colorlet{rwfst}{oigreen}
\colorlet{rwsnd}{oipurple}
\tikzset{
  rwline/.style={draw=black!65, line width=0.3pt},
  rwsplit/.style={draw=black!30, line width=0.2pt},
  rwplus/.style={draw=black!80, line width=0.5pt},
  rwblock/.style={draw=black!85, line width=0.7pt},
  rwlbl/.style={anchor=south, inner sep=2pt, font=\scriptsize},
  rwop/.style={font=\small},
}
\providecommand{\rwArrow}[2]{\node[rwop] at (#1,#2) {$\rightarrow$};}
\providecommand{\rwGrid}[1]{%
  \foreach \k in {0,...,#1} {\draw[rwline] (\k,0) -- (\k,#1); \draw[rwline] (0,\k) -- (#1,\k);}}
\providecommand{\rwCell}[3]{\fill[#3] (#1,#2) rectangle ++(1,1);}
\providecommand{\rwSum}[4]{%
  \fill[#3] (#1,#2) rectangle ++(0.5,1); \fill[#4] (#1+0.5,#2) rectangle ++(0.5,1);
  \draw[rwplus] (#1+0.5,#2) -- ++(0,1);}
\providecommand{\rwProd}[4]{%
  \fill[#3] (#1,#2) rectangle ++(1,0.5); \fill[#4] (#1,#2+0.5) rectangle ++(1,0.5);
  \draw[rwsplit] (#1,#2+0.5) -- ++(1,0);}
\providecommand{\rwF}{{2,0,1,0,1,0,1,0,2}}
\providecommand{\rwG}{{1,2,0,0,1,2,2,0,1}}
\begin{figure*}[t]
\centering
\footnotesize
\resizebox{\textwidth}{!}{%
\begin{subfigure}[b]{5.6cm}
\centering
\begin{tikzpicture}
  \useasboundingbox (0,0.3) rectangle (5.5,-1.5);
  \begin{scope}[x=3mm, y=-3mm]
    \foreach \r in {0,...,4} {\foreach \c in {0,...,4} {
      \pgfmathtruncatemacro\pr{18+11*\r}
      \pgfmathtruncatemacro\pc{18+11*\c}
      \rwSum{\c}{\r}{rwrow!\pr!white}{rwcol!\pc!white}
    }}
    \rwGrid{5}
    \node[rwlbl] at (2.5,0) {$\comp{p + q}$};
  \end{scope}
  \rwArrow{1.75}{-0.75}
  \begin{scope}[shift={(2.0,0)}, x=3mm, y=-3mm]
    \foreach \r in {0,...,4} {\foreach \c in {0,...,4} {
      \pgfmathtruncatemacro\pr{18+11*\r}
      \rwCell{\c}{\r}{rwrow!\pr!white}
    }}
    \rwGrid{5}
    \node[rwlbl] at (2.5,0) {$\comp{p}$};
  \end{scope}
  \node[rwop] at (3.75,-0.75) {$+$};
  \begin{scope}[shift={(4.0,0)}, x=3mm, y=-3mm]
    \foreach \r in {0,...,4} {\foreach \c in {0,...,4} {
      \pgfmathtruncatemacro\pc{18+11*\c}
      \rwCell{\c}{\r}{rwcol!\pc!white}
    }}
    \rwGrid{5}
    \node[rwlbl] at (2.5,0) {$\comp{q}$};
  \end{scope}
\end{tikzpicture}
\subcaption{Entry sum}
\label{fig:rw-sum}
\end{subfigure}%
\hspace{6mm}%
\begin{subfigure}[b]{5.6cm}
\centering
\begin{tikzpicture}
  \useasboundingbox (0,0.3) rectangle (5.5,-1.5);
  \begin{scope}[x=3mm, y=-3mm]
    \foreach \r in {0,...,4} {\foreach \c in {0,...,4} {
      \pgfmathtruncatemacro\pp{18+11*(\r+1)}
      \pgfmathtruncatemacro\pc{18+11*\c}
      \fill[rwrow!\pp!white] (\c,\r) rectangle ++(1,0.5);
      \fill[rwnone!24!white] (\c,\r+0.5) rectangle ++(0.5,0.5);
      \fill[rwcol!\pc!white] (\c+0.5,\r+0.5) rectangle ++(0.5,0.5);
      \draw[rwsplit] (\c,\r+0.5) -- ++(1,0);
      \draw[rwplus] (\c+0.5,\r+0.5) -- ++(0,0.5);
    }}
    \rwGrid{5}
    \node[rwlbl] at (2.5,0) {$\comp{p \cdot (q + r)}$};
  \end{scope}
  \rwArrow{1.75}{-0.75}
  \begin{scope}[shift={(2.0,0)}, x=3mm, y=-3mm]
    \foreach \r in {0,...,4} {\foreach \c in {0,...,4} {
      \pgfmathtruncatemacro\pp{18+11*(\r+1)}
      \rwProd{\c}{\r}{rwrow!\pp!white}{rwnone!24!white}
    }}
    \rwGrid{5}
    \node[rwlbl] at (2.5,0) {$\comp{p \cdot q}$};
  \end{scope}
  \node[rwop] at (3.75,-0.75) {$+$};
  \begin{scope}[shift={(4.0,0)}, x=3mm, y=-3mm]
    \foreach \r in {0,...,4} {\foreach \c in {0,...,4} {
      \pgfmathtruncatemacro\pp{18+11*(\r+1)}
      \pgfmathtruncatemacro\pc{18+11*\c}
      \rwProd{\c}{\r}{rwrow!\pp!white}{rwcol!\pc!white}
    }}
    \rwGrid{5}
    \node[rwlbl] at (2.5,0) {$\comp{p \cdot r}$};
  \end{scope}
\end{tikzpicture}
\subcaption{Distribution, then entry sum}
\label{fig:rw-dist}
\end{subfigure}%
\hspace{6mm}%
\begin{subfigure}[b]{4.6cm}
\centering
\begin{tikzpicture}
  \useasboundingbox (0,0.3) rectangle (4.2,-1.5);
  \begin{scope}[x=3mm, y=-3mm]
    \foreach \r in {0,...,4} {\foreach \c in {0,...,4} {
      \pgfmathtruncatemacro\on{mod(\r+\c,2)==0}
      \ifnum\on=1 \rwProd{\c}{\r}{rwnone!48!white}{rwnone!24!white} \fi
    }}
    \rwGrid{5}
    \node[rwlbl] at (2.5,0) {$\comp{k \cdot s}$};
  \end{scope}
  \rwArrow{1.75}{-0.75}
  \begin{scope}[shift={(2.05,-0.6)}, x=3mm, y=-3mm]
    \rwCell{0}{0}{rwnone!48!white}
    \draw[rwline] (0,0) rectangle (1,1);
    \node[rwlbl] at (0.5,0) {$k$};
  \end{scope}
  \node[rwop] at (2.5,-0.75) {$\cdot$};
  \begin{scope}[shift={(2.65,0)}, x=3mm, y=-3mm]
    \foreach \r in {0,...,4} {\foreach \c in {0,...,4} {
      \pgfmathtruncatemacro\on{mod(\r+\c,2)==0}
      \ifnum\on=1 \rwCell{\c}{\r}{rwnone!24!white} \fi
    }}
    \rwGrid{5}
    \node[rwlbl] at (2.5,0) {$\comp{s}$};
  \end{scope}
\end{tikzpicture}
\subcaption{Constant extraction}
\label{fig:rw-const}
\end{subfigure}}

\vspace{6pt}
\resizebox{\textwidth}{!}{%
\begin{subfigure}[b]{5.6cm}
\centering
\begin{tikzpicture}
  \useasboundingbox (0,0.3) rectangle (5.5,-1.5);
  \begin{scope}[x=3mm, y=-3mm]
    \foreach \r in {0,...,3} {
      \pgfmathtruncatemacro\p{18+11*(\r+1)}
      \rwProd{\r+1}{\r}{rwrow!\p!white}{rwnone!24!white}
    }
    \rwGrid{5}
    \node[rwlbl] at (2.5,0) {$\comp{u(i) \cdot s}$};
  \end{scope}
  \rwArrow{1.75}{-0.75}
  \begin{scope}[shift={(2.0,0)}, x=3mm, y=-3mm]
    \foreach \r in {0,...,4} {
      \pgfmathtruncatemacro\p{18+11*(\r+1)}
      \rwCell{\r}{\r}{rwrow!\p!white}
    }
    \rwGrid{5}
    \node[rwlbl] at (2.5,0) {$\comp{u(i)\,[i \mathrel{\texttt{==}} j]}$};
  \end{scope}
  \node[rwop] at (3.75,-0.75) {$\mmul$};
  \begin{scope}[shift={(4.0,0)}, x=3mm, y=-3mm]
    \foreach \r in {0,...,3} {\rwCell{\r+1}{\r}{rwnone!24!white}}
    \rwGrid{5}
    \node[rwlbl] at (2.5,0) {$\comp{s}$};
  \end{scope}
\end{tikzpicture}
\subcaption{Scale extraction for row}
\label{fig:rw-weight-row}
\end{subfigure}%
\hspace{6mm}%
\begin{subfigure}[b]{5.6cm}
\centering
\begin{tikzpicture}
  \useasboundingbox (0,0.3) rectangle (5.5,-1.5);
  \begin{scope}[x=3mm, y=-3mm]
    \foreach \c in {0,...,3} {
      \pgfmathtruncatemacro\p{18+11*(\c+1)}
      \rwProd{\c}{\c+1}{rwnone!24!white}{rwcol!\p!white}
    }
    \rwGrid{5}
    \node[rwlbl] at (2.5,0) {$\comp{s \cdot v(j)}$};
  \end{scope}
  \rwArrow{1.75}{-0.75}
  \begin{scope}[shift={(2.0,0)}, x=3mm, y=-3mm]
    \foreach \c in {0,...,3} {\rwCell{\c}{\c+1}{rwnone!24!white}}
    \rwGrid{5}
    \node[rwlbl] at (2.5,0) {$\comp{s}$};
  \end{scope}
  \node[rwop] at (3.75,-0.75) {$\mmul$};
  \begin{scope}[shift={(4.0,0)}, x=3mm, y=-3mm]
    \foreach \c in {0,...,4} {
      \pgfmathtruncatemacro\p{18+11*(\c+1)}
      \rwCell{\c}{\c}{rwcol!\p!white}
    }
    \rwGrid{5}
    \node[rwlbl] at (2.5,0) {$\comp{v(j)\,[i \mathrel{\texttt{==}} j]}$};
  \end{scope}
\end{tikzpicture}
\subcaption{Scale extraction for column}
\label{fig:rw-weight-col}
\end{subfigure}%
\hspace{6mm}%
\begin{subfigure}[b]{4.6cm}
\centering
\begin{tikzpicture}
  \useasboundingbox (0,0.3) rectangle (4.2,-1.5);
  \begin{scope}[x=1.6667mm, y=-1.6667mm]
    \foreach \R in {0,...,8} {\foreach \C in {0,...,8} {
      \pgfmathtruncatemacro\f{\rwF[3*int(\R/3)+int(\C/3)]}
      \pgfmathtruncatemacro\g{\rwG[3*mod(\R,3)+mod(\C,3)]}
      \ifnum\f>0 \ifnum\g>0
        \pgfmathtruncatemacro\pf{25+20*\f}
        \pgfmathtruncatemacro\pg{25+20*\g}
        \rwProd{\C}{\R}{rwfst!\pf!white}{rwsnd!\pg!white}
      \fi\fi
    }}
    \rwGrid{9}
    \foreach \k in {0,3,6,9} {\draw[rwblock] (\k,0) -- (\k,9); \draw[rwblock] (0,\k) -- (9,\k);}
    \node[rwlbl] at (4.5,0) {$\comp{f(i_1, j_1) \cdot g(i_2, j_2)}$};
  \end{scope}
  \rwArrow{1.75}{-0.75}
  \begin{scope}[shift={(2.0,-0.3)}, x=3mm, y=-3mm]
    \foreach \R in {0,...,2} {\foreach \C in {0,...,2} {
      \pgfmathtruncatemacro\f{\rwF[3*\R+\C]}
      \ifnum\f>0 \pgfmathtruncatemacro\pf{25+20*\f}\rwCell{\C}{\R}{rwfst!\pf!white} \fi
    }}
    \rwGrid{3}
    \node[rwlbl] at (1.5,0) {$\comp{f}$};
  \end{scope}
  \node[rwop] at (3.1,-0.75) {$\otimes$};
  \begin{scope}[shift={(3.3,-0.3)}, x=3mm, y=-3mm]
    \foreach \R in {0,...,2} {\foreach \C in {0,...,2} {
      \pgfmathtruncatemacro\g{\rwG[3*\R+\C]}
      \ifnum\g>0 \pgfmathtruncatemacro\pg{25+20*\g}\rwCell{\C}{\R}{rwsnd!\pg!white} \fi
    }}
    \rwGrid{3}
    \node[rwlbl] at (1.5,0) {$\comp{g}$};
  \end{scope}
\end{tikzpicture}
\subcaption{Kronecker split}
\label{fig:rw-kron}
\end{subfigure}}
\setlength{\abovecaptionskip}{1ex}%
\caption{Visualization of a selection of the linear-algebraic rewrite rules employed by MARBLE\@.}
\label{fig:rewrite-rules}
\end{figure*}
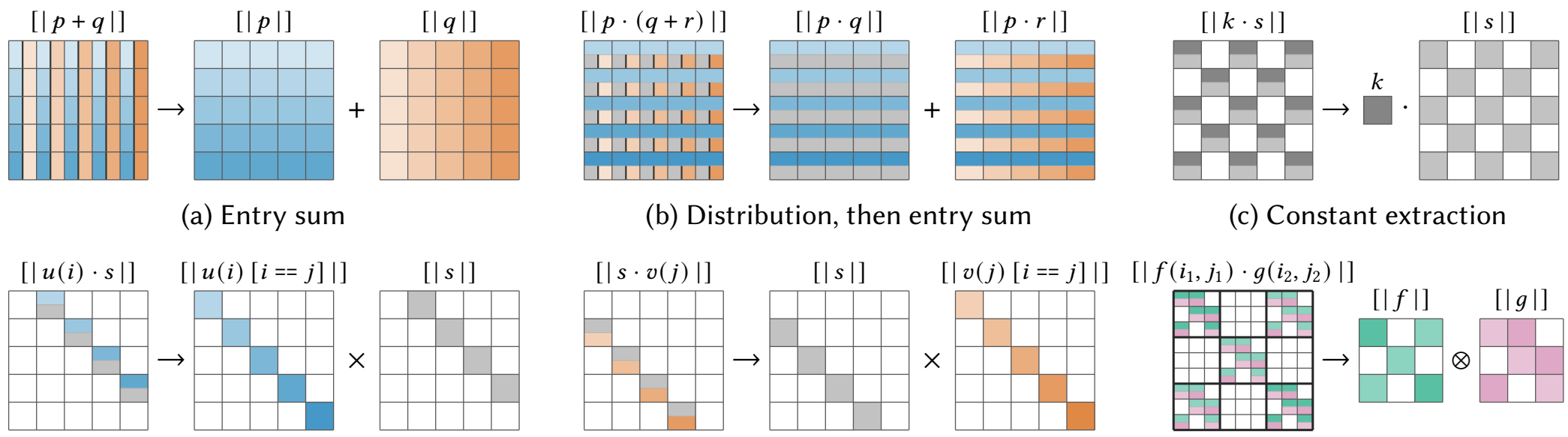

Every rewrite rule is a normalizing \emph{reduction rule}, an undirected \emph{exploration rule}, or both.
An example of a reduction rule is entry sum, which the compiler always performs when possible to simplify the program.
By contrast, factoring (to undo distributivity) is an exploration rule.
Such rules do not always lead to normal forms or decrease the program's complexity, and so MARBLE uses equality saturation~\citep{tateEqualitySaturationNew2009} to explore the space of possible symbolic matrix forms.
This process stops when it finds no new forms or when the search exhausts a fixed budget.

\paragraph{Circuit Selection}
Rewriting yields a set of candidate symbolic matrix forms, such as $\comp{i} + \comp{j}$ and $\comp{i \cdot 1 \cdot 1} + J \mmul D_j$ for $M_\textrm{simple}$.
MARBLE determines the structure and cost of the quantum circuit for each candidate by matching sub-expressions against a database of known circuits.

As shown in Figure~\ref{fig:msimple-pieces}, for $M_\textrm{simple} = D_i \mmul J + J \mmul D_j$, MARBLE recognizes that $J$ can be block-encoded by preparing a uniform superposition, while the diagonal $D_i$ and $D_j$ can be encoded using a comparison circuit.
The matrix $\mmul$ and $+$ operators compose their operands, following semantics given by Cobble~\citep{yuanCobbleCompilingBlock2026}.
If some sub-expression has no match in the database, MARBLE can realize it via an expensive fallback analogous to FABLE~\citep{campsFABLEFastApproximate2022} called the \emph{entry loader} --- which guarantees that every candidate symbolic matrix form has a circuit.

To choose the best candidate, MARBLE symbolically computes the resource cost of each circuit in terms of iterations $\alpha$ and gates and qubits per iteration, parameterized by the matrix size $N$.
For the candidate in Figure~\ref{fig:msimple-pieces}, MARBLE determines that $J$ has $\alpha = N$ and $D$ has $\alpha = N-1$, meaning that the circuit takes $\alpha = (N-1) \cdot N + N \cdot (N-1) = 2N(N-1)$ iterations.
Alternatively, the entry loader can build $\comp{i+j}$ directly with the same $\alpha$, but incurs additional cost of $O(N^2)$ gates.

The symbolic analysis determines that the candidate in Figure~\ref{fig:msimple-pieces} has both the lowest $\alpha$ and the lowest total time cost.
Throughout all previous stages, the compiler has left $N$ as symbolic to ensure that processing time does not scale with $N$.
In the final step of the pipeline, MARBLE instantiates the user-provided $N = 2^{100}$, and writes out the explicit circuit into an OpenQASM file.

\paragraph{Cost Certification}
One circuit does not tell the user how well the encoding scales with problem size or whether a better circuit exists.
Even worse, at $N = 2^{100}$, it is infeasible to simulate the circuit or enumerate many circuits.
To enable scalable and trustworthy reasoning about cost, MARBLE instead leverages its symbolic representation to analytically guarantee whether the:
\begin{enumerate}[leftmargin=17pt, topsep=2pt]
  \item \emph{Circuit is efficient:} the number of qubits and gates per iteration grow as $O(\polylog(N))$ where $N$ is the dimension of the matrix.
  In the $M_\textrm{simple}$ example, the circuit in Figure~\ref{fig:msimple-pieces} is efficient.
  \item \emph{Subnormalization is tight:} the ratio of $\alpha$ to its theoretical lower bound (the spectral norm of the matrix) is low.
  For $M_\textrm{simple}$, this ratio converges to a constant that is no greater than 2.07.
  \item \emph{Spectral norm is bounded:} this lower bound itself grows as $O(\polylog(N))$, a property of the target matrix rather than its encoding.
  MARBLE deduces that $M_\textrm{simple}$ is not bounded, alerting the user to the impossibility of a block encoding with poly-logarithmic total time cost.
\end{enumerate}

Together, these three properties give a developer confidence that a quantum algorithm that uses an encoded matrix can deliver promised asymptotic advantage, while informing the user of the source of any inefficiency so that they may easily re-evaluate their proposed application.

\subsection{Return to Running Example}
We now return to the lattice Boltzmann collision matrix $F_1$, whose MARBLE program was shown at the top of page~\hyperref[prog:f1]{\pageref*{prog:f1}}.
Using the principles developed in the $M_\textrm{simple}$ example, we summarize how rewriting, circuit selection, and cost certification operate for this more complex application.

\paragraph{Algebraic Rewriting}
Just like $M_\textrm{simple}$, the entry comprehension for $F_1$ is first subject to reduction rules.
Using condensed notation for brevity, the rewrite rules of Figure~\ref{fig:rewrite-rules} transform $F_1$ as follows:
\begingroup\small
\begin{align}
  \comp{w(i)\,(1 + 3\,c(i) \cdot c(j)) - \delta_{ij}} &\xrightarrow{\text{distribution}} \comp{w(i)} + 3 \textstyle\sum_{k=1}^{3} \comp{w(i)\,c(i_k)\,c(j_k)} - \comp{\delta_{ij}} \nonumber \\[-0.5ex]
  &\hspace{-5em}\xrightarrow{\text{scale extraction}} \comp{w(i)\,\delta_{ij}} \mmul J + 3 \textstyle\sum_{k=1}^{3} \comp{w(i)\,c(i_k)\,\delta_{ij}} \mmul J \mmul \comp{c(j_k)\,\delta_{ij}} - \comp{\delta_{ij}} \nonumber \\[-0.5ex]
  &\hspace{-5em}\xrightarrow{\text{Kronecker split}} (D \otimes D \otimes D) \mmul J + 3 \textstyle\sum_{k=1}^{3} W_k \mmul J \mmul C_k - I \otimes I \otimes I, \label{eq:f1-reduced}
\end{align}\endgroup
where the named matrix $J$ is the $27 \times 27$ all-ones matrix, $D = \operatorname{diag}(\tfrac{1}{4}, \tfrac{1}{4}, 1)$, and $C = \operatorname{diag}(1, -1, 0)$. $C_k$ denotes $I \otimes I \otimes I$ with $C$ substituted at position $1 \le k \le 3$, and $W_k = (D \otimes D \otimes D) \mmul C_k$.

The first two steps are similar to the $M_\textrm{simple}$ example. The Kronecker split rule splits a matrix indexed by tuples into a Kronecker product of smaller matrices. For example, the entries of the $27 \times 27$ matrix $\comp{w(i)\,\delta_{ij}}$ can be refactored as Kronecker products of the $3 \times 3$ matrix $D$.

After performing all possible reductions and finding the symbolic matrix forms in Equation~\eqref{eq:f1-reduced}, MARBLE uses them to seed equality saturation over exploration rules.
For example, it considers a rule to alternatively factor $\comp{w(i)\,c(i_k)\,\delta_{ij}} = D \mmul C$ as a product of two diagonal matrices.
At the end of equality saturation, MARBLE obtains a large set of candidate symbolic matrix forms.

\paragraph{Circuit Selection}
Unlike $M_\textrm{simple}$, $F_1$ yields a range of candidates with no global winner:
\begin{enumerate}[topsep=2pt]
\item The first is the final expression in Eq.~\eqref{eq:f1-reduced}, whose circuit is depicted in Fig.~\ref{fig:f1-marble}.
Its gate cost per iteration is the lowest because most of its components are simple rank-1 matrices.
\item A second one builds diagonal matrices differently, using more rotations and fewer qubits.
\item A third candidate explicitly computes values of $F_1$ and compresses its structure to omit the identity matrix term, thereby minimizing subnormalization at the cost of gates.
\end{enumerate}

\begin{wraptable}{r}{0.44\columnwidth}
\vspace*{-2.75ex}
\captionsetup{skip=4pt}
\caption{Block encodings for one-cell $F_1$ matrix.}
\label{tab:overview-tab}
\footnotesize
\setlength{\tabcolsep}{2.3pt}
\begin{tabular}{@{}lrrrrr@{}}
\toprule
Candidate & $\alpha$ & $T$ & $R$ & $\alpha\,(T+R)$ & $Q$ \\
\midrule
Eq.~\eqref{eq:f1-reduced} as in Fig.~\ref{fig:f1-marble} & {7.2} & {60} & {16} & \textbf{547} & {20} \\
Eq.~\eqref{eq:f1-reduced}, fewer qubits & 7.2 & 48 & 142 & 1,368 & \textbf{12} \\
{Value compression} & \textbf{5.79} & {232} & {99} & {1,917} & {18} \\
\midrule
\citeauthor{kuklinskiEfficientBlockencodingsRequire2025} & 146 & \textasciitilde{}196 & 22 & \textasciitilde{}31,800 & 27 \\
FABLE & 64 & 0 & 4,096 & 262,144 & 13 \\
\bottomrule
\end{tabular}\vspace*{-3ex}
\end{wraptable}

Table~\ref{tab:overview-tab} lists their $T$ gates, arbitrary non-Clifford rotation gates ($R$), total time cost, and qubits ($Q$).
It compares these candidates against the handwritten baseline of \citet{kuklinskiEfficientBlockencodingsRequire2025} as well as FABLE\@.
MARBLE's best output obtains a $58.2\times$ reduction in total time cost over the best prior baseline.

\paragraph{Cost Certification}
In Section~\ref{sec:evaluation}, we present results for the $N$-cell generalized $F_1$ matrix, where no handwritten baselines exist and FABLE times out.
For this family, MARBLE automatically certifies that its output block encoding is \emph{efficient} (cost per iteration is poly-logarithmic in $N$) and \emph{tight} (ratio of subnormalization to spectral norm approaches one).
Moreover, it certifies that the matrix family is \emph{bounded} --- the spectral norm of the target $F_1$ matrix is itself poly-logarithmic in $N$.

With the help of MARBLE, a programmer can use the synthesized circuit to realize the quantum algorithm for fluid flow simulation, confident that its asymptotic advantage holds as promised.

\section{Language}
\label{sec:language}

In this section, we present the MARBLE language.
First, we define the core syntax of entry comprehensions.
Next, we extend it with additional operators to develop the symbolic matrix intermediate representation used by the compiler, and define the language semantics.
Then, we show how the language provides guarantees about multiplicity, which is key to symbolic reasoning for cost.

\subsection{Core Syntax of Entry Comprehensions}
\label{sec:syntax}

\begin{wrapfigure}{r}{0.46\textwidth}
\vspace{-3ex}
\footnotesize\centering
\resizebox{\linewidth}{!}{$
\begin{array}{@{}l@{\hspace{-2ex}}r@{\ }c@{\ }l@{}}
\textbf{Name} & x, i, j & & \hspace{-3.5ex}\text{ variable} \quad\ p \text{ parameter} \quad\ t \text{ table} \\
\textbf{Literal} & n & & \hspace{-3ex}\in \mathbb{N} \quad\ \ell \in \mathbb{C}\\
\textbf{Program} & P & ::= & d \cdots d\ \ M \\
\textbf{Declaration} & d & ::= & \kw{param}\ p \mid \kw{table}\ t = (\ell, \ldots, \ell) \\
\textbf{Size} & m & ::= & n \mid p \mid n^{n} \mid n^{p} \\
\textbf{Index} & e & ::= & x \mid n \mid e + e \mid e \cdot e \mid [e \mathrel{\texttt{==}} e] \\
\textbf{Value} & s & ::= & e \mid t[e] \mid s + s \mid s \cdot s \\
\textbf{Matrix} & M & ::= & \comp{s \mid i, j : \kw{Range}\ m}
\end{array}
$}
\setlength{\abovecaptionskip}{1ex}
\caption{Core syntax of the MARBLE language.}
\vspace{-3ex}
\label{fig:syntax-core}
\end{wrapfigure}

Figure~\ref{fig:syntax-core} shows the core syntax of MARBLE, which is simplified from the overview for clarity.
For example, we elide function definitions as syntactic sugar that is elaborated into the core.

\paragraph{Declarations}
A program consists of zero or more \emph{declarations} followed by an \emph{entry comprehension} of the matrix to be encoded.
Each declaration is either of a \emph{parameter} $p$, which is a symbolic value such as the size $N$ of the matrix, or of a \emph{coefficient table} of complex numbers that can be looked up by index.
The entry comprehension $\comp{s \mid i, j : \kw{Range}\ m}$ consists of an entry formula $s$ that may refer to the row and column indices $i$ and $j$ as variables.

\paragraph{Types and Expressions}
A scalar expression $s$ of type $\tV$ can be any complex number.
A value that is syntactically an integer, such as a column index $j+1$, has the subtype $\tI$.
An integer value that is statically known to be between 0 and $m-1$ inclusive has the further subtype $\kw{Range}\ m$.
The static size $m$ of a range must be a combination of literals $n$ and declared parameters $p$.

Expressions include variables, literal numbers, and arithmetic operations.
The Iverson bracket $[e \mathrel{\texttt{==}} e]$ evaluates to 1 if its integer operands are equal.
The table access $t[e]$ looks up the index $e$ from the coefficient table $t$, and returns either a complex $\tV$ or an integer $\tI$.

An $\tI$ gives the structure of a matrix, whereas a $\tV$ gives its data.
Structure is the compact pattern of how data is laid out across the matrix.
For example, an entry formula $[i \mathrel{\texttt{==}} j + 1]$ defines a shift matrix, in which every column $j$ has data present at the row $j+1$.
To efficiently compile a large matrix without materializing all of its entries, MARBLE must symbolically manipulate integer indices such as $j+1$ without instantiating parameters such as the matrix size $N$ or considering any specific data value.
The language thus permits an integer $\tI$ to be promoted to a $\tV$, but not the other way around, which would prevent static analysis of the matrix structure.

\subsection{Syntax and Semantics of Symbolic Matrix Forms}
\label{sec:semantics}

Figure~\ref{fig:syntax-ir} extends the core syntax with two features, tuples as indices and arithmetic operations over matrices, needed for the symbolic matrix forms used as the intermediate representation by
\begin{wrapfigure}{r}{0.51\textwidth}
\vspace{-2ex}
\footnotesize\centering
\resizebox{\linewidth}{!}{$
\begin{array}{@{}l@{\ \ }r@{\ }c@{\ }l@{}}
\textbf{Index} & e & ::= & \cdots \mid (e, e) \mid e.n \\
\textbf{Matrix} & M & ::= & \cdots \mid \bar{M} \mid s \cdot M \mid M + M \mid M \mmul M \mid M \otimes M
\end{array}
$}
\setlength{\abovecaptionskip}{1ex}
\caption{Extended syntax of symbolic matrix forms.}
\vspace{-3ex}
\label{fig:syntax-ir}
\end{wrapfigure}
the MARBLE compiler.
The syntax enables creation of tuples and field access by index.
It introduces named matrices $\bar{M}$ standing for e.g.\ the all-ones matrix $J$ or the identity matrix $I$, as well as scalar-matrix product and matrix addition, product, and Kronecker product.
Each matrix operation is well-formed if the row and column domains of operands match mathematically.

\paragraph{Denotational Semantics} The meaning of a program depends on the values of its parameters and the entries in its coefficient tables.
An \textit{environment} $\rho$ fixes this choice by giving each parameter a natural number and each table its entries.
A variable of type $\kw{Range}\ m$ ranges over $0, \ldots, \rho(m)-1$, and a variable of a tuple type ranges over the product of these sets in lexicographic order.

We define $\den{\cdot}$ to be the \textit{denotation} of a term under $\rho$.
Each scalar expression has a complex number as denotation, while a matrix expression has a matrix defined entry-wise.
The function $\den{\cdot}$ is defined as follows, omitting standard rules for literals, scalar arithmetic, and tuples.
\begin{gather*}
  \begin{aligned}
    \den{[e_1 \mathrel{\texttt{==}} e_2]} &= 1 \text{ if } \den{e_1} = \den{e_2}, \text{ else } 0
    & \den{M_1 + M_2} &= \den{M_1} + \den{M_2} \\
    \den{t[e]} &= \text{the entry of } t \text{ at } \den{e}
    & \den{M_1 \mmul M_2} &= \den{M_1}\,\den{M_2} \\
    \den{s \cdot M} &= \den{s}\cdot\den{M}
    & \den{\comp{s \mid i, j}}[u, v] &= \mathcal{D}_{\rho}(s[i \mapsto u,\, j \mapsto v])
  \end{aligned} \\
  \hspace{-4em}\den{M_1 \otimes M_2}[u, v] = \den{M_1}[u.1, v.1] \cdot \den{M_2}[u.2, v.2] \quad\qquad \den{\bar{M}} = \bar{M}
\end{gather*}

The first five rules are largely self-explanatory.
The entry comprehension is the one form that binds variables, so its entry at row $u$ and column $v$ is the formula $s$ evaluated with $i$ set to $u$ and $j$ set to $v$.
The Kronecker product is defined component-wise by the tuple indices $u$ and $v$.
Finally, a named matrix $\bar{M}$, e.g.\ the identity matrix $I$, is defined to have itself as denotation.

\subsection{Enforcement of Static Multiplicity}
\label{sec:static-multiplicity}

The central goal of the language design is for every well-formed program to be representable by a quantum circuit of poly-logarithmic size.
Doing so requires structure in how entries are computed from row and column indices.
Only if the number of distinct computations is poly-logarithmic can an efficient circuit exist for the program; we call this number the \emph{multiplicity} of the program.

\begin{example}
The multiplicity of $M_\textrm{simple}[i, j] = i+j$ is one, as each entry is computed via the same expression.
Its quantum circuit then needs only to perform logic to add $i$ and $j$ in logarithmic time.
By contrast, a diagonal matrix whose $i$-th row stores some arbitrary constant, which must be looked up from a table with one entry for each row, has multiplicity $N$.
Its implementation must perform this lookup via an explicit multiplexer circuit~\citep{babbushEncodingElectronicSpectra2018} taking time $O(N)$.
\end{example}

The MARBLE language guarantees that every well-formed program has low multiplicity --- in particular, the multiplicity is $O(1)$ and independent of $N$, a conservative condition that we call \emph{static} multiplicity.
Intuitively, this property holds on all entry comprehensions that contain only arithmetic operations because a static formula, e.g.\ $i + j$, cannot depend on $N$.
The troublesome cases are when the formula invokes a table lookup -- which MARBLE restricts by requiring the length of each explicitly declared \texttt{table} to be a constant independent of $N$ -- or when a formula recursively invokes itself, which the type checker prohibits.
In conjunction, these static checks ensure that the multiplicity of each well-formed program is a compile-time constant.

A key practical advantage of static as opposed to merely poly-logarithmic multiplicity is that the compiler can uniformly operate over symbolic formulas in $N$, without performing instance-specific reasoning.
For $M_\textrm{simple}$, the compiler can derive the symbolic matrix form $D_i \mmul J + J \mmul D_j$ and its subnormalization $\alpha = 2N(N-1)$ without even knowing the concrete value $N = 2^{100}$ or explicitly materializing any of these entries.
The compiler maintains only small, efficient symbolic representations of the program until the very last step of writing out an explicit circuit.

Without static multiplicity, these symbolic formulas in $N$ and compact program representations do not exist.
A program whose multiplicity depends on $N$ corresponds to circuit decompositions and costs that also change with $N$, meaning that the compiler would have to expensively enumerate a large fraction of the $2^{100}$ entries of the matrix to synthesize a circuit.
In Section~\ref{sec:evaluation}, we provide evidence that static multiplicity captures a wide range of matrices relevant to practical applications while enabling the compiler to execute near-instantaneously for even very large matrices.

\section{Compilation}
\label{sec:compilation}

In this section, we present how MARBLE transforms the symbolic form of a matrix into a quantum circuit.
To recognize the structure of a matrix, the compiler synthesizes \emph{mask formulas} (Sec.~\ref{sec:mask-formula}).
It applies a set of \emph{algebraic rewrite rules} (Sec.~\ref{sec:rewrite-rules}) to produce candidate representations.
Then, it performs \emph{circuit selection} (Sec.~\ref{sec:selection}) to symbolically compute costs and emit the best circuit.

\subsection{Mask Formulas to Separate Structure and Data}
\label{sec:mask-formula}

Synthesizing an efficient circuit without materializing the matrix requires separating its structure from the arbitrary complex values at each entry.
MARBLE captures this structure as a \emph{mask formula} $\comp{u(i) \cdot g(i,j) \cdot v(j)}$ consisting of a \emph{mask} $g(i, j)$ that characterizes the pattern of nonzero entries and two functions $u(i)$ and $v(j)$ known as the \emph{row} and \emph{column scale}.
Each component contains only the terms in the entry comprehension that depend on only the respective index.

\vspace{-.25ex}%
\begin{example}
Sec.~\ref{sec:overview} already showed the mask formula for $\comp{i}$, i.e.\ the matrix $M[i,j] = i$.
Now consider the quantum Fourier transform $\mathrm{QFT}[i,j] = 1/\smash{\sqrt{N}}\exp({2\pi \iota\, ij/N})$ where $\iota$ is the imaginary unit.
This matrix is dense, and each entry depends on both the row and column, leaving the row and column scales as 1.
By contrast, consider the exponential matrix $\comp{e^{i+j}} = \comp{e^i\,e^j}$ which is also dense but splits into row and column terms.
The mask formulas for these two matrices are
{\setlength{\belowdisplayskip}{3pt} \setlength{\belowdisplayshortskip}{0pt}
\setlength{\abovedisplayskip}{3pt} \setlength{\abovedisplayshortskip}{0pt}
\[
  \comp{\underbrace{1}_{\text{\raisebox{0.4ex}[0pt][0pt]{row scale}}} \cdot \underbrace{\smash[b]{\,\mathrm{phase}(i \cdot j, N)\,}}_{\text{\raisebox{0.4ex}[0pt][0pt]{mask}}} \cdot \underbrace{1}_{\text{\raisebox{0.4ex}[0pt][0pt]{column scale}}}}
  \qquad \text{and} \qquad
  \comp{\underbrace{e^{i}}_{\text{\raisebox{0.4ex}[0pt][0pt]{row scale}}} \cdot \underbrace{1}_{\text{\raisebox{0.4ex}[0pt][0pt]{mask}}} \cdot \underbrace{e^{j}}_{\text{\raisebox{0.4ex}[0pt][0pt]{column scale}}}},
\]}%
where terms that do not depend on both $i$ and $j$ are factored out --- often leaving the mask as 1.
\end{example}%
\vspace{-.25ex}%

The advantage of this transformation is that the scale terms are always diagonal matrices.
Any mask formula equals $D_u \mmul \comp{g} \mmul D_v$ (see Figure~\ref{fig:msimple-mask}), where the compiler can emit known circuits for the diagonal terms and focus analysis on the mask.
Whenever the mask $g = 1$ as for the $M_\textrm{simple}$ and $F_1$ examples, the mask itself reduces to the all-ones matrix $J$, and the rewriting is complete.

\subsection{Semantics-Preserving Algebraic Rewrite Rules}
\label{sec:rewrite-rules}

If needed, the compiler further decomposes the matrix using rewrite rules.
Table~\ref{tab:appendix-rules} in Appendix~\ref{sec:appendix-rules} presents all 31 rules used by MARBLE\@; we explain their structure and illustrate specific cases.

Each rewrite rule $\ell \rightarrow r$ has two sides, which are symbolic matrix forms that have metavariables to be matched, and a \textit{side condition} $\varphi$ under which the two sides mathematically denote the same matrix.
In MARBLE, an entry comprehension describes a family of matrices, one for each of its parameters (Sec.~\ref{sec:syntax}) including the dimension $N$.
We call the set $\mathcal{D}$ of valid dimensions $N \ge 1$ the \emph{size domain} of the program.
A rewrite rule is \textit{sound} if, for every well-formed instance of its left side and at every size in $\mathcal{D}$ where $\varphi$ holds, its right side is also well-formed and both sides denote the same matrix.
The compiler applies a rule only where it can prove $\varphi$ statically over all of $\mathcal{D}$.

\begin{example}
The \emph{Kronecker split} rule $\comp{f(i_1, j_1) \cdot g(i_2, j_2)} \rightarrow \comp{f} \otimes \comp{g}$ in Figure~\ref{fig:rw-kron} is mathematically sound when no factor in the entry formula depends on both $(i_1, j_1)$ and $(i_2, j_2)$.
In the weight term of $F_1$, $w(i) = \bar{w}_{i_1} \bar{w}_{i_2} \bar{w}_{i_3}$ and $\delta_{ij} = [i_1 \mathrel{\texttt{==}} j_1]\,[i_2 \mathrel{\texttt{==}} j_2]\,[i_3 \mathrel{\texttt{==}} j_3]$. Grouping gives
\[
  \comp{w(i)\,\delta_{ij}} = \comp{\bar{w}_{i_1} [i_1 \mathrel{\texttt{==}} j_1]} \otimes \comp{\bar{w}_{i_2} [i_2 \mathrel{\texttt{==}} j_2]} \otimes \comp{\bar{w}_{i_3} [i_3 \mathrel{\texttt{==}} j_3]} = D \otimes D \otimes D,
\]
as in Eq.~\eqref{eq:f1-reduced}, by the semantics of $\otimes$ (Sec.~\ref{sec:semantics}) read from right to left.
The compiler discharges the condition by checking each factor, where some e.g.\ $[i_1 \mathrel{\texttt{==}} j_2]$ would block the rewrite.
In Table~\ref{tab:appendix-rules}, most side conditions are read off the formula this way; the few that bound an index, such as $p, q \ge 0$ in the \emph{exponential split} rule $\comp{e^{i+j}} \rightarrow \comp{e^i\,e^j}$, are proved from the ranges of the indices.
\end{example}

\paragraph{Soundness}
The compiler soundly rewrites a program without ever changing its semantics.
In principle, soundness could only be violated by a rule that is not an algebraic identity or is applied when its side condition is false.
To eliminate the first case, we used Lean~4 to construct mechanized proofs that each rewrite rule in Table~\ref{tab:appendix-rules} preserves the mathematical semantics of the matrix and the well-formedness of the symbolic matrix form.
To guard against the second possibility, MARBLE applies a rule only after symbolically verifying the side condition $\varphi$ over the entire size domain $\mathcal{D}$, independently of the size $N$.
Finally, as a redundant check, MARBLE simulates the output circuit at small dimensions and checks numerical equality against the materialized matrix values.

\paragraph{Reduction}
Each rewrite serves as a directed \emph{reduction} rule, an undirected \emph{exploration} rule, or both.
The compiler always applies reduction rules in a fixed order, until the program reaches a \emph{normal form} in which every term is a mask formula.
Every reduction rule decreases a measure of program size.
Thus, reduction terminates and reaches a normal form assuming a fixed order.

\begin{example}
One measure of size is the number of subexpressions that use both $i$ and $j$.
The \emph{entry sum} rule (Figure~\ref{fig:rw-sum}) monotonically lowers this value by splitting $\comp{i + j}$ into $\comp{i} + \comp{j}$.
Side conditions prevent a rule from undoing progress.
For example, scale extraction (Figure~\ref{fig:rw-weight-row}) applies only off the diagonal, since on $D$ it would yield $D \mmul I$ and apply to $D$ again forever.
\end{example}

\colorlet{egrow}{oiblue}
\colorlet{egcol}{oiverm}
\colorlet{egadd}{oigreen}
\tikzset{
  egop/.style={draw=black!75, rounded corners=1.5pt, inner sep=1.5pt, minimum size=10pt, fill=white, font=\scriptsize},
  egpic/.style={draw=black!75, rounded corners=1.5pt, inner sep=0pt, fill=white},
  egnew/.style={draw=egadd!80!black, text=egadd!60!black},
  egclass/.style={draw=black!45, dashed, rounded corners=3.5pt, inner sep=2.2pt},
  egnewclass/.style={draw=egadd!70!black, dashed, rounded corners=3.5pt, inner sep=2.2pt},
  egedge/.style={-{Stealth[length=3.5pt]}, black!55},
  egnewedge/.style={-{Stealth[length=3.5pt]}, egadd!80!black},
  egnote/.style={font=\scriptsize\itshape, text=black!55, inner sep=1pt},
  egname/.style={font=\scriptsize, inner sep=1pt},
  egline/.style={draw=black!60, line width=0.25pt},
  egplus/.style={draw=black!80, line width=0.4pt},
}
\newcommand{\egGrid}[2]{%
  \foreach \k in {0,...,#1} {\draw[egline] (0,\k) -- (#2,\k);}
  \foreach \k in {0,...,#2} {\draw[egline] (\k,0) -- (\k,#1);}}
\newcommand{\egPic}[6]{%
  \begin{scope}[shift={($(#1.center)+(-#4,#5)$)}, x=1.5mm, y=-1.5mm]
    #6
    \egGrid{#2}{#3}
  \end{scope}}
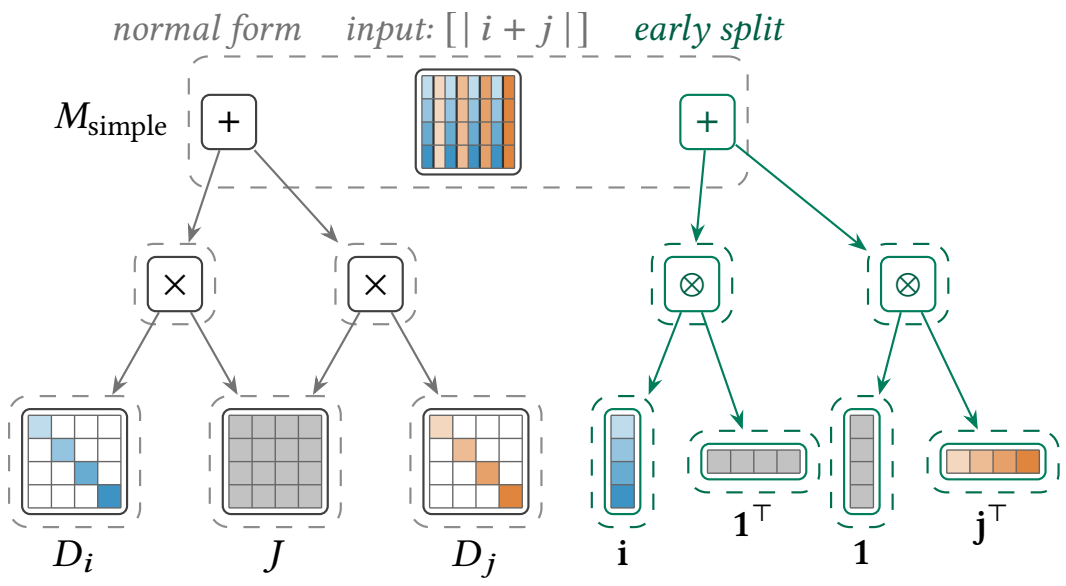
\begin{wrapfigure}{r}{0.5\textwidth}
\vspace{-2.5ex}
\centering
\resizebox{\linewidth}{!}{%
\begin{tikzpicture}
\node[egop] (nf) at (-1.55, 0) {$+$};
\node[egpic, minimum width=6.8mm, minimum height=6.8mm] (m) at (0, 0) {};
\egPic{m}{4}{4}{3mm}{3mm}{%
  \foreach \r in {0,...,3} {\foreach \c in {0,...,3} {
    \pgfmathtruncatemacro{\pr}{25+17*\r}
    \pgfmathtruncatemacro{\pc}{25+17*\c}
    \fill[egrow!\pr!white] (\c,\r) rectangle ++(0.5,1);
    \fill[egcol!\pc!white] (\c+0.5,\r) rectangle ++(0.5,1);
    \draw[egplus] (\c+0.5,\r) -- ++(0,1);
  }}}
\node[egop, egnew] (es) at (1.55, 0) {$+$};
\node[egclass, fit=(nf)(m)(es)] (root) {};
\node[egnote, above=0pt, anchor=south] at ($(nf |- root.north)+(-0.15,0)$) {normal form};
\node[egnote, above=0pt] at (m |- root.north) {input: $\comp{i + j}$};
\node[egnote, above=0pt, anchor=south, text=egadd!60!black] at ($(es |- root.north)+(0,0)$) {early split};
\node[egname, left=2pt of root] {$M_\textrm{simple}$};
\node[egop] (dj) at (-1.9, -1.05) {$\mmul$};
\node[egclass, fit=(dj)] (cdj) {};
\node[egop] (jd) at (-0.6, -1.05) {$\mmul$};
\node[egclass, fit=(jd)] (cjd) {};
\node[egpic, minimum width=6.8mm, minimum height=6.8mm] (di) at (-2.55, -2.2) {};
\egPic{di}{4}{4}{3mm}{3mm}{%
  \foreach \r in {0,...,3} {\pgfmathtruncatemacro{\p}{25+17*\r} \fill[egrow!\p!white] (\r,\r) rectangle ++(1,1);}}
\node[egclass, fit=(di)] (cdi) {};
\node[egpic, minimum width=6.8mm, minimum height=6.8mm] (jj) at (-1.25, -2.2) {};
\egPic{jj}{4}{4}{3mm}{3mm}{\fill[black!24!white] (0,0) rectangle (4,4);}
\node[egclass, fit=(jj)] (cj) {};
\node[egpic, minimum width=6.8mm, minimum height=6.8mm] (dd) at (0.05, -2.2) {};
\egPic{dd}{4}{4}{3mm}{3mm}{%
  \foreach \c in {0,...,3} {\pgfmathtruncatemacro{\p}{25+17*\c} \fill[egcol!\p!white] (\c,\c) rectangle ++(1,1);}}
\node[egclass, fit=(dd)] (cdd) {};
\node[egname, below=1pt of cdi] {$D_i$};
\node[egname, below=1pt of cj] {$J$};
\node[egname, below=1pt of cdd] {$D_j$};
\node[egop, egnew] (ik) at (1.45, -1.05) {$\otimes$};
\node[egnewclass, fit=(ik)] (cik) {};
\node[egop, egnew] (jk) at (2.85, -1.05) {$\otimes$};
\node[egnewclass, fit=(jk)] (cjk) {};
\node[egpic, egnew, minimum width=2.3mm, minimum height=6.8mm] (icol) at (1.0, -2.2) {};
\egPic{icol}{4}{1}{0.75mm}{3mm}{%
  \foreach \r in {0,...,3} {\pgfmathtruncatemacro{\p}{25+17*\r} \fill[egrow!\p!white] (0,\r) rectangle ++(1,1);}}
\node[egnewclass, fit=(icol)] (cicol) {};
\node[egpic, egnew, minimum width=6.8mm, minimum height=2.3mm] (onerow) at (1.85, -2.2) {};
\egPic{onerow}{1}{4}{3mm}{0.75mm}{\fill[black!24!white] (0,0) rectangle (4,1);}
\node[egnewclass, fit=(onerow)] (conerow) {};
\node[egpic, egnew, minimum width=2.3mm, minimum height=6.8mm] (onecol) at (2.55, -2.2) {};
\egPic{onecol}{4}{1}{0.75mm}{3mm}{\fill[black!24!white] (0,0) rectangle (1,4);}
\node[egnewclass, fit=(onecol)] (conecol) {};
\node[egpic, egnew, minimum width=6.8mm, minimum height=2.3mm] (jrow) at (3.4, -2.2) {};
\egPic{jrow}{1}{4}{3mm}{0.75mm}{%
  \foreach \c in {0,...,3} {\pgfmathtruncatemacro{\p}{25+17*\c} \fill[egcol!\p!white] (\c,0) rectangle ++(1,1);}}
\node[egnewclass, fit=(jrow)] (cjrow) {};
\node[egname, below=1pt of cicol] {$\mathbf{i}$};
\node[egname, below=1pt of conerow] {$\mathbf{1}^{\top}$};
\node[egname, below=1pt of conecol] {$\mathbf{1}$};
\node[egname, below=1pt of cjrow] {$\mathbf{j}^{\top}$};
\draw[egedge] (nf) -- (cdj);
\draw[egedge] (nf) -- (cjd);
\draw[egedge] (dj) -- (cdi);
\draw[egedge] (dj) -- (cj);
\draw[egedge] (jd) -- (cj);
\draw[egedge] (jd) -- (cdd);
\draw[egnewedge] (es) -- (cik);
\draw[egnewedge] (es) -- (cjk);
\draw[egnewedge] (ik) -- (cicol);
\draw[egnewedge] (ik) -- (conerow);
\draw[egnewedge] (jk) -- (conecol);
\draw[egnewedge] (jk) -- (cjrow);
\end{tikzpicture}}%
\setlength{\abovecaptionskip}{1.5ex}
\caption{The final e-graph of $M_\textrm{simple}$ after all exploration rules, with a dashed box around each e-class.}
\label{fig:egraph-simple}
\vspace{-2ex}
\end{wrapfigure}
\paragraph{Exploration}
Though fully reduced, the normal form may not possess the cheapest circuit.
The compiler further explores potentially better forms via undirected exploration rules with no termination guarantee.
For efficiency, it uses equality saturation~\citep{tateEqualitySaturationNew2009}, seeding the e-graph with the normal form.
Whenever the left side of some exploration rule matches some sub-expression in the e-graph, the compiler adds the right side to the corresponding e-class.
This process repeats until saturation or up to a fixed bound on the size of the graph.

\begin{example}
Figure~\ref{fig:egraph-simple} depicts the final e-graph of $M_\textrm{simple}$.
Starting from the normal form, the compiler uses the \emph{early split} rule to write each summand of $\comp{i + j}$ as a column times a row vector.
These vectors then have no sum left to split, and this graph of 12 e-classes is saturated.
For the $F_1$ example, three exploration rules add 28 forms, among them the factoring $D \mmul C$ of Section~\ref{sec:overview}.
\end{example}

\subsection{Cost-Aware Circuit Selection}
\label{sec:selection}

The compiler next evaluates each symbolic matrix form identified by the exploration phase as a candidate block encoding by selecting a circuit for each component.
For this step, the compiler uses a database of known circuit constructions, presented in full in Table~\ref{tab:appendix-constructions} in Appendix~\ref{sec:appendix-constructions}.

\colorlet{selrow}{oiblue}
\colorlet{selcol}{oiverm}
\tikzset{
  selwire/.style={draw=black, thick, line cap=round},
  selhad/.style={draw=black, thick, fill=white, inner sep=0pt, minimum size=4.5mm, font=\scriptsize},
  selgate/.style={draw, thick, inner sep=2.5pt, minimum width=7mm, minimum height=5mm, font=\scriptsize},
  selgateJ/.style={selgate, draw=black!50, fill=black!4, text=black!60},
  selgateDi/.style={selgate, draw=selrow, fill=selrow!6, text=selrow},
  selgateDj/.style={selgate, draw=selcol, fill=selcol!6, text=selcol},
  selprod/.style={draw=none, inner sep=3pt, outer sep=0pt},
  selopen/.style={circle, draw=black, thick, fill=white, inner sep=0pt, outer sep=0pt, minimum size=0.12cm},
  selfull/.style={circle, draw=black, thick, fill=black, inner sep=0pt, outer sep=0pt, minimum size=0.12cm},
  selline/.style={draw=black!65, line width=0.3pt},
  selop/.style={font=\small},
  selname/.style={font=\scriptsize, inner sep=1pt},
  selcost/.style={font=\footnotesize, inner sep=1pt},
  selrowlabel/.style={font=\scriptsize\itshape, text=black!55, inner sep=1pt, anchor=south, rotate=90},
  selbracket/.style={draw=black!75, line width=0.4pt},
}
\newcommand{\selProd}[2]{%
  \node[selprod, fit=#2] (#1) {};
  \path let \p1=(#1.south west), \p2=(#1.north east),
    \n1={2*(\x2-\x1)+2*(\y2-\y1)-(8-2*pi)*3pt},
    \n2={\n1/round(\n1/6pt)} in
    [draw=black, thick, rounded corners=3pt, dash pattern=on \n2/2 off \n2/2, dash phase=\n2/4]
    (#1.north) -- (#1.north east) -- (#1.south east) -- (#1.south west) -- (#1.north west) -- cycle;}
\newcommand{\selGrid}[1]{%
  \foreach \k in {0,...,#1} {\draw[selline] (0,\k) -- (#1,\k); \draw[selline] (\k,0) -- (\k,#1);}}
\newcommand{\selPic}[3]{%
  \begin{scope}[shift={(#1,#2)}, x=1.6mm, y=-1.6mm]
    #3
    \selGrid{5}
  \end{scope}}
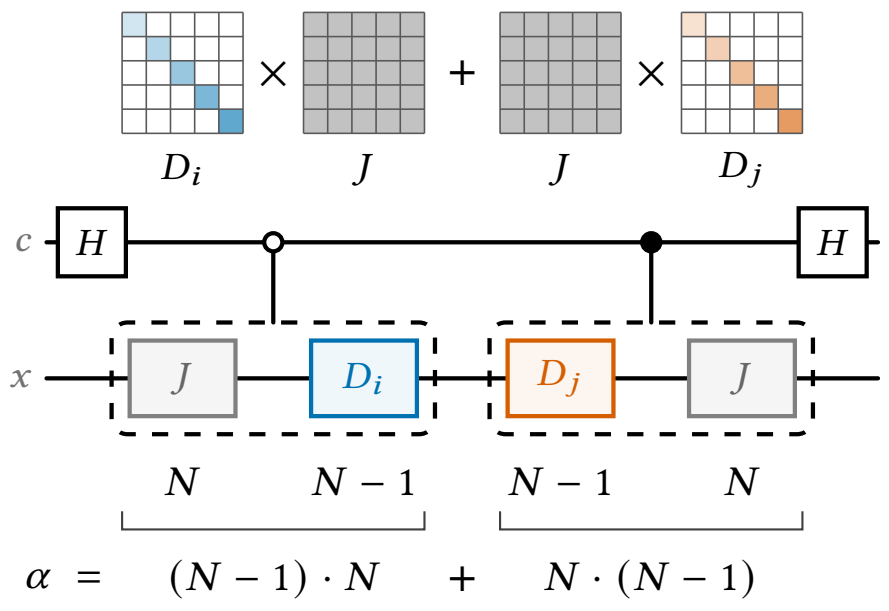
\begin{wrapfigure}{r}{.415\textwidth}
\vspace{-1ex}
\centering
\footnotesize
\begin{tikzpicture}
\selPic{1.4}{3.65}{%
  \foreach \r in {0,...,4} {\pgfmathtruncatemacro{\p}{18+11*\r} \fill[selrow!\p!white] (\r,\r) rectangle ++(1,1);}}
\node[selop] at (2.4, 3.25) {$\mmul$};
\selPic{2.6}{3.65}{\fill[black!24!white] (0,0) rectangle (5,5);}
\node[selop] at (3.65, 3.25) {$+$};
\selPic{3.9}{3.65}{\fill[black!24!white] (0,0) rectangle (5,5);}
\node[selop] at (4.9, 3.25) {$\mmul$};
\selPic{5.1}{3.65}{%
  \foreach \c in {0,...,4} {\pgfmathtruncatemacro{\p}{18+11*\c} \fill[selcol!\p!white] (\c,\c) rectangle ++(1,1);}}
\node[selname, yshift=0.5ex] at (1.8, 2.55) {$D_i$};
\node[selname, yshift=0.5ex] at (3.0, 2.55) {$J$};
\node[selname, yshift=0.5ex] at (4.3, 2.55) {$J$};
\node[selname, yshift=0.5ex] at (5.5, 2.55) {$D_j$};
\begin{scope}[yshift=-1ex]
\node[selname, text=black!55, anchor=east] at (0.85, 2.25) {$c$};
\node[selname, text=black!55, anchor=east] at (0.85, 1.35) {$x$};
\draw[selwire] (0.9, 2.25) -- (6.4, 2.25);
\draw[selwire] (0.9, 1.35) -- (6.4, 1.35);
\node[selhad] (had) at (1.2, 2.25) {$H$};
\node[selhad] (hadd) at (6.1, 2.25) {$H$};
\node[selgateJ] (g1) at (1.8, 1.35) {$J$};
\node[selgateDi] (g2) at (3.0, 1.35) {$D_i$};
\node[selgateDj] (g3) at (4.3, 1.35) {$D_j$};
\node[selgateJ] (g4) at (5.5, 1.35) {$J$};
\selProd{p1}{(g1)(g2)}
\selProd{p2}{(g3)(g4)}
\draw[selwire] (2.4, 2.25) -- (p1.north);
\draw[selwire] (4.9, 2.25) -- (p2.north);
\node[selopen] at (2.4, 2.25) {};
\node[selfull] at (4.9, 2.25) {};
\node[selcost, anchor=east, yshift=1ex] at (1.3, -0.1) {$\alpha \;=$};
\node[selcost, yshift=1ex] at (1.8, 0.55) {$N$};
\node[selcost, yshift=1ex] at (3.0, 0.55) {$N - 1$};
\node[selcost, yshift=1ex] at (4.3, 0.55) {$N - 1$};
\node[selcost, yshift=1ex] at (5.5, 0.55) {$N$};
\draw[selbracket, yshift=1ex] (1.4, 0.35) -- (1.4, 0.23) -- (3.4, 0.23) -- (3.4, 0.35);
\draw[selbracket, yshift=1ex] (3.9, 0.35) -- (3.9, 0.23) -- (5.9, 0.23) -- (5.9, 0.35);
\node[selcost, yshift=1ex] at (2.4, -0.1) {$(N - 1) \cdot N$};
\node[selop, yshift=1ex] at (3.65, -0.1) {$+$};
\node[selcost, yshift=1ex] at (4.9, -0.1) {$N \cdot (N - 1)$};
\end{scope}
\end{tikzpicture}
\setlength{\abovecaptionskip}{1.5ex}%
\caption{Circuit selection for $M_\textrm{simple}$.}
\label{fig:compilation}
\vspace{-3ex}
\end{wrapfigure}

Each circuit construction has a subnormalization $\alpha$ and gate and qubit counts parameterized by the dimension $N$ and other aspects of the matrix, as listed in Table~\ref{tab:appendix-alpha} in Appendix~\ref{sec:appendix-alpha}.
Instead of blindly generating large circuits, the compiler symbolically computes the cost of each candidate and selects the best one according to a user-given target metric such as total time cost.

Moreover, every circuit construction is independent of the concrete dimension $N$, so the compiler can synthesize correct block encodings for an unbounded family of matrices from a fixed database of constructions, with no explicit circuits required from the programmer.

\begin{example}
To illustrate circuit selection for named matrices, Figure~\ref{fig:compilation} depicts the simplified circuit for the $M_\textrm{simple}$ example.
The all-ones matrix $J = \comp{1}$ matches the \emph{rank-one} construction $\comp{c\,u(i)\,v(j)}$ with $u = v = \mathbf{1}$, which prepares $u$ and $v$ as the uniform state $U_N |0\rangle = \frac{1}{\sqrt{N}} \sum_{i < N} |i\rangle$:
{\setlength{\belowdisplayskip}{0pt} \setlength{\belowdisplayshortskip}{0pt}
\setlength{\abovedisplayskip}{0pt} \setlength{\abovedisplayshortskip}{0pt}
\[
  J = N \cdot U_N |0\rangle\langle 0| U_N^{\dagger}, \quad \text{with} \ \ U_N = H \ \ \text{and} \ \ H |0\rangle\langle 0| H = \tfrac{1}{2} \begin{bmatrix} 1 & 1 \\ 1 & 1 \end{bmatrix} = \tfrac{1}{2} J \ \ \text{at } N = 2,%
\]}%
so the $J$ box of Figure~\ref{fig:msimple-pieces} runs $U_N^{\dagger}$, post-selects the register on $0$, and runs $U_N$, with $\alpha = \|u\| \|v\| = N$. The diagonal $D_i$ matches the \emph{comparator} construction with $c = 1$, $s = 0$, and $b = 0$, in which the $D_i$ box prepares a counter in the uniform state over $N - 1$ values and post-selects on a comparison with $i$. The fraction $i / (N - 1)$ of the counter remains, and at $N = 4$ this sub-circuit encodes
\[
  \tfrac{1}{3} \operatorname{diag}(0, 1, 2, 3) = D_i / (N - 1), \quad \text{with}\ \ \alpha = N - 1 .
\]
\end{example}

\paragraph{Combinators}
A symbolic matrix form contains not only primitive matrices but also algebraic $+$, $\mmul$, and $\otimes$ on matrices.
Each combinator is realized via the circuit semantics given by Cobble~\citep{yuanCobbleCompilingBlock2026}.
Restated briefly, the circuit for a sum prepares a choice qubit in superposition, runs each operand under control or anti-control of that qubit, and then undoes the preparation of the choice qubit.
A product runs both operands in sequence, and a Kronecker product runs both in parallel.
The formulas that recursively compute the subnormalization of each combinator are:
\begin{align*}
  \alpha(p_1 + p_2) = \alpha(p_1) + \alpha(p_2), \quad \alpha(s \cdot p) = |s| \cdot \alpha(p), \quad \alpha(p_1 \mmul p_2) = \alpha(p_1 \otimes p_2) = \alpha(p_1) \cdot \alpha(p_2).
\end{align*}

\begin{example}
For $M_\textrm{simple}$, Figure~\ref{fig:compilation} depicts the composition of the circuits of $J$ and $D_i$ via the $+$ and $\mmul$ combinators, and shows the computation of the subnormalization of the entire circuit.
\end{example}

Because each circuit construction has a matching cost formula, the compiler can symbolically compute the cost of a symbolic matrix form for any dimension $N$ without generating the circuit.
Indeed, the value of many algebraic rewrite rules is in exposing a low-cost circuit construction.

\begin{example}
The \emph{scale extraction} and \emph{Kronecker split} rules rewrite $F_1$ as the four rank-1 terms of Equation~\eqref{eq:f1-reduced}.
Naively using the $+$ combinator at this point would result in a subnormalization of $\alpha = 6.2 + 3 \cdot 5.06 + 1 = 22.4$.
Instead, the \emph{fused sum} circuit construction builds rank-1 terms with mutually orthogonal rows and columns together.
The resulting subnormalization is the maximum rather than the sum of the components: $\alpha = \max(6.2, 5.06) + 1 = 7.2$, the value in Table~\ref{tab:overview-tab}.
\end{example}

Although the compiler computes a cost for each symbolic matrix form that is generic in the dimension $N$, the optimal candidate may vary as $N$ changes.
The compiler filters out all candidates whose cost strictly exceeds some other candidate at every $N$.
At the end, given a concrete value of $N$ provided by the user, the compiler emits the OpenQASM for the lowest-cost circuit.

\begin{theorem}[Compilation Time]
\label{thm:compile-time}
Given a well-formed program instantiated at dimension $N$ that has an efficient circuit, MARBLE completes compilation and emits a circuit in time $O(\polylog(N))$.
\end{theorem}

\begin{proof}
Most compiler stages run in time that depends only on the size of the program, which does not increase with $N$ by the property of static multiplicity.
For example, reduction terminates in a number of steps dependent only on the structure of the program, and exploration stops at a fixed bound.
When deciding a side condition $\varphi$ of a rewrite rule, the compiler operates over $N$ symbolically and checks each condition in time polynomial in the logarithmic bit width of $N$.

The compiler only examines the specific value of $N$ when writing out the final circuit (of poly-logarithmic size by premise), and when comparing the costs of two candidate representations of a matrix at a specific $N$.
This analysis also performs arithmetic on integers of width $O(\log N)$.
\end{proof}
\FloatBarrier

\section{Cost Certification}
\label{sec:certification}

After compilation, MARBLE automatically certifies three properties about the cost of a program.
In conjunction, they imply that the output circuit realizes an efficient block encoding of the matrix and can thus be used to satisfy the complexity requirements of quantum algorithms.

\subsection{Efficiency of Circuit}
\label{sec:efficient}

We say that a block encoding has an \emph{efficient} circuit when its numbers of gates and qubits scale as $O(\polylog(N))$ in the dimension $N$ of the matrix.
Intuitively, this property states that each iteration of the circuit is efficient to run on a quantum computer, even if many iterations are needed.

\paragraph{Mechanism}
MARBLE certifies this property for a chosen representation of the program when it can prove the bound symbolically over $N$.
As described in Sec.~\ref{sec:compilation}, the selection stage of the compiler ends with a set of circuits and corresponding cost formulas for gates, qubits, and subnormalization parameterized by $N$.
The compiler then checks whether they scale as $O(\polylog(N))$ by reasoning over sums, products, logarithms, and other operations such as division that can be handled by standard computer algebra.
If no proof is achieved, MARBLE conservatively outputs \emph{unknown}.

\begin{example}
One circuit candidate for $M_\textrm{simple}$ uses the \emph{entry loader} fallback approach, taking $1 + 2\lceil \log_2 N \rceil$ qubits and a gate cost that contains the term $2^{2\lceil \log_2 N \rceil}$.
MARBLE first analyzes the qubit formula $1 + (2 \cdot \lceil \log_2 N \rceil)$ which is clearly $O(\polylog(N))$.
By contrast, in the gate formula, $2^{2\lceil \log_2 N \rceil}$ is $\Omega(N^2)$.
This candidate is not efficient; the compiler selects another if available.
\end{example}

\subsection{Tightness of Subnormalization}
\label{sec:tight}

We call a block encoding \emph{asymptotically tight} when its subnormalization $\alpha$ converges to a constant multiple of the spectral norm $\sigma_1$ of the matrix, which is the lower bound of $\alpha$ (see Sec.~\ref{sec:background}).
We call this constant ratio \emph{tightness}, and we call a block encoding \emph{tight} when the ratio is one.
A tightness of one implies that the number of iterations that the block-encoding circuit must be repeated for the algorithm to produce a correct result is the best that is mathematically possible.

\paragraph{Mechanism}
Because it is intractable to directly compute the spectral norm $\sigma_1$ of a matrix with dimension on the order of $N = 2^{100}$, the compiler instead looks for a lower bound $L(N) \le \sigma_1(N)$ such that $\alpha \le c \cdot L$ implies $\alpha \le c \cdot \sigma_1$. Conveniently, any vector $x$ gives rise to such an $L$, as
\begin{equation}
  {|x^\dagger M x|} / {\|x\|^2} \;\le\; {\|x\|\,\|Mx\|} / {\|x\|^2} \;\le\; \sigma_1 = \|M\| ,
\end{equation}
where the first inequality is Cauchy-Schwarz, and the second inequality holds because the spectral norm is the largest value of $\|Mx\| / \|x\|$ over all $x$.
The compiler chooses appropriate families of vectors parameterized by $N$ to ensure that the sums in this fraction have closed forms.

\begin{example}
Taking $x$ to be the all-ones vector, we have $x^\dagger M_\textrm{simple}\, x = \sum_{i,j} (i + j) = N^2(N-1)$ and $\|x\|^2 = N$, giving a lower bound $L(N) = N(N-1)$. The candidate circuit for $M_\textrm{simple}$ in Fig.~\ref{fig:msimple-pieces} has $\alpha = 2N(N-1)$, which is 2$\times$ this lower bound.
Since a ratio can approach its limit from above, MARBLE picks $c$ slightly above the limit of $\alpha / L$ as $N$ grows, here $c = 2.0625$, and symbolically proves $\alpha \le c \cdot L$ for every $N \ge 8$. It concludes that this candidate is asymptotically tight.
\end{example}

To identify a precise $c$, the compiler tries multiple candidate lower bounds, such as by combining two or three vectors.
The largest lower bound yields the smallest and most precise tightness $c$.
On the other hand, the compiler can deduce that $\alpha$ for a given candidate circuit grows asymptotically faster than $\sigma_1$ as $N$ increases by finding another candidate whose $\alpha'$ grows strictly slower than $\alpha$.
Because $\alpha'$ is also an upper bound on $\sigma_1$, this block encoding cannot be asymptotically tight.

\subsection{Boundedness of Spectral Norm}
\label{sec:bounded}

Tightness states that the subnormalization cost of the block encoding is close to its lower bound, but not that this bound itself is efficient.
We say that a matrix is \emph{bounded} when its spectral norm $\sigma_1$ scales as $O(\polylog(N))$ --- a mathematical property of the target matrix itself rather than its block-encoding implementation.
Together with tightness of the block encoding, this property states that the number of iterations of the circuit needed is efficient in the dimension of the matrix.

\paragraph{Mechanism}
MARBLE analyzes this property by reusing the lower bounds $L$ on $\sigma_1$ computed in the previous subsection.
If any $L$ grows faster than $O(\polylog(N))$, then so does $\sigma_1$.
In the other direction, the subnormalization $\alpha$ of every candidate block encoding is an upper bound on $\sigma_1$, so if the compiler discovers any candidate whose $\alpha$ is $O(\polylog(N))$, then the matrix is bounded.

\begin{example}
No entry of a matrix can be larger in magnitude than $\sigma_1$, and so the entry of $M_\textrm{simple}$ at the first row and the last column gives $L = N - 1$, which grows faster than $O(\polylog(N))$.
Thus, $M_\textrm{simple}$ is not bounded and possesses no block encoding with poly-logarithmic total time cost.
\end{example}

\section{Evaluation}
\label{sec:evaluation}

In this section, we evaluate the performance of MARBLE via the following research questions:

\begin{enumerate}[leftmargin=2.35em]
  \item [\emph{RQ1.}] How do MARBLE's results compare to published hand-written circuits for the same matrices?
  \item [\emph{RQ2.}] Can MARBLE construct block-encoding circuits for matrices where none existed before?
  \item [\emph{RQ3.}] How does MARBLE perform against general-purpose encodings of arbitrary matrices?
  \item [\emph{RQ4.}] How does MARBLE's compile time scale with the matrix dimension $N$\@?
\end{enumerate}

\subsection{RQ1: Comparison Against Published Block Encodings}

\input{figures/rq1-gain}

\begin{enumerate}[leftmargin=2.35em]
  \item [\emph{RQ1.}] How do MARBLE's results compare to published hand-written circuits for the same matrices?
\end{enumerate}

\paragraph{Benchmarks}
To answer this question, we collected 39 existing block-encoding constructions from the literature of quantum algorithms for scientific computing, spanning lattice Boltzmann methods~\citep{penuelDetailedAssessmentCalculating2025,kuklinskiEfficientBlockencodingsRequire2025,jenningsEndtoEndQuantumAlgorithm2026,uenoDemonstrationQuantumCircuit2026}, finite elements~\citep{sureshQuantumBlockEncodings2026}, finite differences~\citep{mahmudMomentStructuredBlockEncodings2026,helleQuantumAdvectiondiffusionSolver2026,kharaziExplicitBlockEncodings2025}, and physical Hamiltonians~\citep{leeModalAnalysisQuantum2025,kanOptimizedQuantumAlgorithms2025,liuEfficientQuantumCircuit2024,childsFirstQuantumSimulation2018,kanResourceoptimizedFaulttolerantSimulation2025a,babbushQuantumSimulationSachdevYeKitaev2018,suFaultTolerantQuantumSimulations2021}.
Each reference provides the mathematical formula of the target matrix, and either lists the achieved subnormalization $\alpha$ or gives sufficient description of the circuit so that the subnormalization can be computed.
Six of the benchmark matrices are not given as a specialized circuit but instead assume the standard Pauli linear-combination-of-unitaries (LCU) circuit construction of~\citet{childsHamiltonianSimulation2012}.

\paragraph{Methodology}
We implemented each matrix as an entry comprehension in MARBLE and synthesized the circuit that achieves the lowest subnormalization $\alpha$.
Following \citet{yuanCobbleCompilingBlock2026,sunderhaufBlockencodingStructuredMatrices2024}, we target $\alpha$ as a metric because it dominates the performance of algorithms in practice as the number of times that the block-encoding circuit must be executed by the algorithm.

To check correctness of the circuit output by MARBLE, we numerically computed each matrix in NumPy, which acts as a reference implementation for small dimensions $N$.
At the same dimensions, we independently computed the unitary matrix represented by each OpenQASM circuit output by MARBLE, and confirmed the numerical equality of each encoded and target matrix.

To contextualize the results, we also computed the theoretical lower bound of $\alpha$, which is the spectral norm of the target matrix.
For all but seven benchmarks, the norm is computed analytically as a part of MARBLE's symbolic cost certification.
Otherwise, we conservatively bounded the norm by the largest singular value of the largest matrix instance that proved classically tractable.

\vspace{-.3ex}%
\paragraph{Results}
In Figure~\ref{fig:rq1-gain}, we present the ratio of the $\alpha$ of the published encoding to the $\alpha$ achieved by MARBLE, where a higher ratio indicates greater speedup.
The right end of each \textcolor{darkgray}{gray} bar represents the best theoretically possible speedup.
We present the full results in Table~\ref{tab:eval} in Appendix~\ref{sec:appendix-eval}.

MARBLE matches or improves $\alpha$ on all but two benchmarks.
On the low end, MARBLE automatically matches the $\alpha$ achieved by the construction of~\citet{jenningsEndtoEndQuantumAlgorithm2026} for lattice BGK collision matrices (row~11), which equals the $\sigma_1$ lower bound.
On the high end, MARBLE achieves a speedup of $10{,}487\times$ over~\citet{penuelDetailedAssessmentCalculating2025} for the D3Q27 Carleman matrix for fluid flow (row~3).

In rows~63 and~66, efficient circuit constructions exist for the matrix~\citep{childsFirstQuantumSimulation2018,babbushQuantumSimulationSachdevYeKitaev2018} but the matrix cannot be represented using MARBLE\@.
These matrices, corresponding to the Heisenberg and SYK models in physics respectively, require the loading of a number of explicit coefficients that grows with the dimension of the system.
In general, no system that assumes static multiplicity to enable efficient compilation, including MARBLE, can represent these matrices.

Reductions in $\alpha$ are generally larger for complex matrices with sub-structure.
For example, the matrix in row~3 has as constituents the matrices in rows~5, 7, and~9, and their savings compound.
For matrices whose formula is relatively simple, the gains are modest and often match baselines, e.g.\ the Laplacian operators in rows~50 and~52 whose $\alpha$ already equals the $\sigma_1$ lower bound.

For every output, MARBLE automatically certifies that the block encoding is efficient (gates per iteration is poly-logarithmic).
It computes the tightness (ratio of $\alpha$ to spectral norm) wherever the spectral norm is known.
Finally, MARBLE correctly deduces that the matrix is bounded (spectral norm is poly-logarithmic) for all cases except rows~45, 48, 49, 52, 53, and~67.
It correctly deduces that those six matrices are not efficiently encodable on a quantum computer, consistent with the analyses of \citet[Section 2.2]{kharaziExplicitBlockEncodings2025} for row~52 and \citet[Page 9]{suFaultTolerantQuantumSimulations2021} for row~67.

\vspace{-.3ex}%
\subsection{RQ2: Construction of Novel Block Encodings}
\label{sec:rq2}

\begin{enumerate}[leftmargin=2.35em]
  \item [\emph{RQ2.}] Can MARBLE construct block-encoding circuits for matrices where none existed before?
\end{enumerate}

For this question, we looked for complex problems whose matrices have known formulas but few, if any, quantum block-encoding circuits.
One such category of problems is \emph{structured Markov processes}, which model real-world processes such as communication traffic and network queues.
Indeed, quantum algorithms for structured Markov processes require block-encoding circuits for specific matrices~\citep{kalantzisEfficientSolutionsGeneral2025}, yet no explicit circuits exist to our knowledge.

We used MARBLE to construct explicit block encodings of these matrices, ranging from the M/M/1 queue to finite-buffer networks with blocking.
For example, the formula for the generator matrix $Q$ for an M/M/1 Markov process is given by~\citet[Eq. (8.6)]{grossFundamentalsQueueingTheory2008} as follows:
{\setlength{\belowdisplayskip}{3pt} \setlength{\belowdisplayshortskip}{0pt}
\setlength{\abovedisplayskip}{3pt} \setlength{\abovedisplayshortskip}{0pt}
\[Q^{\top}_{ij} = \lambda\,[i\mathrel{\texttt{==}}j+1] + \mu\,[i\mathrel{\texttt{==}}j-1] - [i\mathrel{\texttt{==}}j]\,\bigl(\lambda\,[j<2^b-1]+\mu\,[j>0]\bigr) \mathrel{\big|} i,j \in \{0, \dots, 2^b - 1\}\]}

This formula can be expressed as an entry comprehension, and MARBLE synthesizes an efficient block encoding for it within 0.11 seconds.
It achieves $\alpha$ within $1.063\times$ of the spectral norm lower bound, which MARBLE automatically confirms is poly-logarithmic in the system dimension $2^b$.

In rows 68--74 in Table~\ref{tab:eval} (Appendix~\ref{sec:appendix-eval}), we present full results for several Markov processes, such as queues with blocking and feedback and ring topologies.
Though our results are necessarily limited in scope, they suggest that MARBLE is flexible enough to handle not only matrices from well-known quantum algorithms but also potential new application domains for researchers.

\subsection{RQ3: Comparison Against General-Purpose Encodings of Arbitrary Matrices}
\label{sec:rq3}

\begin{figure}[t]
\centering
\begin{tikzpicture}
\begin{groupplot}[group style={group size=3 by 3, horizontal sep=24pt, vertical sep=16pt},
  width=100pt, height=46pt, scale only axis, xmode=log, ymode=log, log basis x=2, xmin=1, xmax=128,
  xtick={1,2,4,8,16,32,64}, xticklabels={1,2,4,8,16,32,64}, xmajorgrids, ymajorgrids, axis x line*=bottom, axis y line*=left, clip=false,
  xticklabel style={font=\tiny}, yticklabel style={font=\tiny}]
\nextgroupplot[ymin=1, ymax=1e+05, ytick={1e0,1e1,1e2,1e3,1e4,1e5}, yticklabels={$10^{0}$,$10^{1}$,$10^{2}$,$10^{3}$,$10^{4}$,$10^{5}$}, yminorticks=false]
\addplot[oiblue, solid, mark=*, mark size=1.3pt] coordinates {(1,5) (2,5) (3,5) (4,5) (5,5) (6,5) (7,5) (8,5) (9,5) (10,5) (11,5) (12,5) (16,5) (20,5) (24,5) (32,5) (48,5) (64,5) (80,5) (100,5)};
\addplot[oiverm, densely dashed, mark=square*, mark size=1.3pt] coordinates {(1,3) (2,10) (3,20) (4,40) (5,80) (6,160) (7,320) (8,640) (9,1280) (10,2560) (11,5120)};
\node[anchor=south east, font=\scriptsize, text=oiblue, inner sep=1pt, yshift=2.5pt] at (axis cs:100,5) {5};
\draw[gray, densely dotted, thick] (axis cs:12,1) -- (axis cs:12,1e+05);
\nextgroupplot[ymin=1, ymax=1e+07, ytick={1e0,1e2,1e4,1e6}, yticklabels={$10^{0}$,$10^{2}$,$10^{4}$,$10^{6}$}, yminorticks=false]
\addplot[oiblue, solid, mark=*, mark size=1.3pt] coordinates {(1,98) (2,162) (3,226) (4,290) (5,354) (6,418) (7,482) (8,546) (9,610) (10,674) (11,738) (12,802) (16,1058) (20,1314) (24,1570) (32,2082) (48,3106) (64,4130) (80,5154) (100,6434)};
\addplot[oiverm, densely dashed, mark=square*, mark size=1.3pt] coordinates {(1,4) (2,16) (3,63) (4,250) (5,1002) (6,3985) (7,15882) (8,63382) (9,252899) (10,1.0088e+06) (11,4.02434e+06)};
\node[anchor=south east, font=\scriptsize, text=oiblue, inner sep=1pt, yshift=2.5pt] at (axis cs:100,6434) {6{,}434};
\draw[gray, densely dotted, thick] (axis cs:12,1) -- (axis cs:12,1e+07);
\nextgroupplot[ymin=1, ymax=1e+11, ytick={1e0,1e3,1e6,1e9}, yticklabels={$10^{0}$,$10^{3}$,$10^{6}$,$10^{9}$}, yminorticks=false]
\addplot[oiblue, solid, mark=*, mark size=1.3pt] coordinates {(1,490) (2,810) (3,1130) (4,1450) (5,1770) (6,2090) (7,2410) (8,2730) (9,3050) (10,3370) (11,3690) (12,4010) (16,5290) (20,6570) (24,7850) (32,10410) (48,15530) (64,20650) (80,25770) (100,32170)};
\addplot[oiverm, densely dashed, mark=square*, mark size=1.3pt] coordinates {(1,12) (2,160) (3,1260) (4,10000) (5,80160) (6,637600) (7,5.08224e+06) (8,4.05645e+07) (9,3.23711e+08) (10,2.58252e+09) (11,2.06046e+10)};
\node[anchor=south east, font=\scriptsize, text=oiblue, inner sep=1pt, yshift=2.5pt] at (axis cs:100,32170) {32{,}170};
\draw[gray, densely dotted, thick] (axis cs:12,1) -- (axis cs:12,1e+11);
\nextgroupplot[ymin=1, ymax=1e+05, ytick={1e0,1e1,1e2,1e3,1e4,1e5}, yticklabels={$10^{0}$,$10^{1}$,$10^{2}$,$10^{3}$,$10^{4}$,$10^{5}$}, yminorticks=false]
\addplot[oiblue, solid, mark=*, mark size=1.3pt] coordinates {(7.49185,18.201) (8.49185,18.201) (9.49185,18.201) (10.4919,18.201) (11.4919,18.201) (12.4919,18.201) (13.4919,18.201) (14.4919,18.201) (15.4919,18.201) (16.4919,18.201) (17.4919,18.201) (18.4919,18.201) (22.4919,18.201) (26.4919,18.201) (30.4919,18.201) (38.4919,18.201) (54.4919,18.201) (70.4919,18.201) (86.4919,18.201) (106.492,18.201)};
\addplot[oiverm, densely dashed, mark=square*, mark size=1.3pt] coordinates {(7.49185,1024) (8.49185,2048) (9.49185,4096) (10.4919,8192)};
\node[anchor=south east, font=\scriptsize, text=oiblue, inner sep=1pt, yshift=2.5pt] at (axis cs:106.5,18.201) {18.2};
\draw[gray, densely dotted, thick] (axis cs:11.49,1) -- (axis cs:11.49,1e+05);
\nextgroupplot[ymin=1e+04, ymax=1e+07, ytick={1e4,1e5,1e6,1e7}, yticklabels={$10^{4}$,$10^{5}$,$10^{6}$,$10^{7}$}, yminorticks=false]
\addplot[oiblue, solid, mark=*, mark size=1.3pt] coordinates {(7.49185,21814) (8.49185,21814) (9.49185,21814) (10.4919,21814) (11.4919,21814) (12.4919,21814) (13.4919,21814) (14.4919,21814) (15.4919,21814) (16.4919,21814) (17.4919,21814) (18.4919,21814) (22.4919,21814) (26.4919,21814) (30.4919,21814) (38.4919,21814) (54.4919,21814) (70.4919,21814) (86.4919,21814) (106.492,21814)};
\addplot[oiverm, densely dashed, mark=square*, mark size=1.3pt] coordinates {(7.49185,65533) (8.49185,262127) (9.49185,1.04851e+06) (10.4919,4.19408e+06)};
\node[anchor=south east, font=\scriptsize, text=oiblue, inner sep=1pt, yshift=2.5pt] at (axis cs:106.5,21814) {21{,}814};
\draw[gray, densely dotted, thick] (axis cs:11.49,1e+04) -- (axis cs:11.49,1e+07);
\nextgroupplot[ymin=1e+04, ymax=1e+11, ytick={1e4,1e6,1e8,1e10}, yticklabels={$10^{4}$,$10^{6}$,$10^{8}$,$10^{10}$}, yminorticks=false]
\addplot[oiblue, solid, mark=*, mark size=1.3pt] coordinates {(7.49185,55317.2) (8.49185,55317.2) (9.49185,55317.2) (10.4919,55317.2) (11.4919,55317.2) (12.4919,55317.2) (13.4919,55317.2) (14.4919,55317.2) (15.4919,55317.2) (16.4919,55317.2) (17.4919,55317.2) (18.4919,55317.2) (22.4919,55317.2) (26.4919,55317.2) (30.4919,55317.2) (38.4919,55317.2) (54.4919,55317.2) (70.4919,55317.2) (86.4919,55317.2) (106.492,55317.2)};
\addplot[oiverm, densely dashed, mark=square*, mark size=1.3pt] coordinates {(7.49185,6.71058e+07) (8.49185,5.36836e+08) (9.49185,4.29469e+09) (10.4919,3.43579e+10)};
\node[anchor=south east, font=\scriptsize, text=oiblue, inner sep=1pt, yshift=2.5pt] at (axis cs:106.5,55317.2) {55{,}317};
\draw[gray, densely dotted, thick] (axis cs:11.49,1e+04) -- (axis cs:11.49,1e+11);
\nextgroupplot[ymin=1, ymax=1e+05, ytick={1e0,1e1,1e2,1e3,1e4,1e5}, yticklabels={$10^{0}$,$10^{1}$,$10^{2}$,$10^{3}$,$10^{4}$,$10^{5}$}, yminorticks=false]
\addplot[oiblue, solid, mark=*, mark size=1.3pt] coordinates {(3,4.24264) (4,5.65685) (5,7.07107) (6,8.48528) (7,9.89949) (8,11.3137) (9,12.7279) (10,14.1421) (11,15.5563) (12,16.9706) (16,22.6274) (20,28.2843) (24,33.9411) (32,45.2548) (48,67.8823) (64,90.5097) (80,113.137) (100,141.421)};
\addplot[oiverm, densely dashed, mark=square*, mark size=1.3pt] coordinates {(3,24) (4,64) (5,160) (6,384) (7,896) (8,2048) (9,4608) (10,10240) (11,22528) (12,49152)};
\node[anchor=south east, font=\scriptsize, text=oiblue, inner sep=1pt, yshift=2.5pt] at (axis cs:100,141.421) {141};
\draw[gray, densely dotted, thick] (axis cs:13,1) -- (axis cs:13,1e+05);
\nextgroupplot[ymin=10, ymax=1e+08, ytick={1e1,1e3,1e5,1e7}, yticklabels={$10^{1}$,$10^{3}$,$10^{5}$,$10^{7}$}, yminorticks=false, xlabel={$\log_2 N$ (matrix is $N \times N$; both axes logarithmic)}]
\addplot[oiblue, solid, mark=*, mark size=1.3pt] coordinates {(3,118) (4,128) (5,206) (6,246) (7,282) (8,264) (9,358) (10,410) (11,446) (12,462) (16,528) (20,754) (24,870) (32,1048) (48,1662) (64,2080) (80,2722) (100,3390)};
\addplot[oiverm, densely dashed, mark=square*, mark size=1.3pt] coordinates {(3,32) (4,91) (5,512) (6,1632) (7,8192) (8,32484) (9,131072) (10,507049) (11,2.09715e+06) (12,8.05173e+06)};
\node[anchor=south east, font=\scriptsize, text=oiblue, inner sep=1pt, yshift=2.5pt] at (axis cs:100,3390) {3{,}390};
\draw[gray, densely dotted, thick] (axis cs:13,10) -- (axis cs:13,1e+08);
\nextgroupplot[ymin=100, ymax=1e+12, ytick={1e2,1e4,1e6,1e8,1e10,1e12}, yticklabels={$10^{2}$,$10^{4}$,$10^{6}$,$10^{8}$,$10^{10}$,$10^{12}$}, yminorticks=false]
\addplot[oiblue, solid, mark=*, mark size=1.3pt] coordinates {(3,300) (4,288) (5,980) (6,1416) (7,1876) (8,1472) (9,3060) (10,4040) (11,4796) (12,5136) (16,6272) (20,13840) (24,18336) (32,25344) (48,66624) (64,100864) (80,178240) (100,276400)};
\addplot[oiverm, densely dashed, mark=square*, mark size=1.3pt] coordinates {(3,768) (4,5824) (5,81920) (6,626688) (7,7.34003e+06) (8,6.65272e+07) (9,6.0398e+08) (10,5.19218e+09) (11,4.72446e+10) (12,3.95759e+11)};
\node[anchor=south east, font=\scriptsize, text=oiblue, inner sep=1pt, yshift=2.5pt] at (axis cs:100,276400) {$2.8 \times 10^{5}$};
\draw[gray, densely dotted, thick] (axis cs:13,100) -- (axis cs:13,1e+12);
\end{groupplot}
\node[anchor=south, rotate=90, align=center, font=\scriptsize\itshape] at ($(group c1r1.west) + (-20pt,0)$) {M/M/1 queue};
\node[anchor=south, rotate=90, align=center, font=\scriptsize\itshape] at ($(group c1r2.west) + (-20pt,0)$) {second-order Carleman\\D2Q9, lattice};
\node[anchor=south, rotate=90, align=center, font=\scriptsize\itshape] at ($(group c1r3.west) + (-20pt,0)$) {transverse-field\\Ising ring};
\node[anchor=south, align=center, font=\footnotesize] at ($(group c1r1.north) + (0,3pt)$) {\textbf{Subnormalization $\boldsymbol\alpha$}};
\node[anchor=south, align=center, font=\footnotesize] at ($(group c2r1.north) + (0,3pt)$) {\textbf{non-Clifford gates}};
\node[anchor=south, align=center, font=\footnotesize] at ($(group c3r1.north) + (0,3pt)$) {$\boldsymbol\alpha\boldsymbol\times{}$\textbf{non-Clifford gates}};
\end{tikzpicture}
\setlength{\abovecaptionskip}{1ex}%
\caption[MARBLE against FABLE as the matrix size $N$ grows]{MARBLE (\tikz[baseline=-0.5ex]{\draw[oiblue](0,0)--(0.5,0);\fill[oiblue](0.25,0)circle(2pt);}) against FABLE (\tikz[baseline=-0.5ex]{\draw[oiverm,densely dashed](0,0)--(0.5,0);\fill[oiverm](0.19,-0.06)rectangle(0.31,0.06);}) as the matrix size $N$ grows.
Lower is better in every panel.
The dotted \textcolor{darkgray}{gray} line marks FABLE timeout (>5 minutes or 8 GB memory).
Note that the encoded matrix is $N \times N$, and the $x$-axis shows $\log_2 N$ on a logarithmic scale, so each $x$-tick doubles $\log_2 N$ and squares $N$.}
\label{fig:rq3}
\vspace{-7pt}%
\end{figure}

\begin{enumerate}[leftmargin=2.35em]
  \item [\emph{RQ3.}] How does MARBLE perform against general-purpose encodings of arbitrary matrices?
\end{enumerate}

\paragraph{Methodology}
In this question, we evaluate against FABLE~\citep{campsFABLEFastApproximate2022}, a leading tool that accepts an arbitrary numerical matrix and produces a quantum circuit for its block encoding.
Unlike MARBLE, FABLE explicitly computes every entry of the matrix and thus requires that the problem instance be classically tractable.
We selected three circuit families from RQ1 and RQ2 and used FABLE and MARBLE to synthesize their block encoding circuits.
We compared these circuits on subnormalization $\alpha$, number of non-Clifford gates, and the product of the two.

\paragraph{Results}
In Figure~\ref{fig:rq3}, we present the subnormalization, gate count, and their product for block-encoding circuits generated by both tools.
The first two columns show MARBLE's least-$\alpha$ circuit, and the third its least-product circuit.
FABLE scales polynomially in matrix dimension $N$, whereas MARBLE scales poly-logarithmically in subnormalization and gates.
Notably, subnormalization scales quickly for FABLE but is constant for MARBLE over two matrix families --- mathematically explained by the spectral norm of those matrix families being independent of the dimension $N$.

\subsection{RQ4: Scaling of Compile Time with Matrix Dimension}
\label{sec:rq4}

\begin{figure}[t]
\centering
\begin{tikzpicture}
\begin{groupplot}[group style={group size=3 by 1, horizontal sep=22pt},
  width=78pt, height=48pt, scale only axis, xmin=0, xmax=100,
  xtick={0,20,40,60,80,100},
  xticklabels={$2^{0}$,$2^{20}$,$2^{40}$,$2^{60}$,$2^{80}$,$2^{100}$}, stack plots=y, xticklabel style={font=\tiny}, yticklabel style={font=\tiny, /pgf/number format/fixed}, scaled y ticks=false, ymajorgrids, grid style={gray!18},
  axis x line*=bottom, axis y line*=left, clip=false, title style={font=\footnotesize, yshift=-6pt, align=center, text depth=0pt},
  legend style={draw=none, font=\scriptsize, text width=50pt, align=left, inner sep=0pt, /tikz/every even column/.append style={column sep=10pt}}, area legend]
\nextgroupplot[ymin=0, ymax=0.26, title={\emph{M/M/1 queue}}, legend to name=bandkey, legend columns=1, reverse legend, ylabel={time (seconds)}]
\addplot[fill=gray!22, draw=none] coordinates {(1,0.0575123) (10,0.0604592) (20,0.0585918) (30,0.0603975) (40,0.0625118) (50,0.0599784) (60,0.0606337) (70,0.058877) (80,0.0600299) (90,0.0662645) (100,0.0604814)} \closedcycle;
\addplot[fill=oiblue!88, draw=none] coordinates {(1,0.0943492) (10,0.0984414) (20,0.0947217) (30,0.0960035) (40,0.0959488) (50,0.0953211) (60,0.0951544) (70,0.0943142) (80,0.0965877) (90,0.0954701) (100,0.094944)} \closedcycle;
\addplot[fill=oiverm!65, draw=none] coordinates {(1,0) (10,0.00184519) (20,0.00387878) (30,0.00378703) (40,0.00165608) (50,0.00472428) (60,0.00691409) (70,0.00753283) (80,0.00887748) (90,0.0143076) (100,0.0115344)} \closedcycle;
\addlegendentry{Python startup}
\addlegendentry{compile program}
\addlegendentry{emit OpenQASM}
\node[anchor=north east, font=\scriptsize, text=black, inner sep=1pt, align=right] at (rel axis cs:0.98,0.98) {largest $N$: 0.17\,s};
\nextgroupplot[ymin=0, ymax=4.0, title={\emph{Carleman D2Q9}}]
\addplot[fill=gray!22, draw=none, forget plot] coordinates {(7.49185,0.0607481) (10.4919,0.0592359) (20.4919,0.0600781) (30.4919,0.0593667) (40.4919,0.0629707) (50.4919,0.0606313) (60.4919,0.0611208) (70.4919,0.0671598) (80.4919,0.0607054) (90.4919,0.0601037) (99.4919,0.0606607)} \closedcycle;
\addplot[fill=oiblue!88, draw=none, forget plot] coordinates {(7.49185,2.49049) (10.4919,2.60846) (20.4919,2.56835) (30.4919,2.55204) (40.4919,2.56361) (50.4919,2.52772) (60.4919,2.56175) (70.4919,2.59548) (80.4919,2.56672) (90.4919,2.56387) (99.4919,2.58752)} \closedcycle;
\addplot[fill=oiverm!65, draw=none, forget plot] coordinates {(7.49185,0.105438) (10.4919,0.00461535) (20.4919,0.0383848) (30.4919,0.0594775) (40.4919,0.0800254) (50.4919,0.0548268) (60.4919,0.0175022) (70.4919,0) (80.4919,0.0812512) (90.4919,0.0565438) (99.4919,0)} \closedcycle;
\node[anchor=north east, font=\scriptsize, text=black, inner sep=1pt, align=right] at (rel axis cs:0.98,0.98) {largest $N$: 2.65\,s};
\nextgroupplot[ymin=0, ymax=0.40, title={\emph{4th-order derivative}}]
\addplot[fill=gray!22, draw=none, forget plot] coordinates {(3,0.0603861) (10,0.060416) (20,0.0621245) (30,0.0579674) (40,0.0612933) (50,0.0642674) (60,0.0615746) (70,0.0590271) (80,0.0588004) (90,0.0595383) (100,0.0597363)} \closedcycle;
\addplot[fill=oiblue!88, draw=none, forget plot] coordinates {(3,0.116904) (10,0.122389) (20,0.126364) (30,0.124724) (40,0.12507) (50,0.122679) (60,0.124568) (70,0.121851) (80,0.124194) (90,0.121118) (100,0.123175)} \closedcycle;
\addplot[fill=oiverm!65, draw=none, forget plot] coordinates {(3,0.00653403) (10,0.0026144) (20,0.00478198) (30,0.00782922) (40,0.0141306) (50,0.0296796) (60,0.0335558) (70,0.0549216) (80,0.0549264) (90,0.0780589) (100,0.0856975)} \closedcycle;
\node[anchor=north east, font=\scriptsize, text=black, inner sep=1pt, align=right] at (rel axis cs:0.98,0.98) {largest $N$: 0.27\,s};
\end{groupplot}
\node[anchor=west] at ($(group c3r1.east) + (6pt,0)$) {\pgfplotslegendfromname{bandkey}};
\end{tikzpicture}
\setlength{\abovecaptionskip}{1ex}%
\caption{Breakdown of MARBLE compile time as the matrix dimension $N$ ($x$-axis) grows.}
\label{fig:rq4-breakdown}
\vspace{-3pt}%
\end{figure}
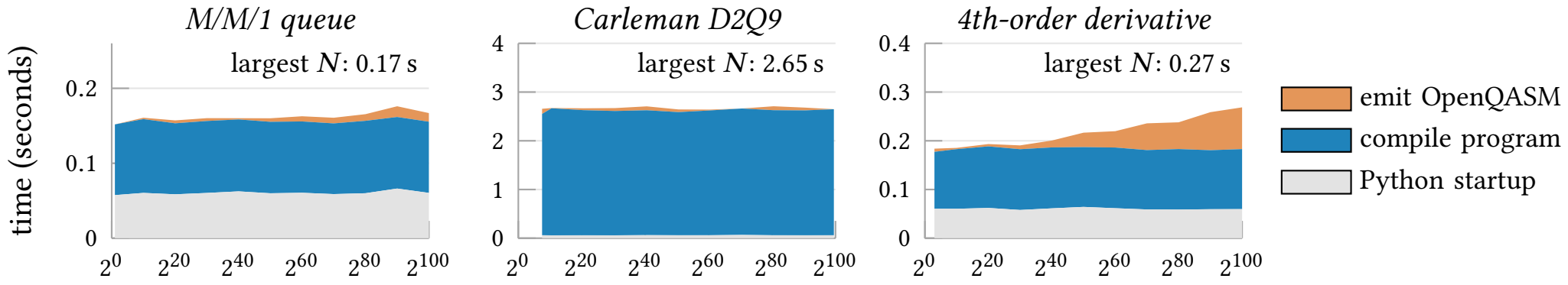

\begin{enumerate}[leftmargin=2.35em]
  \item [\emph{RQ4.}] How does MARBLE's compile time scale with the matrix dimension $N$?
\end{enumerate}

\paragraph{Methodology}
We measured the compile time of MARBLE as a sum of two components: 1) the time taken from parsing to immediately before emitting the OpenQASM circuit, and 2) the time taken to emit the quantum circuit.
We make this distinction because whereas most of the compiler executes in time independent of $N$, the final compilation step instantiates $N$ concretely and takes time to write out a circuit that is directly proportional to the number of gates in the circuit.

\paragraph{Results}
In Figure~\ref{fig:rq4-breakdown}, we break down compile time against $N$ for three illustrative benchmarks: M/M/1, which is a typical program; Carleman D2Q9, an example of a particularly slow compilation; and 4th-order derivative, whose circuit grows quickly with $N$.
Indeed, the time before emitting the circuit remains constant across all values of $N$ from 4 to $2^{100}$.
For cases where the number of gates grows with $N$, the time taken to emit the circuit naturally also grows.
Regardless, total compile time stays under 5 seconds for any matrix size for which file I/O is not a bottleneck.

For comparison, Figure~\ref{fig:rq3} shows that FABLE does not scale to matrices with dimension greater than \textasciitilde{}$2^{12} \times 2^{12}$, because its approach fundamentally must enumerate the whole matrix.

\section{Limitations and Next Steps}
\label{sec:limitations}

\paragraph{Quantum Gate Set}
MARBLE counts gates in the Clifford+$T$+$R_z$ gate set.
Because an arbitrary-angle $R_z$ rotation gate cannot be synthesized exactly from the standard Clifford+$T$ gate set without incurring approximation error $\epsilon$, the system either leaves rotation gates in its output or estimates their cost in Clifford+$T$, assuming each $R_z$ gate costs $O(\log(1/\epsilon))$ $T$ gates.
A longer-term solution would be to design compiler optimizations that target an arbitrary quantum gate set.

\paragraph{Cost Parameters}
The two primary hardware resource costs of block-encoding circuits that are tracked and optimized by MARBLE are subnormalization $\alpha$ and total time cost $\alpha\times\text{gates}$.
However, developers may also wish to optimize for other costs such as the number of qubits or spacetime volume (product of qubits and gates).
An important direction is therefore to extend the system to customizable cost functions that enable the developers to define their target cost of interest.

\paragraph{Rewrite Rules}
The quality of optimization depends on the quantity of rewrite rules, which are currently hand-written and incomplete.
For example, no rule currently factors a permutation out of a matrix to reduce gate costs.
To enable sustainable and sound development, one idea is to enable the user to provide instance-specific rewrites instead of rebuilding the compiler from scratch.
A more ambitious system could automatically verify user-provided rules in a proof assistant such as Lean and accumulate vetted rules into a shared library for future applications and users.

\paragraph{Exploration Budget}
The stage of the compiler that applies exploration rules is not guaranteed to terminate, because every new symbolic matrix form that is produced by some rewrite rule can match some other rule indefinitely.
For practical purposes, we configured a fixed cap of $10^5$ nodes in the intermediate representation, after which the compiler moves on to circuit selection.
This upper bound can impact whether MARBLE can find the best possible block encoding of a matrix in principle, and the bound is user-customizable for a developer who may wish to explore more optimizations.
In the longer term, one idea is to dynamically use the remaining search budget on the most promising candidates rather than capping every search path uniformly.

\section{Related Work}
\label{sec:related}

\paragraph{Block Encodings}
Software tools exist that provide a library of circuit constructions to encode specific matrices or compositions of matrices.
Cobble~\citep{yuanCobbleCompilingBlock2026}, Unitaria~\citep{deimlUnitariaQuantumLinear2026}, Qualtran~\citep{harriganExpressingAnalyzingQuantum2024}, and Qrisp~\citep{petricBlockencodingsProgrammingAbstractions2026} each provide combinators for sums, products, and polynomials of user-encoded matrices.
These systems also provide pre-built circuits for specific matrix families, which are derived from the literature on hand-derived block encodings.
For example, \citet{sunderhaufBlockencodingStructuredMatrices2024} derive Toeplitz and tridiagonal matrices, \citet{kharaziExplicitBlockEncodings2025} provide differential operators, \citet{suFaultTolerantQuantumSimulations2021} provide chemistry Hamiltonians, and \citet{jenningsEndtoEndQuantumAlgorithm2026,penuelDetailedAssessmentCalculating2025,uenoDemonstrationQuantumCircuit2026} derive fluid dynamics systems.

Other tools accept arbitrary numerical matrices but at cost polynomial, not poly-logarithmic, in the dimension of the matrix.
FABLE~\cite{campsFABLEFastApproximate2022} accepts a numerical matrix and builds a circuit of $O(N^2)$ gates from its entries, truncating rotations with small angles.
S-FABLE and LS-FABLE~\cite{kuklinskiSFABLELSFABLEFast2024}, dictionary-based encoding~\cite{yangDictionarybasedBlockEncoding2025}, and coherent-permutation encoding~\cite{settyBlockEncodingSparse2026} operate similarly given sparse matrices.

In all existing work, the programmer must either 1) hand-write gate-based circuits for primitive block encodings before they can use automated tools to compose matrices, 2) invoke only a library of pre-constructed circuits for specific matrix families, or 3) abandon poly-logarithmic scaling as required for quantum advantage in general.
By contrast, MARBLE accepts a mathematical formula that defines a general matrix in publication-style notation and synthesizes an efficient circuit.

\paragraph{Classical Matrix Compilers}
Generating a program from a matrix formula by algebraic rewriting has a long history in classical computing.
SPIRAL~\cite{puschelSpiralGeneratorPlatformAdapted2004} rewrites formulas for signal transforms into fast implementations, while TACO~\cite{kjolstadTensorAlgebraCompiler2017} compiles tensor algebra written in index notation.
StructTensor~\cite{ghorbaniCompilingStructuredTensor2023} computes the structure of a tensor expression symbolically at compile time --- a close analogue to our symbolic matrix form.

MARBLE relies on equality saturation~\cite{tateEqualitySaturationNew2009} to preserve the inputs and outputs of each algebraic rewrite for a cost model to then choose.
\citet{yangEqualitySaturationQuantum2026} previously explored equality saturation for quantum circuit optimization at the level of gate cancellation and replacement.
By contrast, our work identifies that the high-level structure of matrix formulas and linear-algebraic identities enables equality saturation to operate on small algebraic expressions rather than massive quantum circuits.
Furthermore, the cost function of subnormalization has no classical counterpart and trades off non-trivially against gate count, calling for the flexibility of equality saturation.

\section{Conclusion}
\label{sec:conclusion}

Quantum computers promise to solve the largest problems in science --- but only if they have the right structure.
In the past, finding that structure has been a craft reserved for experts who search by hand, one problem at a time, for results that get published as equations instead of code.

We introduce a programming language and compiler to synthesize an efficient quantum circuit from a formula for the entries of a matrix, enabling scientists to reach far more problems than can be encoded by hand.
Our abstractions of static multiplicity and symbolic matrix forms convert the long-held intuition that quantum advantage requires structure into a constructive property that a compiler can check, much as type systems did for the correctness of classical programs.

We hope that this work leads to principled approaches for the automated discovery of quantum algorithms and programs.
In the coming future when experts have more matrices and rewrites than time to study them, MARBLE turns individual circuit discoveries into sound compilation principles as well as prototyping and verification tools that grow more capable over time.
By building systems to reason about applications rather than circuits, our goal is to enable any scientist who can write down the mathematics of a problem to understand its potential in the quantum world.

\medskip
\noindent
\textbf{\textsf{AI Use}}\quad
All ideas of this work, including the language, the compiler organization, the rewrite rules, and the cost reasoning, are due to the authors.
The authors used Claude Opus 5.5 by Anthropic to help implement the compiler, mechanize proofs, collect benchmarks, and set up experiments.

\bibliographystyle{ACM-Reference-Format}
\bibliography{marble-references}


\begin{thebibliography}{53}


\ifx \showCODEN    \undefined \def \showCODEN     #1{\unskip}     \fi
\ifx \showISBNx    \undefined \def \showISBNx     #1{\unskip}     \fi
\ifx \showISBNxiii \undefined \def \showISBNxiii  #1{\unskip}     \fi
\ifx \showISSN     \undefined \def \showISSN      #1{\unskip}     \fi
\ifx \showLCCN     \undefined \def \showLCCN      #1{\unskip}     \fi
\ifx \shownote     \undefined \def \shownote      #1{#1}          \fi
\ifx \showarticletitle \undefined \def \showarticletitle #1{#1}   \fi
\ifx \showURL      \undefined \def \showURL       {\relax}        \fi
\providecommand\bibfield[2]{#2}
\providecommand\bibinfo[2]{#2}
\providecommand\natexlab[1]{#1}
\providecommand\showeprint[2][]{arXiv:#2}

\bibitem[Babbush et~al\mbox{.}(2019)]%
        {babbushQuantumSimulationSachdevYeKitaev2018}
\bibfield{author}{\bibinfo{person}{Ryan Babbush}, \bibinfo{person}{Dominic~W.
  Berry}, {and} \bibinfo{person}{Hartmut Neven}.}
  \bibinfo{year}{2019}\natexlab{}.
\newblock \showarticletitle{Quantum simulation of the Sachdev-Ye-Kitaev model
  by asymmetric qubitization}.
\newblock \bibinfo{journal}{\emph{Physical Review A}} \bibinfo{volume}{99},
  \bibinfo{number}{4} (\bibinfo{year}{2019}), \bibinfo{pages}{040301}.
\newblock
\href{https://doi.org/10.1103/PhysRevA.99.040301}{doi:\nolinkurl{10.1103/PhysRevA.99.040301}}


\bibitem[Babbush et~al\mbox{.}(2018)]%
        {babbushEncodingElectronicSpectra2018}
\bibfield{author}{\bibinfo{person}{Ryan Babbush}, \bibinfo{person}{Craig
  Gidney}, \bibinfo{person}{Dominic~W. Berry}, \bibinfo{person}{Nathan Wiebe},
  \bibinfo{person}{Jarrod McClean}, \bibinfo{person}{Alexandru Paler},
  \bibinfo{person}{Austin Fowler}, {and} \bibinfo{person}{Hartmut Neven}.}
  \bibinfo{year}{2018}\natexlab{}.
\newblock \showarticletitle{Encoding Electronic Spectra in Quantum Circuits
  with Linear T Complexity}.
\newblock \bibinfo{journal}{\emph{Physical Review X}} \bibinfo{volume}{8},
  \bibinfo{number}{4} (\bibinfo{year}{2018}), \bibinfo{pages}{041015}.
\newblock
\href{https://doi.org/10.1103/PhysRevX.8.041015}{doi:\nolinkurl{10.1103/PhysRevX.8.041015}}


\bibitem[Berry et~al\mbox{.}(2014)]%
        {berry2014}
\bibfield{author}{\bibinfo{person}{Dominic~W. Berry},
  \bibinfo{person}{Andrew~M. Childs}, \bibinfo{person}{Richard Cleve},
  \bibinfo{person}{Robin Kothari}, {and} \bibinfo{person}{Rolando~D. Somma}.}
  \bibinfo{year}{2014}\natexlab{}.
\newblock \showarticletitle{Exponential improvement in precision for simulating
  sparse Hamiltonians}. In \bibinfo{booktitle}{\emph{ACM SIGACT Symposium on
  Theory of Computing}}. \bibinfo{pages}{283--292}.
\newblock
\href{https://doi.org/10.1145/2591796.2591854}{doi:\nolinkurl{10.1145/2591796.2591854}}


\bibitem[Bl{\"o}te and Deng(2002)]%
        {bloteClusterMonteCarlo2002}
\bibfield{author}{\bibinfo{person}{Henk W.~J. Bl{\"o}te} {and}
  \bibinfo{person}{Youjin Deng}.} \bibinfo{year}{2002}\natexlab{}.
\newblock \showarticletitle{Cluster Monte Carlo Simulation of the Transverse
  Ising Model}.
\newblock \bibinfo{journal}{\emph{Physical Review E}} \bibinfo{volume}{66},
  \bibinfo{number}{6} (\bibinfo{year}{2002}), \bibinfo{pages}{066110}.
\newblock
\href{https://doi.org/10.1103/PhysRevE.66.066110}{doi:\nolinkurl{10.1103/PhysRevE.66.066110}}


\bibitem[Bluvstein et~al\mbox{.}(2026)]%
        {bluvsteinFaulttolerantNeutralatomArchitecture2026}
\bibfield{author}{\bibinfo{person}{Dolev Bluvstein},
  \bibinfo{person}{Alexandra~A. Geim}, \bibinfo{person}{Sophie~H. Li},
  \bibinfo{person}{Simon~J. Evered}, \bibinfo{person}{J.~Pablo
  Bonilla~Ataides}, \bibinfo{person}{Gefen Baranes}, \bibinfo{person}{Andi Gu},
  \bibinfo{person}{Tom Manovitz}, \bibinfo{person}{Muqing Xu},
  \bibinfo{person}{Marcin Kalinowski}, \bibinfo{person}{Shayan Majidy},
  \bibinfo{person}{Christian Kokail}, \bibinfo{person}{Nishad Maskara},
  \bibinfo{person}{Elias~C. Trapp}, \bibinfo{person}{Luke~M. Stewart},
  \bibinfo{person}{Simon Hollerith}, \bibinfo{person}{Hengyun Zhou},
  \bibinfo{person}{Michael~J. Gullans}, \bibinfo{person}{Susanne~F. Yelin},
  \bibinfo{person}{Markus Greiner}, \bibinfo{person}{Vladan Vuleti{\'c}},
  \bibinfo{person}{Madelyn Cain}, {and} \bibinfo{person}{Mikhail~D. Lukin}.}
  \bibinfo{year}{2026}\natexlab{}.
\newblock \showarticletitle{A Fault-Tolerant Neutral-Atom Architecture for
  Universal Quantum Computation}.
\newblock \bibinfo{journal}{\emph{Nature}} \bibinfo{volume}{649},
  \bibinfo{number}{8095} (\bibinfo{year}{2026}), \bibinfo{pages}{39--46}.
\newblock
\href{https://doi.org/10.1038/s41586-025-09848-5}{doi:\nolinkurl{10.1038/s41586-025-09848-5}}


\bibitem[Camps et~al\mbox{.}(2024)]%
        {campsExplicitQuantumCircuits2024}
\bibfield{author}{\bibinfo{person}{Daan Camps}, \bibinfo{person}{Lin Lin},
  \bibinfo{person}{Roel Van~Beeumen}, {and} \bibinfo{person}{Chao Yang}.}
  \bibinfo{year}{2024}\natexlab{}.
\newblock \showarticletitle{Explicit Quantum Circuits for Block Encodings of
  Certain Sparse Matrices}.
\newblock \bibinfo{journal}{\emph{SIAM J. Matrix Anal. Appl.}}
  \bibinfo{volume}{45}, \bibinfo{number}{1} (\bibinfo{year}{2024}),
  \bibinfo{pages}{801--827}.
\newblock
\href{https://doi.org/10.1137/22M1484298}{doi:\nolinkurl{10.1137/22M1484298}}


\bibitem[Camps and Van~Beeumen(2022)]%
        {campsFABLEFastApproximate2022}
\bibfield{author}{\bibinfo{person}{Daan Camps} {and} \bibinfo{person}{Roel
  Van~Beeumen}.} \bibinfo{year}{2022}\natexlab{}.
\newblock \showarticletitle{FABLE: Fast Approximate Quantum Circuits for
  Block-Encodings}. In \bibinfo{booktitle}{\emph{IEEE International Conference
  on Quantum Computing and Engineering}}. \bibinfo{pages}{104--113}.
\newblock
\href{https://doi.org/10.1109/QCE53715.2022.00029}{doi:\nolinkurl{10.1109/QCE53715.2022.00029}}


\bibitem[Childs et~al\mbox{.}(2018)]%
        {childsFirstQuantumSimulation2018}
\bibfield{author}{\bibinfo{person}{Andrew~M. Childs}, \bibinfo{person}{Dmitri
  Maslov}, \bibinfo{person}{Yunseong Nam}, \bibinfo{person}{Neil~J. Ross},
  {and} \bibinfo{person}{Yuan Su}.} \bibinfo{year}{2018}\natexlab{}.
\newblock \showarticletitle{Toward the First Quantum Simulation with Quantum
  Speedup}.
\newblock \bibinfo{journal}{\emph{Proceedings of the National Academy of
  Sciences}} \bibinfo{volume}{115}, \bibinfo{number}{38}
  (\bibinfo{year}{2018}), \bibinfo{pages}{9456--9461}.
\newblock
\href{https://doi.org/10.1073/pnas.1801723115}{doi:\nolinkurl{10.1073/pnas.1801723115}}


\bibitem[Childs and Wiebe(2012)]%
        {childsHamiltonianSimulation2012}
\bibfield{author}{\bibinfo{person}{Andrew~M. Childs} {and}
  \bibinfo{person}{Nathan Wiebe}.} \bibinfo{year}{2012}\natexlab{}.
\newblock \showarticletitle{Hamiltonian Simulation Using Linear Combinations of
  Unitary Operations}.
\newblock \bibinfo{journal}{\emph{Quantum Information and Computation}}
  \bibinfo{volume}{12}, \bibinfo{number}{11--12} (\bibinfo{year}{2012}),
  \bibinfo{pages}{901--924}.
\newblock
\href{https://doi.org/10.26421/qic12.11-12-1}{doi:\nolinkurl{10.26421/qic12.11-12-1}}


\bibitem[Dagotto and Moreo(1989)]%
        {dagottoPhaseDiagramFrustrated1989}
\bibfield{author}{\bibinfo{person}{Elbio Dagotto} {and}
  \bibinfo{person}{Adriana Moreo}.} \bibinfo{year}{1989}\natexlab{}.
\newblock \showarticletitle{Phase Diagram of the Frustrated Spin-1/2 Heisenberg
  Antiferromagnet in 2 Dimensions}.
\newblock \bibinfo{journal}{\emph{Physical Review Letters}}
  \bibinfo{volume}{63}, \bibinfo{number}{19} (\bibinfo{year}{1989}),
  \bibinfo{pages}{2148--2151}.
\newblock
\href{https://doi.org/10.1103/PhysRevLett.63.2148}{doi:\nolinkurl{10.1103/PhysRevLett.63.2148}}


\bibitem[Deiml et~al\mbox{.}(2026)]%
        {deimlUnitariaQuantumLinear2026}
\bibfield{author}{\bibinfo{person}{Matthias Deiml}, \bibinfo{person}{Oliver
  H{\"u}ttenhofer}, \bibinfo{person}{Ram Mosco}, \bibinfo{person}{Jakob~S.
  Kottmann}, {and} \bibinfo{person}{Daniel Peterseim}.}
  \bibinfo{year}{2026}\natexlab{}.
\newblock \bibinfo{title}{Unitaria: Quantum Linear Algebra via Block
  Encodings}.
\newblock
\showeprint[arxiv]{2605.10768}~[quant-ph]
\href{https://doi.org/10.48550/arXiv.2605.10768}{doi:\nolinkurl{10.48550/arXiv.2605.10768}}


\bibitem[Ghorbani et~al\mbox{.}(2023)]%
        {ghorbaniCompilingStructuredTensor2023}
\bibfield{author}{\bibinfo{person}{Mahdi Ghorbani}, \bibinfo{person}{Mathieu
  Huot}, \bibinfo{person}{Shideh Hashemian}, {and} \bibinfo{person}{Amir
  Shaikhha}.} \bibinfo{year}{2023}\natexlab{}.
\newblock \showarticletitle{Compiling Structured Tensor Algebra}. In
  \bibinfo{booktitle}{\emph{ACM SIGPLAN Conference on Object-Oriented
  Programming, Systems, Languages, and Applications}}.
  \bibinfo{pages}{204--233}.
\newblock
\href{https://doi.org/10.1145/3622804}{doi:\nolinkurl{10.1145/3622804}}


\bibitem[Gily{\'e}n et~al\mbox{.}(2019)]%
        {gilyenQuantumSingularValue2019}
\bibfield{author}{\bibinfo{person}{Andr{\'a}s Gily{\'e}n},
  \bibinfo{person}{Yuan Su}, \bibinfo{person}{Guang~Hao Low}, {and}
  \bibinfo{person}{Nathan Wiebe}.} \bibinfo{year}{2019}\natexlab{}.
\newblock \showarticletitle{Quantum Singular Value Transformation and beyond:
  Exponential Improvements for Quantum Matrix Arithmetics}. In
  \bibinfo{booktitle}{\emph{ACM SIGACT Symposium on Theory of Computing}}.
  \bibinfo{pages}{193--204}.
\newblock
\href{https://doi.org/10.1145/3313276.3316366}{doi:\nolinkurl{10.1145/3313276.3316366}}


\bibitem[Goings et~al\mbox{.}(2022)]%
        {goingsReliablyAssessingElectronic2022}
\bibfield{author}{\bibinfo{person}{Joshua~J. Goings}, \bibinfo{person}{Alec
  White}, \bibinfo{person}{Joonho Lee}, \bibinfo{person}{Christofer~S.
  Tautermann}, \bibinfo{person}{Matthias Degroote}, \bibinfo{person}{Craig
  Gidney}, \bibinfo{person}{Toru Shiozaki}, \bibinfo{person}{Ryan Babbush},
  {and} \bibinfo{person}{Nicholas~C. Rubin}.} \bibinfo{year}{2022}\natexlab{}.
\newblock \showarticletitle{Reliably Assessing the Electronic Structure of
  Cytochrome P450 on Today's Classical Computers and Tomorrow's Quantum
  Computers}.
\newblock \bibinfo{journal}{\emph{Proceedings of the National Academy of
  Sciences}} \bibinfo{volume}{119}, \bibinfo{number}{38}
  (\bibinfo{year}{2022}), \bibinfo{pages}{e2203533119}.
\newblock
\href{https://doi.org/10.1073/pnas.2203533119}{doi:\nolinkurl{10.1073/pnas.2203533119}}


\bibitem[{Google Quantum AI}(2025)]%
        {google2025}
\bibfield{author}{\bibinfo{person}{{Google Quantum AI}}.}
  \bibinfo{year}{2025}\natexlab{}.
\newblock \showarticletitle{Quantum error correction below the surface code
  threshold}.
\newblock \bibinfo{journal}{\emph{Nature}}  \bibinfo{volume}{638}
  (\bibinfo{year}{2025}), \bibinfo{pages}{920--926}.
\newblock
\href{https://doi.org/10.1038/s41586-024-08449-y}{doi:\nolinkurl{10.1038/s41586-024-08449-y}}


\bibitem[Gross et~al\mbox{.}(2008)]%
        {grossFundamentalsQueueingTheory2008}
\bibfield{author}{\bibinfo{person}{Donald Gross}, \bibinfo{person}{John~F.
  Shortle}, \bibinfo{person}{James~M. Thompson}, {and} \bibinfo{person}{Carl~M.
  Harris}.} \bibinfo{year}{2008}\natexlab{}.
\newblock \bibinfo{booktitle}{\emph{Fundamentals of Queueing Theory}
  (\bibinfo{edition}{4th} ed.)}.
\newblock \bibinfo{publisher}{Wiley-Interscience}.
\newblock
\href{https://doi.org/10.1002/9781118625651}{doi:\nolinkurl{10.1002/9781118625651}}


\bibitem[Harrigan et~al\mbox{.}(2024)]%
        {harriganExpressingAnalyzingQuantum2024}
\bibfield{author}{\bibinfo{person}{Matthew~P. Harrigan}, \bibinfo{person}{Tanuj
  Khattar}, \bibinfo{person}{Charles Yuan}, \bibinfo{person}{Anurudh Peduri},
  \bibinfo{person}{Noureldin Yosri}, \bibinfo{person}{Fionn~D. Malone},
  \bibinfo{person}{Ryan Babbush}, {and} \bibinfo{person}{Nicholas~C. Rubin}.}
  \bibinfo{year}{2024}\natexlab{}.
\newblock \bibinfo{title}{Expressing and Analyzing Quantum Algorithms with
  Qualtran}.
\newblock
\showeprint[arxiv]{2409.04643}~[quant-ph]
\href{https://doi.org/10.48550/arXiv.2409.04643}{doi:\nolinkurl{10.48550/arXiv.2409.04643}}


\bibitem[Harrow et~al\mbox{.}(2009)]%
        {harrowQuantumAlgorithmLinear2009}
\bibfield{author}{\bibinfo{person}{Aram~W. Harrow}, \bibinfo{person}{Avinatan
  Hassidim}, {and} \bibinfo{person}{Seth Lloyd}.}
  \bibinfo{year}{2009}\natexlab{}.
\newblock \showarticletitle{Quantum Algorithm for Linear Systems of Equations}.
\newblock \bibinfo{journal}{\emph{Physical Review Letters}}
  \bibinfo{volume}{103}, \bibinfo{number}{15} (\bibinfo{year}{2009}),
  \bibinfo{pages}{150502}.
\newblock
\href{https://doi.org/10.1103/PhysRevLett.103.150502}{doi:\nolinkurl{10.1103/PhysRevLett.103.150502}}


\bibitem[Helle et~al\mbox{.}(2026)]%
        {helleQuantumAdvectiondiffusionSolver2026}
\bibfield{author}{\bibinfo{person}{Gard~Olav Helle}, \bibinfo{person}{Tommaso
  Benacchio}, \bibinfo{person}{Anna~Bomme Ousager}, {and}
  \bibinfo{person}{Jørgen~Ellegaard Andersen}.}
  \bibinfo{year}{2026}\natexlab{}.
\newblock \bibinfo{title}{A quantum advection-diffusion solver using the
  quantum singular value transform}.
\newblock
\showeprint[arxiv]{2512.22163}~[quant-ph]
\href{https://doi.org/10.48550/arXiv.2512.22163}{doi:\nolinkurl{10.48550/arXiv.2512.22163}}


\bibitem[H{\"o}lscher et~al\mbox{.}(2026)]%
        {holscherEndtoendQuantumAlgorithm2026}
\bibfield{author}{\bibinfo{person}{Leonhard H{\"o}lscher},
  \bibinfo{person}{Oliver Ahrend}, \bibinfo{person}{Lukas Karch},
  \bibinfo{person}{Carlotta L'Estocq}, \bibinfo{person}{Marc~Marfany Andreu},
  \bibinfo{person}{Tobias Stollenwerk}, \bibinfo{person}{Frank~K Wilhelm},
  {and} \bibinfo{person}{Julia Kowalski}.} \bibinfo{year}{2026}\natexlab{}.
\newblock \showarticletitle{End-to-End Quantum Algorithm for Topology
  Optimization in Structural Mechanics}.
\newblock \bibinfo{journal}{\emph{Quantum Science and Technology}}
  \bibinfo{volume}{11}, \bibinfo{number}{2} (\bibinfo{year}{2026}),
  \bibinfo{pages}{025029}.
\newblock
\href{https://doi.org/10.1088/2058-9565/ae4cfe}{doi:\nolinkurl{10.1088/2058-9565/ae4cfe}}


\bibitem[Jennings et~al\mbox{.}(2026)]%
        {jenningsEndtoEndQuantumAlgorithm2026}
\bibfield{author}{\bibinfo{person}{David Jennings}, \bibinfo{person}{Kamil
  Korzekwa}, \bibinfo{person}{Matteo Lostaglio}, \bibinfo{person}{Richard
  Ashworth}, \bibinfo{person}{Emanuele Marsili}, {and} \bibinfo{person}{Stephen
  Rolston}.} \bibinfo{year}{2026}\natexlab{}.
\newblock \showarticletitle{End-to-End Quantum Algorithm for Nonlinear Fluid
  Dynamics with Bounded Quantum Advantage}.
\newblock \bibinfo{journal}{\emph{PRX Quantum}} \bibinfo{volume}{7},
  \bibinfo{number}{3} (\bibinfo{year}{2026}), \bibinfo{pages}{033060}.
\newblock
\href{https://doi.org/10.1103/xysy-q3fp}{doi:\nolinkurl{10.1103/xysy-q3fp}}


\bibitem[Kalantzis et~al\mbox{.}(2025)]%
        {kalantzisEfficientSolutionsGeneral2025}
\bibfield{author}{\bibinfo{person}{Vasileios Kalantzis},
  \bibinfo{person}{Mark~S. Squillante}, {and} \bibinfo{person}{Shashanka
  Ubaru}.} \bibinfo{year}{2025}\natexlab{}.
\newblock \bibinfo{title}{On Efficient Solutions of General Structured Markov
  Processes in Quantum Computational Environments}.
\newblock
\showeprint[arxiv]{2404.17959}~[quant-ph]
\href{https://doi.org/10.48550/arXiv.2404.17959}{doi:\nolinkurl{10.48550/arXiv.2404.17959}}


\bibitem[Kan et~al\mbox{.}(2026)]%
        {kanOptimizedQuantumAlgorithms2025}
\bibfield{author}{\bibinfo{person}{Angus Kan}, \bibinfo{person}{Jessica
  Lemieux}, \bibinfo{person}{Olga Okrut}, {and} \bibinfo{person}{Burak
  Şahinoğlu}.} \bibinfo{year}{2026}\natexlab{}.
\newblock \showarticletitle{Optimized quantum algorithms for simulating the
  Schwinger effect}.
\newblock \bibinfo{journal}{\emph{Physical Review D}} \bibinfo{volume}{113},
  \bibinfo{number}{11} (\bibinfo{year}{2026}), \bibinfo{pages}{114503}.
\newblock
\href{https://doi.org/10.1103/ps29-tshk}{doi:\nolinkurl{10.1103/ps29-tshk}}


\bibitem[Kan and Symons(2025)]%
        {kanResourceoptimizedFaulttolerantSimulation2025a}
\bibfield{author}{\bibinfo{person}{Angus Kan} {and} \bibinfo{person}{Benjamin
  C.~B. Symons}.} \bibinfo{year}{2025}\natexlab{}.
\newblock \showarticletitle{Resource-Optimized Fault-Tolerant Simulation of the
  Fermi-Hubbard Model and High-Temperature Superconductor Models}.
\newblock \bibinfo{journal}{\emph{npj Quantum Information}}
  \bibinfo{volume}{11}, \bibinfo{number}{1} (\bibinfo{year}{2025}),
  \bibinfo{pages}{138}.
\newblock
\href{https://doi.org/10.1038/s41534-025-01091-0}{doi:\nolinkurl{10.1038/s41534-025-01091-0}}


\bibitem[Kharazi et~al\mbox{.}(2025)]%
        {kharaziExplicitBlockEncodings2025}
\bibfield{author}{\bibinfo{person}{Tyler Kharazi}, \bibinfo{person}{Ahmad~M.
  Alkadri}, \bibinfo{person}{Jin-Peng Liu}, \bibinfo{person}{Kranthi~K.
  Mandadapu}, {and} \bibinfo{person}{K.~Birgitta Whaley}.}
  \bibinfo{year}{2025}\natexlab{}.
\newblock \showarticletitle{Explicit Block Encodings of Boundary Value Problems
  for Many-Body Elliptic Operators}.
\newblock \bibinfo{journal}{\emph{Quantum}}  \bibinfo{volume}{9}
  (\bibinfo{year}{2025}), \bibinfo{pages}{1764}.
\newblock
\href{https://doi.org/10.22331/q-2025-06-04-1764}{doi:\nolinkurl{10.22331/q-2025-06-04-1764}}


\bibitem[Kjolstad et~al\mbox{.}(2017)]%
        {kjolstadTensorAlgebraCompiler2017}
\bibfield{author}{\bibinfo{person}{Fredrik Kjolstad}, \bibinfo{person}{Shoaib
  Kamil}, \bibinfo{person}{Stephen Chou}, \bibinfo{person}{David Lugato}, {and}
  \bibinfo{person}{Saman Amarasinghe}.} \bibinfo{year}{2017}\natexlab{}.
\newblock \showarticletitle{The Tensor Algebra Compiler}. In
  \bibinfo{booktitle}{\emph{ACM SIGPLAN Conference on Object-Oriented
  Programming, Systems, Languages, and Applications}}. \bibinfo{pages}{1--29}.
\newblock
\href{https://doi.org/10.1145/3133901}{doi:\nolinkurl{10.1145/3133901}}


\bibitem[Kuklinski and Rempfer(2024)]%
        {kuklinskiSFABLELSFABLEFast2024}
\bibfield{author}{\bibinfo{person}{Parker Kuklinski} {and}
  \bibinfo{person}{Benjamin Rempfer}.} \bibinfo{year}{2024}\natexlab{}.
\newblock \bibinfo{title}{S-FABLE and LS-FABLE: Fast Approximate Block-Encoding
  Algorithms for Unstructured Sparse Matrices}.
\newblock
\showeprint[arxiv]{2401.04234}~[quant-ph]
\href{https://doi.org/10.48550/arXiv.2401.04234}{doi:\nolinkurl{10.48550/arXiv.2401.04234}}


\bibitem[Kuklinski et~al\mbox{.}(2025)]%
        {kuklinskiEfficientBlockencodingsRequire2025}
\bibfield{author}{\bibinfo{person}{Parker Kuklinski}, \bibinfo{person}{Benjamin
  Rempfer}, \bibinfo{person}{Justin Elenewski}, {and} \bibinfo{person}{Kevin
  Obenland}.} \bibinfo{year}{2025}\natexlab{}.
\newblock \bibinfo{title}{Efficient Block-Encodings Require Structure}.
\newblock
\showeprint[arxiv]{2509.19667}~[quant-ph]
\href{https://doi.org/10.48550/arXiv.2509.19667}{doi:\nolinkurl{10.48550/arXiv.2509.19667}}


\bibitem[Lee and Kanno(2025)]%
        {leeModalAnalysisQuantum2025}
\bibfield{author}{\bibinfo{person}{Yasunori Lee} {and} \bibinfo{person}{Keita
  Kanno}.} \bibinfo{year}{2025}\natexlab{}.
\newblock \showarticletitle{Modal Analysis on Quantum Computers via
  Qubitization}. In \bibinfo{booktitle}{\emph{IEEE International Conference on
  Quantum Computing and Engineering}}. \bibinfo{pages}{2170--2180}.
\newblock
\href{https://doi.org/10.1109/QCE65121.2025.00237}{doi:\nolinkurl{10.1109/QCE65121.2025.00237}}


\bibitem[Liu et~al\mbox{.}(2025)]%
        {liuEfficientQuantumCircuit2024}
\bibfield{author}{\bibinfo{person}{Diyi Liu}, \bibinfo{person}{Weijie Du},
  \bibinfo{person}{Lin Lin}, \bibinfo{person}{James~P. Vary}, {and}
  \bibinfo{person}{Chao Yang}.} \bibinfo{year}{2025}\natexlab{}.
\newblock \showarticletitle{An efficient quantum circuit for block encoding a
  pairing Hamiltonian}.
\newblock \bibinfo{journal}{\emph{Journal of Computational Science}}
  \bibinfo{volume}{85} (\bibinfo{year}{2025}), \bibinfo{pages}{102480}.
\newblock
\href{https://doi.org/10.1016/j.jocs.2024.102480}{doi:\nolinkurl{10.1016/j.jocs.2024.102480}}


\bibitem[Liu et~al\mbox{.}(2021)]%
        {liuEfficientQuantumAlgorithm2021}
\bibfield{author}{\bibinfo{person}{Jin-Peng Liu},
  \bibinfo{person}{Herman~{\O}ie Kolden}, \bibinfo{person}{Hari~K. Krovi},
  \bibinfo{person}{Nuno~F. Loureiro}, \bibinfo{person}{Konstantina Trivisa},
  {and} \bibinfo{person}{Andrew~M. Childs}.} \bibinfo{year}{2021}\natexlab{}.
\newblock \showarticletitle{Efficient Quantum Algorithm for Dissipative
  Nonlinear Differential Equations}.
\newblock \bibinfo{journal}{\emph{Proceedings of the National Academy of
  Sciences}} \bibinfo{volume}{118}, \bibinfo{number}{35}
  (\bibinfo{year}{2021}), \bibinfo{pages}{e2026805118}.
\newblock
\href{https://doi.org/10.1073/pnas.2026805118}{doi:\nolinkurl{10.1073/pnas.2026805118}}


\bibitem[Mahmud and Herrman(2026)]%
        {mahmudMomentStructuredBlockEncodings2026}
\bibfield{author}{\bibinfo{person}{Jishnu Mahmud} {and}
  \bibinfo{person}{Rebekah Herrman}.} \bibinfo{year}{2026}\natexlab{}.
\newblock \bibinfo{title}{Moment-Structured Block Encodings of Periodic
  Finite-Difference Operators}.
\newblock
\showeprint[arxiv]{2607.11596}~[quant-ph]
\href{https://doi.org/10.48550/arXiv.2607.11596}{doi:\nolinkurl{10.48550/arXiv.2607.11596}}


\bibitem[Martyn et~al\mbox{.}(2021)]%
        {martynGrandUnificationQuantum2021}
\bibfield{author}{\bibinfo{person}{John~M. Martyn}, \bibinfo{person}{Zane~M.
  Rossi}, \bibinfo{person}{Andrew~K. Tan}, {and} \bibinfo{person}{Isaac~L.
  Chuang}.} \bibinfo{year}{2021}\natexlab{}.
\newblock \showarticletitle{Grand Unification of Quantum Algorithms}.
\newblock \bibinfo{journal}{\emph{PRX Quantum}} \bibinfo{volume}{2},
  \bibinfo{number}{4} (\bibinfo{year}{2021}), \bibinfo{pages}{040203}.
\newblock
\href{https://doi.org/10.1103/PRXQuantum.2.040203}{doi:\nolinkurl{10.1103/PRXQuantum.2.040203}}


\bibitem[Nguyen et~al\mbox{.}(2022)]%
        {nguyenBlockencodingDenseFullrank2022}
\bibfield{author}{\bibinfo{person}{Quynh~T. Nguyen}, \bibinfo{person}{Bobak~T.
  Kiani}, {and} \bibinfo{person}{Seth Lloyd}.} \bibinfo{year}{2022}\natexlab{}.
\newblock \showarticletitle{Block-Encoding Dense and Full-Rank Kernels Using
  Hierarchical Matrices: Applications in Quantum Numerical Linear Algebra}.
\newblock \bibinfo{journal}{\emph{Quantum}}  \bibinfo{volume}{6}
  (\bibinfo{year}{2022}), \bibinfo{pages}{876}.
\newblock
\href{https://doi.org/10.22331/q-2022-12-13-876}{doi:\nolinkurl{10.22331/q-2022-12-13-876}}


\bibitem[Nielsen and Chuang(2010)]%
        {nielsenQuantumComputationQuantum2010}
\bibfield{author}{\bibinfo{person}{Michael~A. Nielsen} {and}
  \bibinfo{person}{Isaac~L. Chuang}.} \bibinfo{year}{2010}\natexlab{}.
\newblock \bibinfo{booktitle}{\emph{Quantum Computation and Quantum
  Information} (\bibinfo{edition}{10th anniversary} ed.)}.
\newblock \bibinfo{publisher}{Cambridge University Press}.
\newblock
\href{https://doi.org/10.1017/CBO9780511976667}{doi:\nolinkurl{10.1017/CBO9780511976667}}


\bibitem[Penuel et~al\mbox{.}(2025)]%
        {penuelDetailedAssessmentCalculating2025}
\bibfield{author}{\bibinfo{person}{John Penuel}, \bibinfo{person}{Amara
  Katabarwa}, \bibinfo{person}{Peter~D. Johnson}, \bibinfo{person}{Parker
  Kuklinski}, \bibinfo{person}{Benjamin Rempfer}, \bibinfo{person}{Collin
  Farquhar}, \bibinfo{person}{Yudong Cao}, {and} \bibinfo{person}{Michael~C.
  Garrett}.} \bibinfo{year}{2025}\natexlab{}.
\newblock \bibinfo{title}{Detailed Assessment of Calculating Drag Force with
  Quantum Computers: Explicit Time-Evolution Precludes Exponential Advantage
  for Nonlinear Differential Equations}.
\newblock
\showeprint[arxiv]{2406.06323}~[quant-ph]
\href{https://doi.org/10.48550/arXiv.2406.06323}{doi:\nolinkurl{10.48550/arXiv.2406.06323}}


\bibitem[Petri{\v c} and Zander(2026)]%
        {petricBlockencodingsProgrammingAbstractions2026}
\bibfield{author}{\bibinfo{person}{Matic Petri{\v c}} {and}
  \bibinfo{person}{Ren{\'e} Zander}.} \bibinfo{year}{2026}\natexlab{}.
\newblock \bibinfo{title}{Block-Encodings as Programming Abstractions: The
  Eclipse Qrisp BlockEncoding Interface}.
\newblock
\showeprint[arxiv]{2604.18276}~[quant-ph]
\href{https://doi.org/10.48550/arXiv.2604.18276}{doi:\nolinkurl{10.48550/arXiv.2604.18276}}


\bibitem[Pfeuty(1970)]%
        {pfeutyOnedimensionalIsingModel1970}
\bibfield{author}{\bibinfo{person}{Pierre Pfeuty}.}
  \bibinfo{year}{1970}\natexlab{}.
\newblock \showarticletitle{The One-Dimensional Ising Model with a Transverse
  Field}.
\newblock \bibinfo{journal}{\emph{Annals of Physics}} \bibinfo{volume}{57},
  \bibinfo{number}{1} (\bibinfo{year}{1970}), \bibinfo{pages}{79--90}.
\newblock
\href{https://doi.org/10.1016/0003-4916(70)90270-8}{doi:\nolinkurl{10.1016/0003-4916(70)90270-8}}


\bibitem[P{\"u}schel et~al\mbox{.}(2004)]%
        {puschelSpiralGeneratorPlatformAdapted2004}
\bibfield{author}{\bibinfo{person}{Markus P{\"u}schel},
  \bibinfo{person}{Jos{\'e} M.~F. Moura}, \bibinfo{person}{Bryan Singer},
  \bibinfo{person}{Jianxin Xiong}, \bibinfo{person}{Jeremy Johnson},
  \bibinfo{person}{David Padua}, \bibinfo{person}{Manuela Veloso}, {and}
  \bibinfo{person}{Robert~W. Johnson}.} \bibinfo{year}{2004}\natexlab{}.
\newblock \showarticletitle{Spiral: A Generator for Platform-Adapted Libraries
  of Signal Processing Algorithms}.
\newblock \bibinfo{journal}{\emph{International Journal of High Performance
  Computing Applications}} \bibinfo{volume}{18}, \bibinfo{number}{1}
  (\bibinfo{year}{2004}), \bibinfo{pages}{21--45}.
\newblock
\href{https://doi.org/10.1177/1094342004041291}{doi:\nolinkurl{10.1177/1094342004041291}}


\bibitem[Rubin et~al\mbox{.}(2024)]%
        {rubin2024}
\bibfield{author}{\bibinfo{person}{Nicholas~C. Rubin},
  \bibinfo{person}{Dominic~W. Berry}, \bibinfo{person}{Alina Kononov},
  \bibinfo{person}{Fionn~D. Malone}, \bibinfo{person}{Tanuj Khattar},
  \bibinfo{person}{Alec White}, \bibinfo{person}{Joonho Lee},
  \bibinfo{person}{Hartmut Neven}, \bibinfo{person}{Ryan Babbush}, {and}
  \bibinfo{person}{Andrew~D. Baczewski}.} \bibinfo{year}{2024}\natexlab{}.
\newblock \showarticletitle{Quantum computation of stopping power for inertial
  fusion target design}.
\newblock \bibinfo{journal}{\emph{Proceedings of the National Academy of
  Sciences}} \bibinfo{volume}{121}, \bibinfo{number}{23}
  (\bibinfo{year}{2024}), \bibinfo{pages}{e2317772121}.
\newblock
\href{https://doi.org/10.1073/pnas.2317772121}{doi:\nolinkurl{10.1073/pnas.2317772121}}


\bibitem[Sato et~al\mbox{.}(2026)]%
        {satoExplicitBlockencodingPartial2026}
\bibfield{author}{\bibinfo{person}{Yuki Sato}, \bibinfo{person}{Jumpei Kato},
  \bibinfo{person}{Hiroshi Yano}, \bibinfo{person}{Kosuke Ito}, {and}
  \bibinfo{person}{Naoki Yamamoto}.} \bibinfo{year}{2026}\natexlab{}.
\newblock \bibinfo{title}{Explicit Block-Encoding for Partial Differential
  Equation-Constrained Optimization}.
\newblock
\showeprint[arxiv]{2511.14420v2}~[quant-ph]
\href{https://doi.org/10.48550/arXiv.2511.14420}{doi:\nolinkurl{10.48550/arXiv.2511.14420}}


\bibitem[Setty(2026)]%
        {settyBlockEncodingSparse2026}
\bibfield{author}{\bibinfo{person}{Abhishek Setty}.}
  \bibinfo{year}{2026}\natexlab{}.
\newblock \showarticletitle{Block Encoding of Sparse Matrices via Coherent
  Permutation}.
\newblock \bibinfo{journal}{\emph{Physical Review Research}}
  \bibinfo{volume}{8}, \bibinfo{number}{3} (\bibinfo{year}{2026}),
  \bibinfo{pages}{033282}.
\newblock
\href{https://doi.org/10.1103/1nts-v6y9}{doi:\nolinkurl{10.1103/1nts-v6y9}}


\bibitem[Su et~al\mbox{.}(2021)]%
        {suFaultTolerantQuantumSimulations2021}
\bibfield{author}{\bibinfo{person}{Yuan Su}, \bibinfo{person}{Dominic~W.
  Berry}, \bibinfo{person}{Nathan Wiebe}, \bibinfo{person}{Nicholas Rubin},
  {and} \bibinfo{person}{Ryan Babbush}.} \bibinfo{year}{2021}\natexlab{}.
\newblock \showarticletitle{Fault-Tolerant Quantum Simulations of Chemistry in
  First Quantization}.
\newblock \bibinfo{journal}{\emph{PRX Quantum}} \bibinfo{volume}{2},
  \bibinfo{number}{4} (\bibinfo{year}{2021}), \bibinfo{pages}{040332}.
\newblock
\href{https://doi.org/10.1103/PRXQuantum.2.040332}{doi:\nolinkurl{10.1103/PRXQuantum.2.040332}}


\bibitem[S{\"u}nderhauf et~al\mbox{.}(2024)]%
        {sunderhaufBlockencodingStructuredMatrices2024}
\bibfield{author}{\bibinfo{person}{Christoph S{\"u}nderhauf},
  \bibinfo{person}{Earl Campbell}, {and} \bibinfo{person}{Joan Camps}.}
  \bibinfo{year}{2024}\natexlab{}.
\newblock \showarticletitle{Block-Encoding Structured Matrices for Data Input
  in Quantum Computing}.
\newblock \bibinfo{journal}{\emph{Quantum}}  \bibinfo{volume}{8}
  (\bibinfo{year}{2024}), \bibinfo{pages}{1226}.
\newblock
\href{https://doi.org/10.22331/q-2024-01-11-1226}{doi:\nolinkurl{10.22331/q-2024-01-11-1226}}


\bibitem[Suresh(2026)]%
        {sureshQuantumBlockEncodings2026}
\bibfield{author}{\bibinfo{person}{Krishnan Suresh}.}
  \bibinfo{year}{2026}\natexlab{}.
\newblock \bibinfo{title}{Quantum Block Encodings for Periodic Two-Phase Finite
  Element Operators: 2D Poisson and 2D Elasticity}.
\newblock
\showeprint[arxiv]{2609.17796}~[cs.CE]
\href{https://doi.org/10.48550/arXiv.2609.17796}{doi:\nolinkurl{10.48550/arXiv.2609.17796}}


\bibitem[Tate et~al\mbox{.}(2009)]%
        {tateEqualitySaturationNew2009}
\bibfield{author}{\bibinfo{person}{Ross Tate}, \bibinfo{person}{Michael Stepp},
  \bibinfo{person}{Zachary Tatlock}, {and} \bibinfo{person}{Sorin Lerner}.}
  \bibinfo{year}{2009}\natexlab{}.
\newblock \showarticletitle{Equality Saturation: A New Approach to
  Optimization}. In \bibinfo{booktitle}{\emph{ACM SIGPLAN Symposium on
  Principles of Programming Languages}}. \bibinfo{pages}{264--276}.
\newblock
\href{https://doi.org/10.1145/1480881.1480915}{doi:\nolinkurl{10.1145/1480881.1480915}}


\bibitem[Ueno et~al\mbox{.}(2026)]%
        {uenoDemonstrationQuantumCircuit2026}
\bibfield{author}{\bibinfo{person}{Kazumasa Ueno}, \bibinfo{person}{Keita
  Kanno}, {and} \bibinfo{person}{Yasunori Lee}.}
  \bibinfo{year}{2026}\natexlab{}.
\newblock \bibinfo{title}{A Demonstration of Quantum Circuit Implementation for
  Obstacle Flow Using Carleman-Linearized Lattice Boltzmann Method}.
\newblock
\showeprint[arxiv]{2605.28135}~[quant-ph]
\href{https://doi.org/10.48550/arXiv.2605.28135}{doi:\nolinkurl{10.48550/arXiv.2605.28135}}


\bibitem[Vaszary et~al\mbox{.}(2026)]%
        {vaszarySolvingNonlinearVlasov2026}
\bibfield{author}{\bibinfo{person}{Tam{\'a}s Vaszary}, \bibinfo{person}{Animesh
  Datta}, \bibinfo{person}{Tom Goffrey}, {and} \bibinfo{person}{Brian
  Appelbe}.} \bibinfo{year}{2026}\natexlab{}.
\newblock \showarticletitle{Solving the Nonlinear Vlasov Equation on a Quantum
  Computer}.
\newblock \bibinfo{journal}{\emph{Quantum}}  \bibinfo{volume}{10}
  (\bibinfo{year}{2026}), \bibinfo{pages}{2206}.
\newblock
\href{https://doi.org/10.22331/q-2026-09-10-2206}{doi:\nolinkurl{10.22331/q-2026-09-10-2206}}


\bibitem[Xu et~al\mbox{.}(2025)]%
        {xuImprovedQuantumLattice2025}
\bibfield{author}{\bibinfo{person}{Li Xu}, \bibinfo{person}{Ming Li},
  \bibinfo{person}{Lei Zhang}, \bibinfo{person}{Hai Sun}, {and}
  \bibinfo{person}{Jun Yao}.} \bibinfo{year}{2025}\natexlab{}.
\newblock \showarticletitle{Improved Quantum Lattice Boltzmann Method for
  Advection-Diffusion Equations with a Linear Collision Model}.
\newblock \bibinfo{journal}{\emph{Physical Review E}} \bibinfo{volume}{111},
  \bibinfo{number}{4} (\bibinfo{year}{2025}), \bibinfo{pages}{045305}.
\newblock
\href{https://doi.org/10.1103/PhysRevE.111.045305}{doi:\nolinkurl{10.1103/PhysRevE.111.045305}}


\bibitem[Yang et~al\mbox{.}(2025)]%
        {yangDictionarybasedBlockEncoding2025}
\bibfield{author}{\bibinfo{person}{Chunlin Yang}, \bibinfo{person}{Zexian Li},
  \bibinfo{person}{Hongmei Yao}, \bibinfo{person}{Zhaobing Fan},
  \bibinfo{person}{Guofeng Zhang}, {and} \bibinfo{person}{Jianshe Liu}.}
  \bibinfo{year}{2025}\natexlab{}.
\newblock \showarticletitle{Dictionary-Based Block Encoding of Sparse Matrices
  with Low Subnormalization and Circuit Depth}.
\newblock \bibinfo{journal}{\emph{Quantum}}  \bibinfo{volume}{9}
  (\bibinfo{year}{2025}), \bibinfo{pages}{1805}.
\newblock
\href{https://doi.org/10.22331/q-2025-07-22-1805}{doi:\nolinkurl{10.22331/q-2025-07-22-1805}}


\bibitem[Yang et~al\mbox{.}(2026)]%
        {yangEqualitySaturationQuantum2026}
\bibfield{author}{\bibinfo{person}{Ganxiang Yang}, \bibinfo{person}{Paige
  Raun}, \bibinfo{person}{Runzhou Tao}, {and} \bibinfo{person}{Ronghui Gu}.}
  \bibinfo{year}{2026}\natexlab{}.
\newblock \showarticletitle{Equality Saturation for Quantum Circuit
  Optimization}. In \bibinfo{booktitle}{\emph{ACM SIGPLAN Conference on
  Programming Language Design and Implementation}}. \bibinfo{pages}{174--199}.
\newblock
\href{https://doi.org/10.1145/3808254}{doi:\nolinkurl{10.1145/3808254}}


\bibitem[Yuan(2026)]%
        {yuanCobbleCompilingBlock2026}
\bibfield{author}{\bibinfo{person}{Charles Yuan}.}
  \bibinfo{year}{2026}\natexlab{}.
\newblock \showarticletitle{Cobble: Compiling Block Encodings for Quantum
  Computational Linear Algebra}. In \bibinfo{booktitle}{\emph{ACM SIGPLAN
  Conference on Programming Language Design and Implementation}}.
  \bibinfo{pages}{200--225}.
\newblock
\href{https://doi.org/10.1145/3808255}{doi:\nolinkurl{10.1145/3808255}}


\bibitem[Zamora et~al\mbox{.}(2026)]%
        {zamoraQuantumLatticeBoltzmann2026}
\bibfield{author}{\bibinfo{person}{Antonio David~Bastida Zamora},
  \bibinfo{person}{Ljubomir Budinski}, \bibinfo{person}{Valtteri Lahtinen},
  {and} \bibinfo{person}{Pierre Sagaut}.} \bibinfo{year}{2026}\natexlab{}.
\newblock \showarticletitle{Quantum Lattice Boltzmann Method for Several Time
  Steps: A Local Carleman Linearization Algorithm}.
\newblock \bibinfo{journal}{\emph{Physical Review E}} \bibinfo{volume}{113},
  \bibinfo{number}{3} (\bibinfo{year}{2026}), \bibinfo{pages}{035307}.
\newblock
\href{https://doi.org/10.1103/kbx3-c35d}{doi:\nolinkurl{10.1103/kbx3-c35d}}


\end{thebibliography}

\newpage
\appendix
\raggedbottom
\section{Full List of Rewrite Rules}
\label{sec:appendix-rules}
{\footnotesize
\renewcommand{\arraystretch}{1.15}
\begin{longtable}{@{}>{\raggedright}p{0.16\linewidth} >{\raggedright}p{0.41\linewidth}
  >{\raggedright\arraybackslash}p{0.22\linewidth} l@{}}
\caption{Algebraic rewrite rules of MARBLE\@. $\Sigma_x$ and $\bigoplus_x$ denote a sum and a
  direct sum over $x$ respectively, and every rule written for one side of $\mmul$ also holds for the other.}
\label{tab:appendix-rules} \\
\toprule
Rule & Rewrite & Side condition & Stage \\
\midrule
\endfirsthead
\toprule
Rule & Rewrite & Side condition & Stage \\
\midrule
\endhead
\bottomrule
\endlastfoot
Pinned-sum collapse & $\Sigma_{x \in \sigma}\, [x \mathrel{\texttt{==}} e] \cdot s \rightarrow [e \in \sigma] \cdot s[x := e]$
  & $e$ does not read $x$ & Reduction \\
Inner-sum product & $\comp{\Sigma_k\, a \cdot b} \rightarrow \comp{a \mid i, k} \mmul \comp{b \mid k, j}$
  & $a$ does not read $j$, $b$ does not read $i$ & Reduction \\
Sum sinking & $\Sigma_x \comp{g \cdot h} \rightarrow \comp{(\Sigma_x\, g) \cdot h}$
  & $h$ does not read $x$, $g$ does not read both $i$ and $j$ & Reduction \\
Sum lifting & $\comp{s \cdot \Sigma_x\, g} \rightarrow \Sigma_x \comp{s \cdot g}$
  & $g$ reads both $i$ and $j$ & Reduction \\
Entry sum & $\comp{p + q} \rightarrow \comp{p} + \comp{q}$
  & the entry reads both $i$ and $j$ & Reduction \\
Distribution & $\comp{p \cdot (q + r)} \rightarrow \comp{p \cdot q + p \cdot r}$
  & $q + r$ reads both $i$ and $j$ & Reduction \\
Exponential split & $\comp{c^{\,p + q}} \rightarrow \comp{c^{\,p} \cdot c^{\,q}}$, and likewise for $e^{p + q}$ and
  $\mathrm{phase}(p + q, m)$
  & $p$ and $q$ read different indices, $c \neq 0$ is a constant or $p, q \ge 0$ & Reduction \\
Constant extraction & $\comp{k \cdot s} \rightarrow k \cdot \comp{s}$
  & $k$ reads no index & Reduction \\
Scale extraction & $\comp{u(i) \cdot s} \rightarrow \comp{u(i)\, [i \mathrel{\texttt{==}} j]} \mmul \comp{s}$, and likewise for $v(j)$
  & the matrix is not diagonal & Both \\
Kronecker split & $\comp{f(i_1, j_1) \cdot g(i_2, j_2)} \rightarrow \comp{f} \otimes \comp{g}$
  & no factor reads both $(i_1, j_1)$ and $(i_2, j_2)$ & Both \\
Index-map extraction & $\comp{s(A(i), j)} \rightarrow \comp{[A(i) - \ell \mathrel{\texttt{==}} k] \mid i, k} \mmul
  \comp{s(k + \ell, j) \mid k, j}$
  & $A(i)$ has a known range starting at $\ell$ and is not $a \cdot i + b$ & Reduction \\
\midrule
Matrix distribution & $(A + B) \mmul C \rightarrow A \mmul C + B \mmul C$, and likewise for $\otimes$ and $\Sigma_x$
  & none & Reduction \\
Sum narrowing & $\Sigma_x (A \mmul B) \rightarrow (\Sigma_x A) \mmul B$, likewise for $\otimes$, and
  $\Sigma_{x \in \sigma} A \rightarrow |\sigma| \cdot A$
  & the operand taken out does not read $x$ & Reduction \\
Selector pushing & $\bigoplus_t (A \mmul B) \rightarrow (\bigoplus_t A) \mmul (\bigoplus_t B)$,
  $\bigoplus_t (A \otimes B) \rightarrow (\bigoplus_t (A \otimes I)) \mmul (\bigoplus_t (I \otimes B))$, and
  $\bigoplus_t \comp{s} \rightarrow \comp{\Sigma_t\, [i_0 \mathrel{\texttt{==}} t]\, [t \mathrel{\texttt{==}} j_0]\, s}$
  & none & Reduction \\
Identity removal & $I \mmul A \rightarrow A$
  & none & Reduction \\
Scale floating & $(s \cdot A) \mmul B \rightarrow s \cdot (A \mmul B)$, $\Sigma_x (s \cdot A) \rightarrow s \cdot \Sigma_x A$,
  and $s \cdot (t \cdot A) \rightarrow (s t) \cdot A$
  & $s$ does not read $x$ & Reduction \\
Interchange & $(A \otimes B) \mmul (C \otimes D) \rightarrow (A \mmul C) \otimes (B \mmul D)$
  & sizes of $A$ and $C$ match, and of $B$ and $D$ & Reduction \\
\midrule
Scale split & $\comp{(p + q) \cdot s} \rightarrow \comp{p \cdot s} + \comp{q \cdot s}$
  & $p + q$ reads only $i$ or only $j$ & Exploration \\
Diagonal factors & $\comp{f \cdot g \cdot [i \mathrel{\texttt{==}} j]} \rightarrow \comp{f\, [i \mathrel{\texttt{==}} j]} \mmul \comp{g\, [i \mathrel{\texttt{==}} j]}$
  & none & Exploration \\
Side cut & $\comp{\mathrm{phase}(p + q, m)} \rightarrow \comp{\mathrm{phase}(p, m) \cdot \mathrm{phase}(q, m)}$, and likewise
  for $(-1){}^{p + q}$
  & none, even where $p$ and $q$ read the same index & Exploration \\
Bitwise brackets & $\comp{[e(i, j) \mathrel{\texttt{==}} 0]} \rightarrow \comp{\prod_{k < w} [e_k(i, j) \mathrel{\texttt{==}} 0]}$, where $e_k$ reads bit $k$
  of $i$ and $j$
  & $e$ is built from $i$ and $j$ by $\&$ and $\mathrm{xor}$, sizes are $2^w$ & Exploration \\
Image extraction & as index-map extraction
  & $A(i)$ is one-to-one with a known range & Exploration \\
\midrule
Bracket signs & $\comp{[a \mathrel{\texttt{==}} b] \cdot s} \rightarrow \comp{\tfrac{1}{2} (1 + (-1){}^{a + b}) \cdot s}$
  & $a, b \in \{0, 1\}$ & Exploration \\
Windows & $S \rightarrow G \mmul (I \otimes K) \mmul G^{\dagger}$, with one block $K$ on each window of $w$ nodes
  & $S$ is a periodic stencil with constant coefficients, and every offset $\delta$ of $S$ has $|\delta| < w$
  & Exploration \\
Early split & $\comp{\sum_t p_t} \rightarrow \sum_t \comp{p_t}$, each $\comp{p_t}$ then Kronecker split
  & some $p_t$ splits & Exploration \\
Order split & $\Sigma_k \Sigma_l\, B(k, l) \rightarrow \Sigma_k\, B(k, k) + \Sigma_k \Sigma_l\, [k < l]\, (B(k, l) + B(l, k))$
  & $k$ and $l$ range over the same set & Exploration \\
Offsets & $\comp{g(i - j)} \rightarrow \Sigma_{d}\, g(d) \cdot \comp{[i \mathrel{\texttt{==}} j + d]}$, where $d$ runs from $1 - N$ to
  $N - 1$
  & $i, j : \kw{Range}\ N$, and the entry reads $i$ and $j$ only through $i - j$ & Exploration \\
Merging & $L \mmul R_1 + L \mmul R_2 \rightarrow L \mmul (R_1 + R_2)$, and likewise for $\otimes$
  & $R_1$ and $R_2$ have fixed sizes & Exploration \\
Common scale & $s \cdot A + s \cdot B \rightarrow s \cdot (A + B)$
  & none & Exploration \\
Like terms & $s_1 \cdot B + s_2 \cdot B \rightarrow (s_1 + s_2) \cdot B$
  & $s_1 + s_2$ is never $0$, or always $0$ and the terms are dropped & Exploration \\
Family weights & $\Sigma_d \comp{w(d) \cdot s} \rightarrow \Sigma_d\, w(d) \cdot \comp{s}$
  & $w$ reads only $d$ & Exploration \\
\end{longtable}
}

\section{Full List of Circuit Constructions}
\label{sec:appendix-constructions}
{\footnotesize
\renewcommand{\arraystretch}{1.0}
\begin{longtable}{@{}>{\raggedright}p{0.19\linewidth} >{\raggedright}p{0.5\linewidth}
  >{\raggedright\arraybackslash}p{0.24\linewidth}@{}}
\caption{Circuit constructions of MARBLE with subnormalizations $\alpha$.
In the table, $c$ is a constant and the data register is assumed to contain $m$ values on $w$ qubits.
Fallback methods are used as a last resort.}
\label{tab:appendix-constructions} \\
\toprule
Construction & Matrices it builds & $\alpha$ \\
\midrule
\endfirsthead
\toprule
Construction & Matrices it builds & $\alpha$ \\
\midrule
\endhead
\bottomrule
\endlastfoot
\multicolumn{3}{@{}l}{\emph{Diagonals}} \\
Constant & $\comp{c\, [i \mathrel{\texttt{==}} j]}$ & $|c|$ \\
Interval & $\comp{c\, [\ell \le i < h]\, [i \mathrel{\texttt{==}} j]}$ & $|c|$ \\
Predicate & $\comp{c\, \prod_k [\varphi_k(i)]\, [i \mathrel{\texttt{==}} j]}$, each $\varphi_k$ built from comparisons with constants by
  and, or and not & $|c|$ \\
Phase & $\comp{c\, \mathrm{phase}(a \cdot i + b, M)\, [i \mathrel{\texttt{==}} j]}$ & $|c|$ \\
Phase polynomial & $\comp{c\, \mathrm{phase}(P(i), M)\, [i \mathrel{\texttt{==}} j]}$, $P$ a polynomial of degree at least 2 & $|c|$ \\
Exponential & $\comp{c\, s^{a \cdot i}\, [i \mathrel{\texttt{==}} j]}$, and likewise for $e^{a \cdot i}$ & $|c| \max_{i < 2^w} |s^{a \cdot i}|$ \\
Comparator & $\comp{c\, (2^s \cdot i + b)\, [i \mathrel{\texttt{==}} j]}$, $b \ge 0$ & $|c|\, (2^s (m - 1) + b)$ \\
Walsh sum & $\comp{c\, g(i)\, [i \mathrel{\texttt{==}} j]}$, $g$ of degree at most 2 in the bits of $i$, with Walsh terms $w_t$
  & $|c| \sum_t |w_t|$ \\
Product over bits & $\comp{c\, \prod_k q(i_k)\, [i \mathrel{\texttt{==}} j]}$, $i_k$ bit $k$ of $i$ & $|c| \prod_k \max_b |q(b)|$ \\
Small register & $\comp{c\, h(k(i))\, [i \mathrel{\texttt{==}} j]}$, $k(i)$ a few bits or a residue of $i$ & $|c|\, \alpha_h$ \\
Table & $\comp{c\, t(i)\, [i \mathrel{\texttt{==}} j]}$ of fixed size & $|c| \max |t|$ \\
\midrule
\multicolumn{3}{@{}l}{\emph{Other matrices}} \\
Map & $\comp{c\, [i \mathrel{\texttt{==}} B(j)]}$, $B$ one-to-one, and its transpose & $|c|$ \\
Map with fibers & $\comp{c\, [A(i) \mathrel{\texttt{==}} j]}$ for $A(i) = i \gg k$, $i \bmod 2^k$ or a projection, and its transpose
  & $|c| \sqrt{F}$, $F$ the fiber size \\
Gather & $\comp{[i \mathrel{\texttt{==}} (e + t(z)) \bmod M]}$ from $(e, z)$, $z$ in a fixed range of $Z$ values, and its transpose
  & $\sqrt{Z}$ \\
Rank one & $\comp{c\, u(i)\, v(j)}$ & $|c|\, \|u\|\, \|v\|$ \\
Characters & $\comp{c\, \mathrm{phase}(a \cdot i \cdot j, N)}$, $N = 2^n$ & $|c| \sqrt{\gcd(a, N) \cdot N}$ \\
Fourier on $\mathbb{Z}_N$ & $\comp{c\, \mathrm{phase}(a \cdot i \cdot j, N)}$, any $N$ & $|c| \sqrt{\gcd(a, N) \cdot N}$ \\
Walsh--Hadamard & $\comp{c\, (-1){}^{\mathrm{parity}(i \,\&\, j)}}$ on $\kw{Range}\ 2^n$ & $|c| \sqrt{2^n}$ \\
Product over bits & $\comp{c\, \prod_k q(i_k, j_k)}$, $i_k$ and $j_k$ bit $k$ of $i$ and $j$ & $\sigma_1$ \\
Dense & any matrix of fixed size & $\sigma_1$ \\
\midrule
\multicolumn{3}{@{}l}{\emph{Built for several pieces at once}} \\
Fused sum & $\sum_k \lambda_k\, u_k v_k^{\top}$ of fixed size
  & $\max_k |\lambda_k|\, \|u_k\|\, \|v_k\|$ where orthogonal, else $\sigma_1$ \\
Compression & a matrix of fixed size, split into its symmetry sectors & $\sigma_1$ \\
Exclusive shifts & $\sum_r [i \bmod 2^b \mathrel{\texttt{==}} r]\, [j \mathrel{\texttt{==}} (i + t_r) \bmod m]$, a permutation & 1 \\
Tight gather & a gather on box cells, each node scaled by one over the square root of its cell count & 1 \\
Symbol & a periodic stencil $\sum_d c_d S^d$ on $\kw{Range}\ 2^n$ & $\max_\theta |\hat{c}(\theta)|$ \\
Two-phase blocks & $\sum_p P_p \otimes B_p$, $P_p$ the projectors onto a box of elements and onto the rest
  & $A \ge \max_p \sigma_1(B_p)$ \\
Boundary classes & $\sum_c P_c \otimes B_c$, $c$ the sides of the grid an element touches & $A \ge \max_c \sigma_1(B_c)$ \\
Anchored family & $\sum_x w(x)\, P_x$, $P_x$ a Pauli word on qubits that depend on $x$ & $\sum_x |w(x)|$ \\
Stars & a sum of $|S|$ stars on a torus, each a sum of anticommuting Pauli words with weights $r$
  & $|S|\, \|r\|_2$ \\
Lift & a translation-invariant sum of Pauli words on a torus, diagonal after a Fourier transform & $\sigma_1$ \\
\midrule
\multicolumn{3}{@{}l}{\emph{Fallbacks}} \\
Per-value loader & any diagonal $\comp{u(i)\, [i \mathrel{\texttt{==}} j]}$ & $A \ge \max |u|$ \\
Entry loader & any $\comp{f(i, j)}$ with $R$ rows and $C$ columns & $\sqrt{R C}\, \alpha_D$, $D$ the diagonal of $f$'s
  values \\
\end{longtable}
}

\section{Full List of Subnormalization Formulas}
\label{sec:appendix-alpha}
{\footnotesize
\renewcommand{\arraystretch}{1.1}
\begin{longtable}{@{}>{\raggedright}p{0.16\linewidth} >{\raggedright}p{0.4\linewidth}
  >{\raggedright\arraybackslash}p{0.36\linewidth}@{}}
\caption{Subnormalization $\alpha$ of each symbolic matrix form in MARBLE\@.
In the table, $|\tau|$ is the number of values of the domain $\tau$, and for $\Sigma\, x : \tau.\ p$ or $\bigoplus x : \tau.\ p$, $\alpha(p)$ is the largest $\alpha$ reached by $p$ over the domain of~$x$.
In the examples, $X$ and $Z$ are Pauli matrices and $X_k$ denotes $X$ on qubit $k$.}
\label{tab:appendix-alpha} \\
\toprule
Candidate & $\alpha$ & Example \\
\midrule
\endfirsthead
\toprule
Candidate & $\alpha$ & Example \\
\midrule
\endhead
\bottomrule
\endlastfoot
$c\langle M \rangle$ & $\alpha_c$, the subnormalization of the construction $c$ (Table~\ref{tab:appendix-constructions})
  & $J$ on $N$ values, $\alpha = N$ \\
$p_1 \mmul p_2$ & $\alpha(p_1) \cdot \alpha(p_2)$ & $X \mmul Z$, $\alpha = 1 \cdot 1$ \\
$p_1 \otimes \cdots \otimes p_n$ & $\alpha(p_1) \cdots \alpha(p_n)$ & $J \otimes X$, $\alpha = N \cdot 1$ \\
$p_1 + p_2$ & $\alpha(p_1) + \alpha(p_2)$ & $X + Z$, $\alpha = 1 + 1$ \\
$s \cdot p$ & $|s| \cdot \alpha(p)$, where $s$ does not read the index of a surrounding sum, and otherwise the candidate
  is $\comp{s\, [i \mathrel{\texttt{==}} j]} \mmul p$
  & $-3 \cdot X$, $\alpha = 3 \cdot 1$ \\
$\Sigma\, x : \tau.\ p$ & $|\tau| \cdot \alpha(p)$ & $X_0 + X_1 + X_2 + X_3$, $\alpha = 4 \cdot 1$ \\
$\Sigma\, x : \tau.\ w(x) \cdot p$ & $\big(\sum_{x} |w(x)|\big) \cdot \alpha(p)$
  & $X_0 + 2 X_1 + 3 X_2 + 4 X_3$, $\alpha = (1 + 2 + 3 + 4) \cdot 1$ \\
$\bigoplus x : \tau.\ p$ & $\alpha(p)$ & $X \oplus X \oplus X$, three copies of $X$ on the diagonal, $\alpha = 1$ \\
$\sum_k E(r_k, s_k) \otimes p_k$ & $\|U\|_2$ or a certified bound just above it, where $E(r, s)$ has a single 1 at
  $(r, s)$ and $U[r_k, s_k] = \alpha(p_k)$
  & $E(0, 0) \otimes X + E(1, 1) \otimes 2Z$, $U = \mathrm{diag}(1, 2)$, $\alpha = 2$ against $1 + 2$ for a plain sum \\
$p + E$, $E$ on a small invariant subspace & $\max(\alpha_C, \alpha(p))$, where $E$ has a fixed size, the subspace that
  $E$ acts on is invariant under the matrix of $p$ and its adjoint, and $\alpha_C$ is the subnormalization of the dense
  construction for the block $C$ of $p + E$ on that subspace
  & $I + E$, $E$ with a single $-2$ at $(0, 0)$, $C = (-1)$, $\alpha = \max(1, 1)$ against $1 + 2$ for a plain sum \\
\end{longtable}
}

\FloatBarrier

\section{Full Evaluation Results}
\label{sec:appendix-eval}

{\scriptsize
\setlength{\tabcolsep}{2.0pt}
\setlength{\LTpost}{3pt}
\begin{longtable}{@{}r>{\raggedright\arraybackslash}p{76pt}>{\raggedright\arraybackslash}p{51pt}>{\raggedright\arraybackslash}p{44pt}rrrcccr@{}}
\caption{Comparison of MARBLE against published block encodings of the same matrices. \\
  \emph{Speedup}: published $\alpha$ divided by MARBLE's $\alpha$.\\
  \emph{Efficient}: gates and qubits in circuit scale as $O(\polylog(N))$.\\
  \emph{Tightness}: $\alpha \le c \cdot \sigma_1$ for given $c$.\\
  \emph{Bounded}: $\sigma_1$ is $O(\polylog(N))$.\\
  \emph{Compile time}: Median of five runs in Docker image (Linux, arm64) on an Apple MacBook Air M5.}
\label{tab:eval}\\
\toprule
\# & Matrix & Instance & Baseline & Their $\alpha$ & Our $\alpha$ & Speedup & Eff. & Tightness & Bnd. & \begin{tabular}[b]{@{}r@{}}Compile\\time (s)\end{tabular} \\
\midrule
\endfirsthead
\toprule
\# & Matrix & Instance & Baseline & Their $\alpha$ & Our $\alpha$ & Speedup & Eff. & Tightness & Bnd. & \begin{tabular}[b]{@{}r@{}}Compile\\time (s)\end{tabular} \\
\midrule
\endhead
\bottomrule
\endfoot
\multicolumn{11}{@{}l}{\textit{Lattice Boltzmann: Carleman matrices}} \\
1 & Third-order Carleman, D3Q27, periodic 3D & $N = 2^{100}$, $\tau = 3/5$ & none & -- & $162.202$ & -- & \cmark & 1.317 & \cmark & 0.72 \\
2 & Third-order Carleman, D3Q27, periodic 3D & $N = 2^{100}$, $\tau = 1$ & none & -- & $97.7788$ & -- & \cmark & 1.323 & \cmark & 0.73 \\
3 & Third-order Carleman, D3Q27, quarter weights & $N = 2^{100}$ & \citeauthor{penuelDetailedAssessmentCalculating2025} & $1{,}916{,}928$ & $182.798$ & $10{,}487\times$ & \cmark & 1.099 & \cmark & 0.67 \\
4 & Third-order Carleman, D1Q3, periodic ring & $N = 2^{100}$, $\tau = 3/5$ & none & -- & $23.3266$ & -- & \cmark & 2.857 & \cmark & 0.32 \\
\midrule
\multicolumn{11}{@{}l}{\textit{Lattice Boltzmann: collision operators}} \\
5 & D3Q27 linear collision $F_1$, one site & fixed size & \citeauthor{kuklinskiEfficientBlockencodingsRequire2025} & $146$ & $5.79009$ & $25.2\times$ & \cmark & 1 & \cmark & 0.13 \\
6 & $F_1$ on a lattice & $N = 2^{100}$ & none & -- & $5.79009$ & -- & \cmark & 1 & \cmark & 0.14 \\
7 & D3Q27 quadratic tensor $F_2$ ($27 \times 729$) & fixed size & \citeauthor{penuelDetailedAssessmentCalculating2025} & $13{,}312$ & $32.0181$ & $415.8\times$ & \cmark & 1 & \cmark & 0.17 \\
8 & $F_2$ on a lattice & $N = 2^{100}$ & none & -- & $32.0181$ & -- & \cmark & 1 & \cmark & 0.17 \\
9 & $F_2$ tiled, $F_3$ ($27 \times 19683$) & fixed size & \citeauthor{penuelDetailedAssessmentCalculating2025} & $106{,}496$ & $166.371$ & $640.1\times$ & \cmark & 1 & \cmark & 0.18 \\
10 & $F_3$ on a lattice & $N = 2^{100}$ & none & -- & $166.371$ & -- & \cmark & 1 & \cmark & 0.19 \\
11 & BGK collision, D2Q9, one site & fixed size, $\omega = 1$ & \citeauthor{jenningsEndtoEndQuantumAlgorithm2026} & $1.5$$^{\dagger}$ & $1.5$ & $1.00\times$ & \cmark & 1 & \cmark & 0.12 \\
12 & BGK collision, D2Q9, one site & fixed size, $\omega = 3/2$ & \citeauthor{jenningsEndtoEndQuantumAlgorithm2026} & $2$$^{\dagger}$ & $2$ & $1.00\times$ & \cmark & 1 & \cmark & 0.12 \\
13 & BGK collision, D3Q15, one site & fixed size, $\omega = 1$ & none & -- & $1.53659$ & -- & \cmark & 1 & \cmark & 0.16 \\
14 & BGK collision, D3Q15, one site & fixed size, $\omega = 3/2$ & none & -- & $2.06246$ & -- & \cmark & 1 & \cmark & 0.18 \\
15 & BGK collision, D3Q19, one site & fixed size, $\omega = 1$ & none & -- & $1.62447$ & -- & \cmark & 1 & \cmark & 0.22 \\
16 & BGK collision, D3Q19, one site & fixed size, $\omega = 3/2$ & none & -- & $2.21051$ & -- & \cmark & 1 & \cmark & 0.25 \\
17 & BGK collision, D3Q27, one site & fixed size, $\omega = 1$ & \citeauthor{jenningsEndtoEndQuantumAlgorithm2026} & $1.83712$$^{\dagger}$ & $1.83712$ & $1.00\times$ & \cmark & 1 & \cmark & 0.11 \\
18 & BGK collision, D3Q27, one site & fixed size, $\omega = 3/2$ & \citeauthor{jenningsEndtoEndQuantumAlgorithm2026} & $2.56039$$^{\dagger}$ & $2.56039$ & $1.00\times$ & \cmark & 1 & \cmark & 0.12 \\
19 & BGK collision, D2Q9, lattice & $N = 2^{100}$, $\omega = 1$ & none & -- & $1.5$ & -- & \cmark & 1 & \cmark & 0.13 \\
20 & BGK collision, D2Q9, lattice & $N = 2^{100}$, $\omega = 3/2$ & none & -- & $2$ & -- & \cmark & 1 & \cmark & 0.14 \\
21 & BGK collision, D3Q15, lattice & $N = 2^{100}$, $\omega = 1$ & none & -- & $1.53659$ & -- & \cmark & 1 & \cmark & 0.20 \\
22 & BGK collision, D3Q15, lattice & $N = 2^{100}$, $\omega = 3/2$ & none & -- & $2.06246$ & -- & \cmark & 1 & \cmark & 0.24 \\
23 & BGK collision, D3Q19, lattice & $N = 2^{100}$, $\omega = 1$ & none & -- & $1.62447$ & -- & \cmark & 1 & \cmark & 0.28 \\
24 & BGK collision, D3Q19, lattice & $N = 2^{100}$, $\omega = 3/2$ & none & -- & $2.21051$ & -- & \cmark & 1 & \cmark & 0.35 \\
25 & BGK collision, D3Q27, lattice & $N = 2^{100}$, $\omega = 1$ & none & -- & $1.83712$ & -- & \cmark & 1 & \cmark & 0.12 \\
26 & BGK collision, D3Q27, lattice & $N = 2^{100}$, $\omega = 3/2$ & none & -- & $2.56039$ & -- & \cmark & 1 & \cmark & 0.13 \\
27 & Second-order Carleman, D2Q9 collision, one site & fixed size, $\omega = 1$ & \citeauthor{zamoraQuantumLatticeBoltzmann2026} & not stated & $12.1395$ & -- & \cmark & 1 & \cmark & 1.5 \\
28 & Second-order Carleman, D2Q9 collision, one site & fixed size, $\omega = 3/2$ & \citeauthor{zamoraQuantumLatticeBoltzmann2026} & not stated & $18.201$ & -- & \cmark & 1.120 & \cmark & 2.2 \\
29 & Second-order Carleman, D2Q9 collision, lattice & $N = 2^{100}$, $\omega = 1$ & none & -- & $12.1395$ & -- & \cmark & 1 & \cmark & 1.8 \\
30 & Second-order Carleman, D2Q9 collision, lattice & $N = 2^{100}$, $\omega = 3/2$ & none & -- & $18.201$ & -- & \cmark & 1.120 & \cmark & 2.6 \\
31 & D2Q9 quadratic tensor ($9 \times 81$) & fixed size & \citeauthor{jenningsEndtoEndQuantumAlgorithm2026} & $6$$^{\dagger}$ & $6$ & $1.00\times$ & \cmark & 1 & \cmark & 0.13 \\
32 & D2Q9 quadratic tensor, lattice & $N = 2^{100}$ & none & -- & $6$ & -- & \cmark & 1 & \cmark & 0.13 \\
33 & D2Q5 advection--diffusion equilibrium, one site & fixed size, $u_x = 0$, $u_y = 0$ & \citeauthor{xuImprovedQuantumLattice2025} & not stated & $0.471405$ & -- & \cmark & 1 & \cmark & 0.10 \\
34 & D2Q5 advection--diffusion equilibrium, one site & fixed size, $u_x = 1/10$, $u_y = -1/5$ & \citeauthor{xuImprovedQuantumLattice2025} & not stated & $0.497214$ & -- & \cmark & 1 & \cmark & 0.10 \\
35 & D2Q5 advection--diffusion equilibrium, lattice & $N = 2^{100}$, $u_x = 0$, $u_y = 0$ & none & -- & $0.471405$ & -- & \cmark & 1 & \cmark & 0.10 \\
\midrule
\multicolumn{11}{@{}l}{\textit{Lattice Boltzmann: Carleman time-marching system}} \\
36 & First-order Carleman, D2Q9, periodic 2D & $N = 4$, $n_t = 2$, $h = 1/2$, $\omega = 1$ & \citeauthor{uenoDemonstrationQuantumCircuit2026} & $32$ & $2.25$ & $14.2\times$ & \cmark & unknown & \cmark & 0.22 \\
37 & First-order Carleman, D2Q9, periodic 2D & $N = 4$, $n_t = 2$, $h = 1/2$, $\omega = 5/4$ & \citeauthor{uenoDemonstrationQuantumCircuit2026} & $32$ & $2.36527$ & $13.5\times$ & \cmark & unknown & \cmark & 0.24 \\
38 & First-order Carleman, D2Q9, periodic 2D & $N = 4$, $n_t = 2$, $h = 1/2$, $\omega = 5/3$ & \citeauthor{uenoDemonstrationQuantumCircuit2026} & $32$ & $2.59824$ & $12.3\times$ & \cmark & unknown & \cmark & 0.24 \\
\midrule
\multicolumn{11}{@{}l}{\textit{Finite elements}} \\
39 & Q4 Poisson stiffness, periodic 2D & $N = 2^{100}$ & \citeauthor{sureshQuantumBlockEncodings2026} & $16/3$ & $4$ & $1.33\times$ & \cmark & 1 & \cmark & 0.14 \\
40 & Q4 Poisson stiffness, two-phase inclusion & $N = 2^{100}$ & \citeauthor{sureshQuantumBlockEncodings2026} & $16$ & $12$ & $1.33\times$ & \cmark & 1.063 & \cmark & 0.63 \\
41 & Q4 plane-stress elasticity, periodic 2D & $N = 8$ & \citeauthor{sureshQuantumBlockEncodings2026} & $6.0989$ & $4.3956$ & $1.39\times$ & \cmark & 1.500 & \cmark & 0.17 \\
42 & Q4 plane-stress elasticity, two-phase inclusion & $N = 8$ & \citeauthor{sureshQuantumBlockEncodings2026} & $19.121$ & $13.1868$ & $1.45\times$ & \cmark & 1.500 & \cmark & 0.66 \\
43 & Q4 plane-stress elasticity, half-beam bending ($m \times m$) & $8 \times 8$ mesh & \citeauthor{holscherEndtoendQuantumAlgorithm2026} & not stated & $4.3956$ & -- & \cmark & unknown & \cmark & 0.21 \\
\midrule
\multicolumn{11}{@{}l}{\textit{Finite differences}} \\
44 & Advection--diffusion, periodic 1D, lattice units & $N = 2^{100}$, $a = 4$, $b = 1$ & \citeauthor{mahmudMomentStructuredBlockEncodings2026} & $6$ & $4.6188$ & $1.30\times$ & \cmark & 1.001 & \cmark & 0.18 \\
45 & Advection--diffusion, unit interval, $h = 1/N$ & $N = 2^{100}$, $a = 4$, $b = 1$ & \citeauthor{mahmudMomentStructuredBlockEncodings2026} & $6.428 \cdot 10^{60}$ & $6.428 \cdot 10^{60}$ & $1.00\times$ & \cmark & 1 & \xmark & 0.10 \\
46 & Fourth-order first derivative, periodic & $N = 2^{100}$ & \citeauthor{helleQuantumAdvectiondiffusionSolver2026} & $3/2$ & $1.37222$ & $1.09\times$ & \cmark & 1.001 & \cmark & 0.21 \\
47 & Sixth-order first derivative, periodic & $N = 2^{100}$ & \citeauthor{helleQuantumAdvectiondiffusionSolver2026} & $11/6$ & $1.58598$ & $1.16\times$ & \cmark & 1.001 & \cmark & 0.24 \\
48 & Fourth-order first derivative, unit interval & $N = 2^{100}$ & \citeauthor{helleQuantumAdvectiondiffusionSolver2026} & $1.901 \cdot 10^{30}$ & $1.901 \cdot 10^{30}$ & $1.00\times$ & \cmark & 1.094 & \xmark & 0.12 \\
49 & Sixth-order first derivative, unit interval & $N = 2^{100}$ & \citeauthor{helleQuantumAdvectiondiffusionSolver2026} & $2.324 \cdot 10^{30}$ & $2.324 \cdot 10^{30}$ & $1.00\times$ & \cmark & 1.156 & \xmark & 0.14 \\
50 & 7-point Laplacian, periodic 3D & $N = 2^{100}$ & \citeauthor{kharaziExplicitBlockEncodings2025} & $12$$^{\dagger}$ & $12$ & $1.00\times$ & \cmark & 1 & \cmark & 0.14 \\
51 & Fourth-order second derivative, periodic & $N = 2^{100}$ & \citeauthor{mahmudMomentStructuredBlockEncodings2026} & $16/3$$^{\dagger}$ & $5.33333$ & $1.00\times$ & \cmark & 1 & \cmark & 0.15 \\
52 & 3-point Laplacian, periodic unit interval & $N = 2^{100}$ & \citeauthor{kharaziExplicitBlockEncodings2025} & $6.428 \cdot 10^{60}$$^{\dagger}$ & $6.428 \cdot 10^{60}$ & $1.00\times$ & \cmark & 1 & \xmark & 0.11 \\
53 & Black--Scholes generator, log-price & $N = 1000$, $r = 1/20$, $\sigma = 1/5$ & \citeauthor{satoExplicitBlockencodingPartial2026} & not stated & $80{,}000$ & -- & \cmark & 1.063 & \xmark & 0.17 \\
\midrule
\multicolumn{11}{@{}l}{\textit{Hamiltonians}} \\
54 & SSH chain on a ring, first quantized & $N = 2^{100}$, $v = 1$, $w = 1/2$ & \citeauthor{leeModalAnalysisQuantum2025} & $2$ & $1.5$ & $1.33\times$ & \cmark & 1 & \cmark & 0.12 \\
55 & Schwinger model hopping with gauge links, ring & 3 sites, $x = 5/2$ & \citeauthor{kanOptimizedQuantumAlgorithms2025} & $15$ & $10$ & $1.50\times$ & \cmark & unknown & \cmark & 0.16 \\
56 & Pairing on $m$ doubly degenerate levels & 3 levels & \citeauthor{liuEfficientQuantumCircuit2024} & $16$ & $6$ & $2.67\times$ & \cmark & unknown & \cmark & 0.12 \\
57 & Transverse-field Ising, ring & 100 spins, $N = 2^{100}$, $J = 1$, $h = 1$ & \citeauthor{pfeutyOnedimensionalIsingModel1970}$^{\ddagger}$ & $200$ & $141.421$ & $1.41\times$ & \cmark & 1.132 & \cmark & 0.11 \\
58 & Transverse-field Ising, $L \times L$ torus & $10 \times 10$ spins, $N = 2^{100}$, $J = 1$, $h = 1$ & \citeauthor{bloteClusterMonteCarlo2002}$^{\ddagger}$ & $300$ & $218.89$ & $1.37\times$ & \cmark & 1.031 & \cmark & 0.12 \\
59 & Transverse-field Ising, $L \times L$ torus & $4 \times 4$ spins, $N = 2^{16}$, $J = 1$, $h = 3.04438$ & \citeauthor{bloteClusterMonteCarlo2002}$^{\ddagger}$ & $80.7101$ & $56.6329$ & $1.43\times$ & \cmark & 1.121 & \cmark & 0.12 \\
60 & $J_1$--$J_2$ Heisenberg, $L \times L$ torus & $4 \times 4$ spins, $N = 2^{16}$, $J_1 = 1$, $J_2 = 1/2$ & \citeauthor{dagottoPhaseDiagramFrustrated1989}$^{\ddagger}$ & $36$ & $12$ & $3.00\times$ & \cmark & 1 & \cmark & 0.19 \\
61 & $J_1$--$J_2$ Heisenberg, $L \times L$ torus & $4 \times 4$ spins, $N = 2^{16}$, $J_1 = 1$, $J_2 = 3/10$ & \citeauthor{dagottoPhaseDiagramFrustrated1989}$^{\ddagger}$ & $31.2$ & $13.6$ & $2.29\times$ & \cmark & 1.308 & \cmark & 0.20 \\
62 & $J_1$--$J_2$ Heisenberg, $L \times L$ torus & $4 \times 4$ spins, $N = 2^{16}$, $J_1 = 1$, $J_2 = 7/10$ & \citeauthor{dagottoPhaseDiagramFrustrated1989}$^{\ddagger}$ & $40.8$ & $16.8$ & $2.43\times$ & \cmark & 1.236 & \cmark & 0.19 \\
63 & Heisenberg ring in a random field & $n$ spins & \citeauthor{childsFirstQuantumSimulation2018} & $O(n)$ & refused & -- & -- & -- & -- & -- \\
64 & Fermi--Hubbard, $L \times L$ torus, Majorana form & $32 \times 32$ sites, $N = 2^{2048}$, $t = 1$, $u = 4$ & \citeauthor{kanResourceoptimizedFaulttolerantSimulation2025a} & $5{,}120$ & $2{,}289.73$ & $2.24\times$ & \cmark & unknown & \cmark & 4.1 \\
65 & Fermi--Hubbard, $L \times L$ torus, Majorana form & $32 \times 32$ sites, $N = 2^{2048}$, $t = 1$, $u = 8$ & \citeauthor{kanResourceoptimizedFaulttolerantSimulation2025a} & $6{,}144$ & $2{,}896.31$ & $2.12\times$ & \cmark & unknown & \cmark & 4.1 \\
66 & SYK model, $q = 4$ & $M$ Majorana modes & \citeauthor{babbushQuantumSimulationSachdevYeKitaev2018} & $\approx M^{5/2} J \sqrt{3!}/96$ & refused & -- & -- & -- & -- & -- \\
67 & First-quantized kinetic energy, 2 particles in 3D & $2^3$ momenta per axis, $N = 2^{18}$ & \citeauthor{suFaultTolerantQuantumSimulations2021} & $48$ & $27$ & $1.78\times$ & \cmark & 1 & \xmark & 0.09 \\
\midrule
\multicolumn{11}{@{}l}{\textit{Markov generators}} \\
68 & M/M/1 queue & $N = 2^{100}$, $\lambda = 1$, $\mu = 3/2$ & none & -- & $5$ & -- & \cmark & 1.063 & \cmark & 0.10 \\
69 & Two queues in tandem with blocking & $N = 2^3$ per queue, $\lambda = 1$, $\mu_1 = 6/5$, $\mu_2 = 9/10$ & none & -- & $6.2$ & -- & \cmark & 1.438 & \cmark & 0.10 \\
70 & Three-queue network, feedback and blocking & $N = 2^2$ per queue, $\lambda = 1$, $r_{12} = 4/5$, $r_{13} = 2/5$, $r_{23} = 7/10$, $r_3 = 1/2$, $r_{31} = 3/10$ & none & -- & $7.4$ & -- & \cmark & 1.250 & \cmark & 0.12 \\
71 & M/E$_3$/1 queue & $N = 2^4$, $\lambda = 4/5$ & none & -- & $7.6$ & -- & \cmark & 1.063 & \cmark & 0.24 \\
72 & M/Cox$_3$/1 queue & $N = 2^4$, $\lambda = 4/5$ & none & -- & $9.48364$ & -- & \cmark & 11.86 & \cmark & 0.31 \\
73 & TASEP on a ring & 100 sites, $N = 2^{100}$, $p = 1$ & none & -- & $141.421$ & -- & \cmark & unknown & \cmark & 0.11 \\
74 & SIS contact process on a ring & 6 sites, $N = 2^6$, $\delta = 1$, $\lambda = 7/10$ & none & -- & $14.5986$ & -- & \cmark & unknown & \cmark & 0.10 \\
\end{longtable}
\noindent $^{\dagger}$\,Equal to $\sigma_1$: no block encoding has a smaller $\alpha$. \qquad
$^{\ddagger}$\,No explicit circuit provided by paper; textbook Pauli LCU approach was assumed.}

\end{document}